\documentclass[times,final]{elsarticle}

\usepackage{jcomp}
\usepackage{framed,multirow}

\usepackage{amssymb}
\usepackage{latexsym}

\usepackage{amsmath}
\DeclareMathOperator{\sign}{sgn} 

\usepackage{url}
\usepackage{xcolor}
\definecolor{newcolor}{rgb}{.8,.349,.1}

\usepackage{amsmath}
\usepackage{algorithm}
\usepackage{algorithmic}
\usepackage{subcaption}

\usepackage{amsthm}
\theoremstyle{remark}
\newtheorem*{remark}{Remark}

\DeclareGraphicsExtensions{.pdf}

\newcommand{\pop}[2]{\frac{\partial{#1}}{\partial{#2}}}
\newcommand\ddfrac[2]{\frac{\displaystyle #1}{\displaystyle #2}}

\newcommand{\bea}{\begin{equation}\begin{aligned}}
\newcommand{\eea}{\end{aligned}\end{equation}}

\journal{Journal of Computational Physics}

\begin{document}

\verso{Alam{\'e} et al.}

\begin{frontmatter}

\title{A variational level set methodology without reinitialization for the prediction of equilibrium interfaces over arbitrary solid surfaces}

\author[1]{Karim \snm{Alam{\'e}}}
\author[1]{Sreevatsa \snm{Anantharamu}}
\author[1]{Krishnan \snm{Mahesh}\corref{cor1}}
\cortext[cor1]{Corresponding author: 
  Tel.: +1-612-624-4175;
  Email: kmahesh@umn.edu;}  

\address[1]{Department of Aerospace Engineering and Mechanics, University of Minnesota, Minneapolis, MN 55455, USA}

\received{19 July 2019}
\accepted{12 December 2019}
\availableonline{xxxx}
\communicated{F. Gibou}

\begin{abstract}
A robust numerical methodology to predict equilibrium interfaces over arbitrary solid surfaces is developed. The kernel of the proposed method is the distance regularized level set equations (DRLSE) with techniques to incorporate the no-penetration and volume-conservation constraints. In this framework, we avoid reinitialization that is typically used in traditional level set methods. This allows for a more efficient algorithm since only one advection equation is solved, and avoids numerical error associated with the re-distancing step. A novel surface tension distribution, based on harmonic mean, is prescribed such that the zero level set has the correct liquid-solid  surface tension value. This leads to a more accurate prediction of the triple contact point location. The method uses second-order central difference schemes which facilitates easy parallel implementation, and is validated by comparing to traditional level set methods for canonical problems. The application of the method in the context of Gibbs free energy minimization, to obtain liquid-air interfaces is validated against existing analytical solutions. The capability of the methodology to predict equilibrium shapes over both structured and realistic rough surfaces is demonstrated. 
\end{abstract}


\end{frontmatter}



\section{Introduction}
\begin{figure}
 \begin{center}
  \includegraphics[scale=0.3]{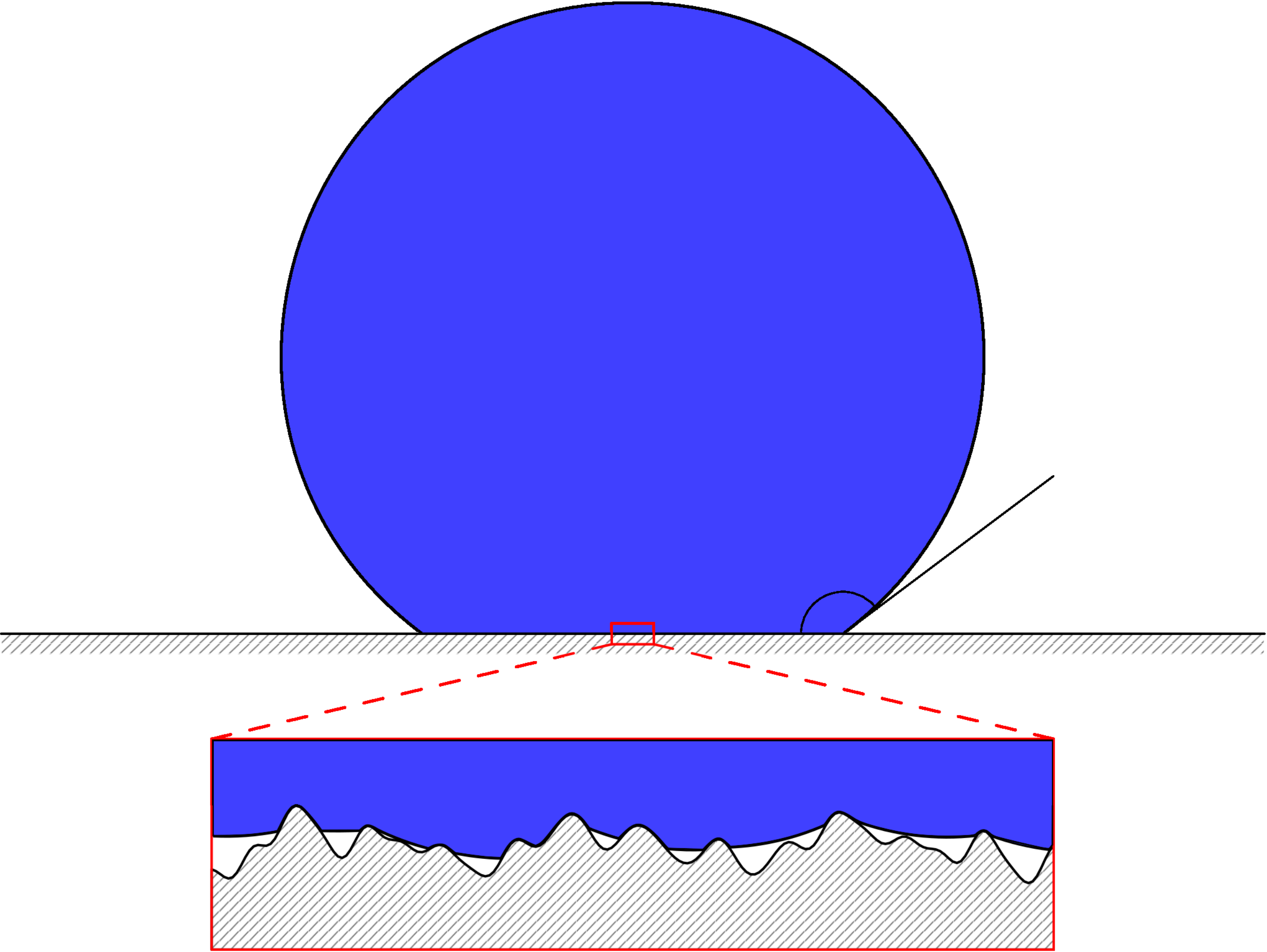}
  \put(-230,170){$(a)$}
  \put(-90,65){$\theta_Y$}
  \includegraphics[scale=0.24]{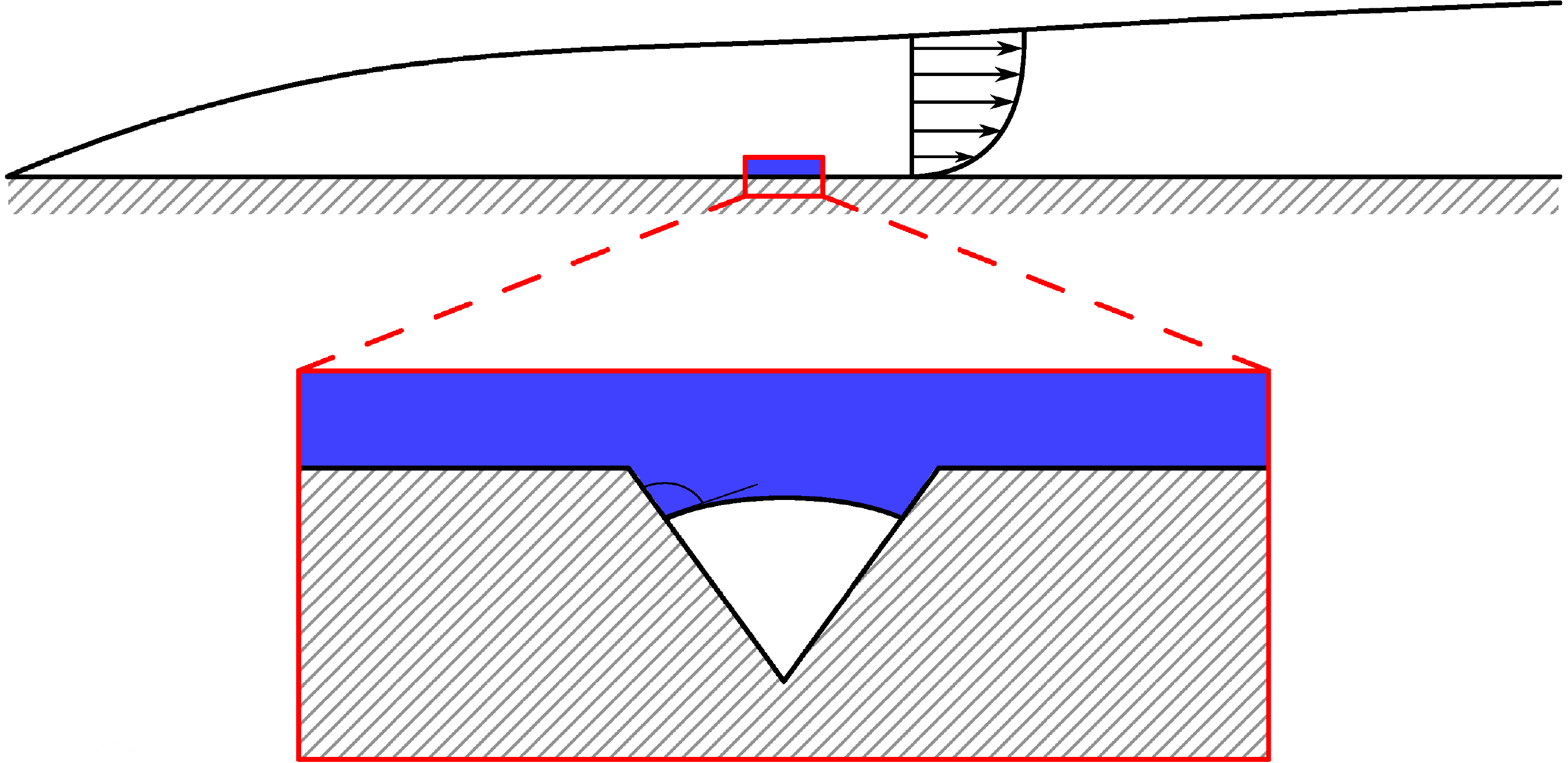}
  \put(-200,170){$(b)$}
  \put(-115,37){$\theta_Y$}
 \end{center}
  \caption{$(a)$ A schematic of a liquid droplet on a solid substrate and a zoomed in view of the liquid-solid contact showing air pockets where the surface roughness is more apparent. $(b)$ a schematic of a boundary layer with a zoom in view depicting a vapor pocket that grows due to a pressure gradient across a surface cavity. Young's contact angle is given by $\theta_Y$.}
 \label{fig:zoomin_roughness}
\end{figure}
The traditional level set (LS) formulation of moving fronts as devised by \cite{osher1988fronts}, has been applied to a wide variety of problems \citep{malladi1995shape,sethian1999level,osher2006level}. This is due to its relative ease of implementation, simplicity in applying Boolean operations, and ability to handle surface pinching or merging without the need for direct user intervention or complicated algorithms that detect when to perform `surgery'. Implementing the traditional level set methods (LSM) requires the use of upwind schemes to maintain numerical stability. As the LS evolves with time, the values of the level set function (LSF) begin to drift away from the original signed distance function (SDF), and the LSF often becomes steep or flat near the zero LS which affects stability. One remedy is to reinitialize the LS equation to restore the LSF to an SDF. This process is problem-dependent; the reinitialization needs to be done occasionally if the approximation of the LS is not sensitive to its approximation of an SDF. Otherwise, the reinitialization needs to be done every iteration which can get computationally expensive. The first reinitialization method was proposed by \cite{chopp1993} which directly computes the SDF. The process is computationally expensive and is restricted to a band of points near the zero LS front. This leads to difficulties in locating and discretizing the front. The method proposed by \cite{sussman1994level} solves the reinitialization equation iteratively but fails when the LSF is far from an SDF. \cite{peng1999pde} proposed a model that addresses the issue; however, numerical error tends to shift the interface slightly from its original position \citep{osher2006level}.

In the context of variational LS formulations, \cite{li2010distance} proposed an energy functional that regularizes the distance function as the LS deviates from an SDF. This is known as the distance regularized level set evolution (DRLSE) where the regularization term acts as a penalty to the deviation, thus maintaining an SDF without the need for an explicit reinitialization step. It was employed in the context of image segmentation but was not extended to solve general level set evolution (LSE) equations. The variational LSM without reinitialization has advantages over traditional LSM which include ease of implementation, higher efficiency due to lower computational cost, and the ability to use central difference schemes instead of upwind schemes. The LSE equations can be categorized into a PDE-based LSM (derived from motion equations) and variational LSM (derived by minimizing a certain energy functional) both of which can lead to similar governing equations while keeping the final steady-state solution the same.

The main goal of this paper is to devise an easy, fast, accurate and robust method for calculating the equilibrium position of a liquid meniscus over an arbitrary rough surface. Having a model to predict the equilibrium position of an interface and its failure criterion is of great importance to numerical simulations and experiments. Figure \ref{fig:zoomin_roughness}$(a)$ shows a schematic of a liquid droplet sitting on a solid substrate with a close-up of the surface and figure \ref{fig:zoomin_roughness}$(b)$ shows a trapped vapor bubble growing out of a cavity due to a pressure gradient present in the boundary layer. Note that the problem can be broken into a macroscopic and a microscopic one.  A macroscopic problem relies on predicting the liquid drop shape at equilibrium while conserving its volume. A microscopic problem deals with finding the equilibrium position of the meniscus on a rough substrate given an external pressure. A robust superhydrophobic surface should be designed to make a non-wetted stable Cassie state \citep{casbax} and increase the critical pressure. In the context of cavitation, predicting nucleation sites and the critical pressure that causes vapor pockets to expand beyond the envelope of the cavity into the liquid is of great importance, where the presence of a hydrophobic surface would cause heterogeneous nucleation sites \citep{brennen2014cavitation}. \cite{carbone2005} introduced an analytical model for a simple cosine substrate and then a numerical model to predict interface shape over a two-dimensional ($2$D) rough periodic substrate defined by a fractal dimension \citep{bottiglione2012role}. Extension to higher dimensions was not explored. \cite{chen2005anisotropy} explored the effect of anisotropy of roughness on wetting using Surface Evolver \citep{brakke1992surface}. The Surface Evolver is three-dimensional ($3$D) but relies on triangulating the interfacial surface. This may require special attention during pinching and breakup where elements need to be added, deleted or merged. Also, the rough substrate was modeled as a simple grooved channel therefore it is not clear whether a more complicated geometry can be incorporated. 

\cite{zhao1998capturing} used the traditional LSM to capture the behavior of bubbles drops on a macroscopic scale, details of the interface shape and contact angles were not compared quantitatively to existing analytical solutions, but their qualitative results were encouraging. Our approach uses a variational level set methodology without reinitialization to capitalize on its advantages introduced earlier. The regularization term is based on the algorithm proposed by \cite{li2010distance} which was only applied in the context of image segmentation to the best of our knowledge. We show that the DRLSE can be used in the context of both PDE-based and variational derived LSM. The algorithm minimizes the Gibbs free energy which is written in a general form that allows for the control of the dissolved gas saturation levels. The results show that the proposed model is effective in handling complicated geometries and can accurately predict interfacial equilibrium positions. 

The paper is organized as follows: the level set formulation, terminology, and notation, in both the traditional and variational sense, is presented in \S\ref{sec:method}. A general framework for the Gibbs free energy equation, based on \cite{xiang2017ultimate}, written in the variational LS framework is introduced in \S\ref{sec:model_gibbs}; it allows for prescribing an external pressure and dissolved gas saturation levels. The algorithm and numerical implementation is presented in \S\ref{sec:num_imp}. It is validated using canonical level set test cases (e.g. curvature-driven flow and motion in the normal direction) in \S\ref{sec:validation}. This is crucial since minimizing the LSF of the Gibbs free energy leads to a governing evolution equation that can be written as a combination of motion in the normal direction with curvature-driven diffusion. The Gibbs energy model is validated with the analytic solution of the Young-Laplace equation and the solution over a cosine substrate derived by \cite{carbone2005}. Results of our numerical experiments to highlight the robustness of the algorithm are shown in \S\ref{sec:results}. The numerical experiments explore the equilibrium shape of a drop/interface over a variety of rough substrates that include longitudinal grooves, posts, and realistically rough surfaces. The paper is summarized in \S\ref{sec:summary}. 

\section{Level Set Formulation}\label{sec:method}
\subsection{Traditional Level Set Formulation}
\label{subsec:traditional_levset}
Traditional level set formulation of moving fronts introduces an implicit level set function $\phi(\mathbf{x},t)$ such that the zero level set contour represents the required interface $\Gamma = \{ \mathbf{x} \, | \, \phi(\mathbf{x},t)=0 \}$. Given the interface $\Gamma(t)$ bounding an open region $\Omega^+$, we wish to compute its evolution under a velocity field $\mathbf{u}(\mathbf{x},t)$. This velocity can be a function of the interfacial geometry (e.g. normals or curvature), position, time, or external physics. The level set function $\phi$ has the following properties:
\begin{align}
 & \phi(\mathbf{x},t) > 0 \, \text{ for $\mathbf{x} \in \Omega^+$} \\
 & \phi(\mathbf{x},t) = 0 \, \text{ for $\mathbf{x} \in \partial \Omega$} \\
 & \phi(\mathbf{x},t) < 0 \, \text{ for $\mathbf{x} \in \Omega^-$} \, ,
\end{align}
where $\Omega^+$ is the interior of the region bounded by $\Gamma$, $\partial \Omega:=\Gamma$, and $\Omega^-$ the exterior region of $\Gamma$ given by its complement of $\Omega^+ \cup \partial \Omega$. The normal to any level set is given by 
\begin{equation}
  \mathbf{N} = \frac{\nabla \phi(\mathbf{x},t)}{|\nabla \phi(\mathbf{x},t)|} \, ,
\end{equation}
and the corresponding curvature of the interface is given by the divergence of its normal defined as:
\begin{equation}
  \kappa = \nabla \cdot \bigg( \frac{\nabla \phi(\mathbf{x},t)}{|\nabla \phi(\mathbf{x},t)|} \bigg) \, .
\end{equation}
The area (volume in $3$D) of the domain enclosed by $\Omega$: 
\begin{equation}
\label{eq:area}
 \text{area} (\Omega) = \int_{\Omega} H \big(\phi(\mathbf{x},t) \big) \, \mathrm{d} \mathbf{x} \, ,
\end{equation}
where $H \big( \phi(\mathbf{x},t) \big)$ is the Heaviside function
\begin{equation}
 H \big( \phi(\mathbf{x},t) \big) = \begin{cases}
			1, & \text{if $\phi(\mathbf{x},t) \ge 0$} \\
			0, & \text{if $\phi(\mathbf{x},t) < 0$} \, .
		       \end{cases}
\end{equation}
The directional derivative of $H \big( \phi(\mathbf{x},t) \big)$ in the normal direction $\mathbf{N}$ is given by the Dirac delta function:  
\begin{equation}
 \hat \delta(\mathbf{x},t) = \nabla H \big( \phi(\mathbf{x},t) \big) \cdot \mathbf{N}
\end{equation}
such that it is non-zero on $\partial \Omega$ where $\phi(\mathbf{x},t)=0$. This can be rewritten as
\begin{equation}
 \hat \delta(\mathbf{x},t) = H' \big( \phi(\mathbf{x},t) \big) \nabla \phi(\mathbf{x},t) \cdot \mathbf{N} = H'(\phi(\mathbf{x},t)) |\nabla \phi(\mathbf{x},t)| \, ,
\end{equation}
where the one-dimensional delta function is defined as $\delta \big( \phi(\mathbf{x},t) \big) = H' \big( \phi(\mathbf{x},t) \big)$ in a distribution sense. The length (surface area in $3$D) of an interface $\Gamma(t)$ can therefore be directly found using the surface integral: 
\begin{equation}
\label{eq:length}
 \text{length} \, (\Gamma) = \int_{\Omega} \delta \big( \phi(\mathbf{x},t) \big) |\nabla \phi(\mathbf{x},t)| \, \mathrm{d} \mathbf{x} \, .
\end{equation}

The motion of the interface is analyzed by convecting all the values of the level set function $\phi(\mathbf{x},t)$ with the velocity $\mathbf{u}(\mathbf{x},t)$ according to a general equation of motion given by: 
\begin{equation}
\label{eq:conv_diff}
 \pop{\phi}{t}+ \mathbf{u} \cdot \nabla \phi = \kappa |\nabla \phi| \, ,
\end{equation}
where the right-hand side determines the rate of diffusion. This will be referred to as a PDE-based LSM. The above equation can be expressed in operator form as: 
\begin{equation}
\label{eq:operator_form}
 \pop{\phi}{t} = L[\phi] \,,
\end{equation}
where $L[\phi]$ is the operator approximating all of the right hand side of the equation that is discretized using an appropriate numerical scheme.

\subsubsection{Reinitializing Signed Distance Functions}
\label{subsec:reinitialization}
As the level set $\phi(\mathbf{x},t)$ evolves with time, the values of $\phi(\mathbf{x},t)$ begin to drift away from the originally signed distance function (SDF). The level sets deteriorate to form shocks, flat or steep shapes. One remedy is to reinitialize the level set equation in order to restore $\phi(\mathbf{x},t)$ back to an SDF. 
\cite{sussman1994level} proposed the reinitialization equation given by: 
\begin{equation}
 \pop{\phi}{t} + S(\phi_o)(|\nabla \phi|-1) = 0 \, ,
 \label{eq:reinit}
\end{equation}
where $S(\phi_o)$ is a sign function that is $1$ in $\Omega^+$, $0$ on $\partial \Omega$ and $-1$ on $\Omega^-$. However, this may cause a circular dependency since the values near the interface in $\Omega^+$ use $\phi(\mathbf{x},t)$ in $\Omega^-$ for the boundary points and vice versa. This balances out if we assume that $\phi$ remains smooth at all times, which is not always the case. If $\phi$ is not smooth, or is steeper on one side, then the circular dependency may lead the reinitialization equation to move the interface to an incorrect location. Therefore, a numerically smeared sign function is typically used instead:
\begin{equation}
 S(\phi_o)=\ddfrac{\phi_o}{\sqrt{\phi^2_o + (\Delta x)^2}} \, ,
\end{equation}
where $S(\phi_o)$ remains constant for the duration on reinitialization, and has better properties if $\phi(\mathbf{x},t)$ is not smooth or if $|\nabla \phi_o|$ is far from $1$. The numerical smearing also reduces its magnitude, causing a slower propagation speed of the information near the interface thereby reducing circular dependencies. In an ideal case, the interface remains stationary during reinitialization. However, numerical error tends to shift the interface slightly \citep{osher2006level}. \cite{gibou2018review} provides a comprehensive review of the most recent advances in the traditional LSM. In summary, reinitialization still comes at a computational cost and the inconsistency between theory and implementation remains.
\subsection{Variational Level Set Formulation}  
\label{subsec:var_levset_form}
An alternative to the evolution of a PDE-based LSM can be directly derived from the problem of minimizing a certain energy functional defined by the level set function. This is known as the variational level set method \citep{zhao1998capturing}. The variational formulation is proposed as follows, let an energy functional $\mathcal{J}(\phi)$ be defined as:
\begin{equation}
\label{eq:gen_efunc}
 \mathcal{J}(\phi) = \mathcal{E}_{int}(\phi) + \mathcal{E}_{ext}(\phi),
\end{equation}
where $\mathcal{E}_{ext}(\phi)$ is the external energy functional related to a physical mechanism to be determined, and $\mathcal{E}_{int}(\phi)$ is the internal energy functional given by:
\begin{equation}
 \mathcal{E}_{int}(\phi) = \alpha \mathcal{R}_p(\phi),
\end{equation}
where $\alpha$ is a positive parameter that controls the effect of $\mathcal{R}_p(\phi)$ which penalizes the deviation of $\phi(\mathbf{x},t)$ from an SDF. To minimize the energy functional $\mathcal{J}(\phi)$, we need to find the steady-state solution of $\phi(\mathbf{x},t)$ by stepping in the direction of steepest descent of $\mathcal{J}(\phi)$. This is given by the negative Fr{\'e}chet derivative $\mathcal{J}'(\phi)$ such that
\begin{equation}
 \phi^{n+1} = \phi^n - \gamma \, \mathcal{J}' (\phi^n),
\end{equation}
where $\gamma$ is the step size taken to be as a `time-step' $dt$. We can write the evolution equation as a gradient flow that minimizes $\mathcal{J}(\phi)$:
\begin{equation}
\label{eq:var_levset}
 \pop{\phi}{t} = - \pop{\mathcal{J}}{\phi} \, ,
\end{equation}
which will be referred to as the variational-based LSM that can also be expressed in operator form as Eq. (\ref{eq:operator_form}): 
\begin{equation}
 \pop{\phi}{t} = L[\phi].
\end{equation}
The above equation highlights the fact that we can obtain the same operator for the traditional and variational formulation of the level set methodology. The key difference is how this operator is discretized and solved; this will be discussed later in \S\ref{sec:num_imp}. 
\subsubsection{Distance Regularization as an Alternative to Reinitialization}
\label{subsec:penalty_regularization}
It is crucial to keep the level set as an approximate SDF during its evolution as discussed in \S\ref{subsec:reinitialization}, where any function $\phi(\mathbf{x},t)$ satisfying $|\nabla \phi| = 1$ is an SDF \citep{arnold2012geometrical}. \cite{li2010distance} proposed an energy functional that regularizes the distance function as the level set deviates from an SDF in the vicinity of the front. The penalty functional is given by: 
\begin{equation}
 \mathcal{R}_p(\phi) = \int_{\Omega} p(|\nabla \phi|) \, \mathrm{d} \mathbf{x} \, ;
\end{equation}
this serves as a metric to characterize how close $\phi(\mathbf{x},t)$ is to an SDF. The energy density $p(|\nabla \phi|)$ is a potential function defined as:
\begin{equation}
 p(|\nabla \phi|) = \begin{cases}
                     \frac{1}{(2\pi)^2}(1-\cos(2\pi |\nabla \phi|))\, ,  & \text{if $|\nabla \phi| < 1$} \\ 
                     \frac{1}{2}(|\nabla \phi| - 1)^2 \, , & \text{if $|\nabla \phi| \ge 1$} \, .
                    \end{cases} 
\end{equation}
It is evident that $p(|\nabla \phi|)$ has two minima at $|\nabla \phi|=0$ and $|\nabla \phi|=1$.
 The function $p(|\nabla \phi|)$ is twice differentiable, this can be verified by finding the roots of its first derivative and checking the sign of the second derivative, $p'(|\nabla \phi|)$ and $p''(|\nabla \phi|)$ respectively, such that
\begin{equation}
  p'(|\nabla \phi|) = \begin{cases}
                     \frac{1}{2\pi}\sin(2\pi |\nabla \phi|) \, ,  & \text{if $|\nabla \phi| < 1$} \\ 
                     |\nabla \phi| - 1 \, ,& \text{if $|\nabla \phi| \ge 1$} 
                    \end{cases} 
\end{equation}
and
\begin{equation}
  p''(|\nabla \phi|) = \begin{cases}
                     \cos(2\pi |\nabla \phi|) \, ,  & \text{if $|\nabla \phi| < 1$} \\ 
                     1 \, , & \text{if $|\nabla \phi| \ge 1$} \, .
                    \end{cases} 
\end{equation}
The roots of $p'(|\nabla \phi|)$ are seen at the zero line intersection for $|\nabla \phi|=0,0.5$ and $1$. The sign of $p''(|\nabla \phi|)$, which is positive for $|\nabla \phi|=0$ and $1$, indicates that they are minimums. The Fr{\'e}chet derivative in the $L^2$ inner product of the regularization energy functional $\mathcal{R}_p(\phi)$ can be shown to be:
\begin{equation}
\label{eq:der_rp}
 \pop{\mathcal{R}_p}{\phi} = - \nabla \cdot (d_p(|\nabla \phi|) \nabla \phi) \, ,
\end{equation}
where $d_p$ is defined as 
\begin{equation}
 d_p(|\nabla \phi|) = p'(|\nabla \phi|)/|\nabla \phi| \, .
\end{equation}
The use of a double-well potential has more favorable properties than using a simple potential function $\frac{1}{2}(|\nabla \phi| - 1)^2$ which was initially introduced by \cite{li2005level}. The advantage is seen in the limiting values  of $|\nabla \phi|$ given by
\begin{equation}
 \lim_{|\nabla \phi| \to 0} d_p(|\nabla \phi|) = \lim_{|\nabla \phi|\to\infty} d_p(|\nabla \phi|) = 1 \, .
\end{equation}
We can verify that the boundedness is lost for the simple potential function which can lead to undesirable effects
\begin{equation}
\lim_{|\nabla \phi| \to 0} |d_p(|\nabla \phi|)| = \infty \, .
\end{equation}
Physically, in the regions where $|\nabla \phi|$ is close to zero, the diffusion rate is positive and arbitrarily large. For this case, the diffusion is backwards, which drastically increases $|\nabla \phi|$ causing oscillations to appear as peaks and valleys in the final level set $\phi$. The oscillations appear at a distance from the zero crossing of the level set but nevertheless may cause a slight shift in the interface location. 
Therefore for a double-well potential, the function $d_p(|\nabla \phi|)$ is always bounded such that
\begin{equation}
 |d_p(|\nabla \phi|)| < 1, \ \text{for all values of} \ |\nabla \phi| \in [0,\infty).
\end{equation}
For $0<|\nabla \phi|<1/2$, the diffusion rate $d_p(|\nabla \phi|)$ is positive, resulting in a forward diffusion which further decreases $|\nabla \phi|$ to zero. For $(1/2)<|\nabla \phi|<1$, the diffusion rate $d_p(|\nabla \phi|)$ is negative, resulting in a backward diffusion which increases $|\nabla \phi|$ towards unity. For $|\nabla \phi|>1$, the diffusion is positive again which leads to forward diffusion, reducing the value of $|\nabla \phi|$ towards $1$.

\section{The Model: Gibbs Free Energy}\label{sec:model_gibbs}
 Consider a thermodynamic system in its general form comprised of a surface, submerged in a bulk liquid consisting of water and dissolved gas, and entrapped air bubbles within the surface cavities consisting of vapor and free gas. The equilibrium liquid-gas interface position is obtained through the minimization of the total Gibbs free energy of a multi-component system \citep{landau1980statistical,patankar2009hydrophobicity} that is given by:
 \begin{equation}
  G_{tot} = \sum_{\iota} (U_{\iota} + p_L V_{\iota} - TS_{\iota}) + G_{int},
 \end{equation}
 where $G_{tot}$ denotes the total Gibbs free energy, $U_{\iota}$ the internal energy of the system, $p_L$ the liquid pressure, $V_{\iota}$ the volume of each phase $\iota$ in the system, $T$ the temperature, and $S_{\iota}$ the entropy of each phase. $G_{int}$ denotes the free energy of all interfaces present in the system,
 \begin{equation}
  G_{int} = \sum_{\iota} (\tau_{LG} A^{\iota}_{LG} + \tau_{SG} A^{\iota}_{SG}) + \tau_{LS} A_{LS},
 \end{equation}
 where $\tau_{LG}$, $\tau_{SG}$ and $\tau_{LS}$ represent the surface tension of the liquid-gas (LG), solid-gas (SG) and liquid-solid (LS) interfaces respectively. Similarly $A_{LS}$, $A_{SG}$ and $A_{LS}$ represent the surface areas of each corresponding phase. It can be shown that the total free energy can be written as
\begin{equation}
\label{eq:gtotal}
 G_{tot} = (p_L - p_G - p_V) V_G + \tau_{LG} (A_{LG} + A_{SG} \cos \theta_Y) + n_G R T \ln \frac{p_G}{s(p_L-p^*_V)} + n_V R T \ln \frac{p_V}{p^*_V} + G_o (p_L,T,s) \, ,
\end{equation}
a form shown by \cite{xiang2017ultimate}. Here, the first term in the right-hand side of Eq.(\ref{eq:gtotal}) is attributed to the bulk phases, where $p_G$ and $p_V$ are the entrapped gas and vapor pressure respectively, and $V_G$ is the volume of the entrapped gas. The second term represents surface tension attributed to the interface simplified using Young's equation, where $\theta_Y$ is Young's contact angle. The third term is attributed to the difference in chemical potential between the free gas in the entrapped air within the cavity and dissolved gas in the liquid phase where $n_G$ is the mole number of entrapped gas, $R$ the ideal gas constant, $T$ the temperature, $s$ the dissolved gas saturation level and $p^*_V$ the saturated vapor pressure. The fourth term is the energy contribution of unsaturated vapor where $n_v$ is the mole number of vapor in the cavity. The last term $G_o$ is the free energy of the wetted Wenzel state \citep{wenzel1936} which is a constant for a given $p_L$, $T$, and $s$ at the reference state where 
\begin{equation}
\label{eq:gwenzel}
 G_o (p_L,T,s) = n_{tot} \, \mu_{DG} + n_{\mathrm{H}_2\mathrm{O}} \, \mu_W + \tau_{LS} A_S.
\end{equation}
In Eq. (\ref{eq:gwenzel}) the total gas mole number of the system is given by $n_{tot}$ and $n_{\mathrm{H}_2\mathrm{O}}$ is the mole number of water and its vapor, $\mu_{DG}$ and $\mu_W$ are the chemical potentials of the dissolved gas and water respectively. 
Note that in order to obtain an equilibrium state, which is an energy minimum of the system, the first-order variation of the total free energy should be set to zero i.e. $\delta G_{tot}=0$. Classical geometry formulas described in \citep{frankel2011geometry,giacomello2012metastable} are used to obtain the expression 
\begin{equation}
 \delta G_{tot}= (p_L - p_G - p_V) \, \delta V_G + \tau_{LG}(\delta A_{LG} + \delta A_{SG} \cos \theta_Y) + R T \ln \frac{p_G}{s(p_L-p^*_V)} \, \delta n_G + R T \ln \frac{p_V}{p^*_V} \, \delta n_G \, .
\end{equation}
The variation with respect to the first and second term determines the shape and location of the interface since it results in the Young-Laplace equation: 
\begin{equation}
\label{eq:young_laplace}
 p_L - p_G - p_V = - \tau_{LG} \, \kappa \, .
\end{equation}
The third term gives the variation with respect to $\delta n_G$, which is the chemical equilibrium condition between the free and dissolved gas in water:
\begin{equation}
\label{eq:diss_gas}
 p_G = s(p_L-p^*_V) \, .
\end{equation}
The fourth term gives the variation with respect to $\delta n_V$, which is the equilibrium equation between vapor and water:
\begin{equation}
\label{eq:sat_pv}
p_V = p^*_V \, .
\end{equation}
 
\subsection{Overview of Gibbs Free Energy in a Variational Level Set Formulation}\label{subsec:gibbs_levset}
Define an energy functional $\mathcal{E}(\phi)$ where $\phi(\mathbf{x},t)>0$ represents the liquid phase, $\phi(\mathbf{x},t)<0$ represents the gas phase and $\phi(\mathbf{x},t)=0$ represents the liquid-gas interface. Define another level set $\psi(\mathbf{x})$ that represents the solid roughness, $\psi(\mathbf{x}) \ge 0$ for the region inside the solid substrate and $\psi(\mathbf{x}) < 0$ outside the solid. Then, $\mathcal{E}(\phi)$ based on the bulk and interfacial energies can be written as:
\begin{equation}
 \mathcal{E}(\phi) = \mathcal{E_B}(\phi) + \mathcal{E_I}(\phi) \, ,
\end{equation}
where $\mathcal{E_B}(\phi)$ and $\mathcal{E_I}(\phi)$ are the bulk and interfacial surface energy respectively. Using Eq. (\ref{eq:gtotal}), the bulk energy due to the pressure difference is given by $\mathcal{E_B}=(p_L-p_G-p_V)V_G$. Denote $\Delta p = p_G+p_V-p_L$ then substitute Eq. (\ref{eq:diss_gas}) and Eq. (\ref{eq:sat_pv}) to get $\Delta p = (1-s)(p_V-p_L)$. 
Using Eq. (\ref{eq:area}), 
\begin{equation}
 \mathcal{E_B}(\phi) = - \int_\Omega  \Delta p H \big(-\psi(\mathbf{x}) \big) H \big( \phi(\mathbf{x},t) \big) \, \mathrm{d} \mathbf{x} \, ,
\end{equation}
where $H(-\psi(\mathbf{x}))$ ensures that the pressure difference is only applied outside the solid region. The energy functional due to the interfacial surface energies from Eq. (\ref{eq:gtotal}) is given by $\mathcal{E_I} = \tau_{LG} (A_{LG} + A_{SG} \cos \theta_Y)$. Note that the contact angle satisfies $|\tau_{LS}-\tau_{SG}|=\tau_{LG} \cos \theta_Y$. If $0 \le \theta_Y \le \pi/2$, then the liquid is said to be wetted, else if $\pi/2 < \theta_Y \le \pi$, then the liquid is said to be non-wetted. The surface tension is represented as a function of $\psi(\mathbf{x})$ such that,
\begin{equation}
\label{eq:binary_tau}
 \tau(\psi) = \begin{cases}
               \tau^+ \, , & \text{for $\psi(\mathbf{x}) \ge 0$} \\ 
               \tau^- \, , & \text{for $\psi(\mathbf{x}) < 0$} \, , 	
              \end{cases}
\end{equation}
where $\tau^+=|\tau_{LS}-\tau_{SG}|$ and $\tau^-=\tau_{LG}$. Using Eq. (\ref{eq:length}), the interfacial surface energy is written as
\begin{equation}
 \mathcal{E_I}(\phi) = \int_\Omega  \tau \big( \psi(\mathbf{x}) \big) \, \delta \big( \phi(\mathbf{x},t) \big) |\nabla \phi(\mathbf{x},t)| \, \mathrm{d} \mathbf{x} \, .
\end{equation}
Since the solid does not vary temporally, substitute $\tau \big( \psi(\mathbf{x}) \big)$ for $\tau(\mathbf{x})$:
\begin{equation}
 \mathcal{E_I}(\phi) = \int_\Omega \tau(\mathbf{x}) \delta \big( \phi(\mathbf{x},t) \big) |\nabla \phi(\mathbf{x},t)| \, \mathrm{d} \mathbf{x}.
\end{equation}
The interface is not allowed to penetrate into the solid, therefore a no penetration constraint is enforced: 
\begin{equation}
 \mathcal{G}_1(\phi) = \int_\Omega H\big ( -\phi(\mathbf{x},t) \big) H \big( \psi(\mathbf{x}) \big) \, \mathrm{d} \mathbf{x} = 0 \, ;
\end{equation}
this prevents any overlap or penetration between the two regions described by $\phi(\mathbf{x},t) \ge 0 $ and $\psi(\mathbf{x}) \ge 0$. The volume of the liquid phase is required to remain conserved, therefore the second constraint imposed is volume conservation given by:
\begin{equation}
 \mathcal{G}_2(\phi) = \int_\Omega H \big( \phi(\mathbf{x},t) \big) H \big(-\psi(\mathbf{x}) \big) \, \mathrm{d} \mathbf{x} = V_o \, ,
\end{equation}
where $V_o$ is the initial volume of the liquid. This is necessary for problems which require the conservation of the volume of the bubble. For problems where we assume an infinite supply of liquid (i.e. a system where the liquid reservoir is much larger than the air cavities), this constraint is not used. In order to minimize $\mathcal{E}(\phi)$, we define the auxiliary energy functional $\mathcal{J}(\phi)$ that contains the Gibbs energy, penalty function and the two constraints, given by:
\begin{equation}
 \mathcal{J}(\phi) = \alpha \mathcal{R}_p (\phi) + \mathcal{E}(\phi) + \mu \mathcal{G}_1(\phi) + \lambda \mathcal {G}_2(\phi) \, ,
\end{equation}
where $\mathcal{R}_p(\phi)$ is the penalty term that enforces $\phi$ to be an SDF in the vicinity of the interface, $\alpha$ is the weighting and $\mu$ and $\lambda$ are the Lagrange multipliers that satisfy the two constraints mentioned above. Then the Fr{\'e}chet derivative in the $L^2$ norm  of $\mathcal{J}(\phi)$ is given by:
\begin{equation}
 \lim_{\epsilon \to 0}\,\frac{1}{\epsilon}\bigg [\mathcal{J}(\phi+\epsilon\chi)-\mathcal{J}(\phi)\bigg] = \bigg \langle \pop{\mathcal{J}}{\phi},\chi \bigg \rangle = \bigg \langle  \pop{\mathcal{R}_p}{\phi},\chi \bigg \rangle + \bigg \langle  \pop{\mathcal{E_I}}{\phi},\chi \bigg \rangle +  
 \bigg \langle  \pop{\mathcal{E_B}}{\phi},\chi \bigg \rangle + 
 \bigg \langle  \pop{\mathcal{G}_1}{\phi},\chi \bigg \rangle + 
 \bigg \langle  \pop{\mathcal{G}_2}{\phi},\chi \bigg \rangle \, .
\end{equation}
The first term in the right-hand side $\bigg \langle \pop{\mathcal{R}_p}{\phi},\chi \bigg \rangle$ is given in Eq. (\ref{eq:der_rp}). The second term can be shown to be:
\begin{equation}
 \bigg \langle  \frac{\partial \mathcal{E_I}}{\partial \phi},\chi \bigg \rangle =
 \int_{\partial \Omega} \delta \big(\phi(\mathbf{x},t)\big) \pop{\phi(\mathbf{x},t)}{n} \frac{\tau (\mathbf{x})}{|\nabla \phi(\mathbf{x},t)|} \, \chi  \, \mathrm{d} \mathbf{s} -
 \int_\Omega \delta \big( \phi(\mathbf{x},t) \big) \nabla \cdot \bigg ( \tau (\mathbf{x}) \frac{\nabla \phi(\mathbf{x},t)}{|\nabla \phi(\mathbf{x},t)|} \bigg ) \, \chi  \, \mathrm{d} \mathbf{x} \, ,
\end{equation}
where the first term is eliminated by taking $\pop{\phi}{n}=0$ at the boundaries $\partial \Omega$. The rest of the terms follow as:
\begin{equation}
 \bigg \langle  \frac{\partial \mathcal{E_B}}{\partial \phi},\chi \bigg \rangle = - \int_\Omega \Delta p \, \delta \big( \phi(\mathbf{x},t) \big) H \big(-\psi(\mathbf{x})\big) \, \chi  \, \mathrm{d} \mathbf{x} \, ,
\end{equation}
\begin{equation}
 \bigg \langle  \frac{\partial \mathcal{G}_1}{\partial \phi},\chi \bigg \rangle = - \int_\Omega \delta \big( \phi(\mathbf{x},t) \big) H \big( \psi(\mathbf{x}) \big) \, \chi  \, \mathrm{d} \mathbf{x} \, ,
\end{equation}
and
\begin{equation}
 \bigg \langle  \frac{\partial \mathcal{G}_2}{\partial \phi},\chi \bigg \rangle = \int_\Omega \delta \big( \phi(\mathbf{x},t) \big) H \big(-\psi(\mathbf{x}) \big) \, \chi  \, \mathrm{d} \mathbf{x} \, .
\end{equation}
The evolution equation given by Eq. (\ref{eq:var_levset}) can now be defined for an infinitesimal step-size as:
\begin{equation}
\label{eq:levset_evol_a}
 \pop{\phi(\mathbf{x},t)}{t} = \alpha \nabla \cdot \bigg (d_p \big( |\nabla \phi(\mathbf{x},t)| \big) \nabla \phi(\mathbf{x},t) \bigg) + \delta \big (\phi(\mathbf{x},t) \big) \, \bigg [ \nabla \cdot \bigg (\tau(\mathbf{x}) \frac{\nabla \phi(\mathbf{x},t)}{|\nabla \phi(\mathbf{x},t)|} \bigg) + \Delta p H(-\psi(\mathbf{x})) + \mu H(\psi(\mathbf{x})) - \lambda H(-\psi(\mathbf{x})) \bigg ] \, .
\end{equation}
To determine the value of the Lagrange multipliers, we enforce the fact that the constraints do not vary in time, i.e. 
\begin{align}
 & \frac{d}{dt} \, \mathcal{G}_1(\phi)  = 0 \\
 & \frac{d}{dt} \, \mathcal{G}_2(\phi)  = 0 \,.
\end{align}
Therefore, the Lagrange multipliers can be obtained as:
\begin{equation}
\label{eq:lag_mu_1}
 \mu = -\ddfrac{\int_\Omega \delta \big( \phi(\mathbf{x},t) \big) \bigg [ \alpha \nabla \cdot \bigg (d_p \big( |\nabla \phi(\mathbf{x},t)| \big) \nabla \phi(\mathbf{x},t) \bigg) + \delta \big( \phi(\mathbf{x},t) \big) \nabla \cdot \bigg( \tau (\mathbf{x}) \frac{\nabla \phi \big(\mathbf{x},t) }{|\nabla \phi(\mathbf{x},t)|} \bigg) \, \bigg] H\big(\psi(\mathbf{x})\big) \, \mathrm{d} \mathbf{x}} {\int_\Omega \delta^2 \big(\phi(\mathbf{x},t) \big) H \big(\psi(\mathbf{x})\big) \, \mathrm{d} \mathbf{x}} 
\end{equation}
and
\begin{equation}
\label{eq:lag_lambda_1}
 \lambda = \ddfrac{\int_\Omega \delta \big(\phi(\mathbf{x},t)\big) \bigg \{ \alpha \nabla \cdot \bigg (d_p \big( |\nabla \phi(\mathbf{x},t)| \big) \nabla \phi(\mathbf{x},t) \bigg) + \delta \big(\phi(\mathbf{x},t)\big) \bigg [\nabla \cdot \bigg( \tau (\mathbf{x}) \frac{\nabla \phi \big(\mathbf{x},t) }{|\nabla \phi(\mathbf{x},t)|} \bigg) + \Delta p \bigg ] \, \bigg \} H\big(-\psi(\mathbf{x})\big) \, \mathrm{d} \mathbf{x}} {\int_\Omega \delta^2 \big(\phi(\mathbf{x},t) \big) H \big(-\psi(\mathbf{x})\big) \, \mathrm{d} \mathbf{x}} \, .
\end{equation}
The above equations use the fact that 
\begin{equation}
 \label{eq:heaviside_product_approx_a}
  H(\psi(\mathbf{x}))H(-\psi(\mathbf{x})) = 0 
\end{equation}
and
\begin{equation}
\label{eq:heaviside_product_approx_b}
  H^2(\psi(\mathbf{x})) = H(\psi(\mathbf{x})) \, .
\end{equation}
Note that at steady state, equilibrium is reached (i.e. $\partial \phi / \partial t = 0$ and $\lambda=0$). Assuming no solid is present (i.e. $\mu=0$) and $\tau(\mathbf{x})=\tau_{LG}$, we get  
\begin{equation}
\label{eq:levset_evol}
 \Delta p  =  \tau_{LG} \, \nabla \cdot \bigg (\frac{\nabla \phi(\mathbf{x},t)}{|\nabla \phi(\mathbf{x},t)|} \bigg) \, ;
\end{equation}
i.e. the Young-Laplace equation in Eq. (\ref{eq:young_laplace}) is recovered.

\section{Numerical Implementation}\label{sec:num_imp}
In practice, the Heaviside function $H(\phi)$ and the delta function $\delta(\phi)$ are slightly smoothed out such that:
\begin{equation}
 H_{\epsilon}(\phi) = \begin{cases}
                        1 \, , & \text{if $\phi > \epsilon$} \\
                        \frac{1}{2} \bigg [1 + \frac{\phi}{\epsilon} + \frac{1}{\pi} \sin \bigg(\frac{\pi \phi}{\epsilon} \bigg) \bigg] \, , & \text{if $|\phi| \le \epsilon$} \\
                        0 \, , & \text{if} \, \phi < -\epsilon                  
                      \end{cases}
\end{equation}
and
\begin{equation}
 \delta_{\epsilon}(\phi) = \begin{cases}
                            0 \, , & \text{if $|\phi| > \epsilon$} \\
                            \frac{1}{2 \epsilon} \bigg [1 + \cos \bigg(\frac{\pi \phi}{\epsilon} \bigg) \bigg] \, , & \text{if $|\phi| \le \epsilon$} \, ,
                           \end{cases}
\end{equation}
where $\epsilon$ is the numerical width of $\delta(\phi)$ and $H(\phi)$ is taken to be slightly larger than the width of a grid cell ($\epsilon=1.5 \Delta x$). 
The DRLSE can be implemented with a simple finite difference scheme. The spatial gradients $\partial \phi / \partial x_i$ are obtained using a central difference scheme and the temporal partial derivative $\partial \phi / \partial t$ is approximated using forward difference. Heun's method is also implemented so that we obtain a second-order approximation with a slightly lower Courant-Friedrichs-Lewy (CFL) restriction. The evolution equation is discretized using the forward Euler as,
\begin{equation}
 \phi^{n+1}_{i,j,k} = \phi^n_{i,j,k} + dt L(\phi^n_{i,j,k}),
\end{equation}
and for Heun's method as 
\begin{align}
 & \bar{\phi}^{n+1}_{i,j,k} = \phi^n_{i,j,k} + dt L(\phi^n_{i,j,k}) \\
 & \phi^{n+1}_{i,j,k} = \frac{1}{2} \phi^n_{i,j,k} + \frac{1}{2} \bar{\phi}^{n+1}_{i,j,k} + dt L(\bar{\phi}^n_{i,j,k}).
\end{align}
The key difference here is the fact that the operator $L[\phi]$ in the variational formulation does not describe a Hamilton-Jacobi equation with a curvature regularization of the traditional formulation but a reaction-diffusion type problem. Here the LSE equation is written out explicitly as: 
\begin{equation}
\label{eq:levset_evol_2}
 \pop{\phi}{t} =\alpha \nabla \cdot (d_p(|\nabla \phi|) \nabla \phi) +  \delta (\phi) \bigg [ \tau \, \nabla \cdot \bigg( \frac{\nabla \phi}{|\nabla \phi|} \bigg)  + \nabla \tau \cdot \frac{\nabla \phi}{|\nabla \phi|} + H(-\psi) \Delta p + \mu H(\psi) - \lambda H(-\psi) \bigg ],
\end{equation}
where there is no need for reinitialization or replacing $\delta(\phi)$ with $|\nabla \phi|$. This allows for the flexibility of initializing $\phi$ as a binary function. The initial condition LS does not necessarily have to be an SDF, $|\nabla \phi|=1$ is only enforced in the vicinity of $\phi=0$ where $\delta(\phi)$ is active. It suffices to define a divergence function $div(\cdot)$ that can be used for both the regularization term and the normals to obtain curvature which is given by $\kappa=\nabla \cdot \mathbf{N}$. Some attention to detail of implementation was required for the Lagrange multipliers. The term in the denominator of the Lagrange multipliers $\mu$ given in Eq. (\ref{eq:lag_mu_1}) causes leakage since the width of $\delta(\phi)$ becomes too narrow; the term $\delta^2(\phi)H^2(\psi)$ is particularly sensitive and is replaced with $\delta(\phi)|\nabla \phi|H(\psi)$. Numerically, the assumptions in Eq. (\ref{eq:heaviside_product_approx_a}) and Eq. (\ref{eq:heaviside_product_approx_b}) do not hold. This is inconsequential when there is no requirement for volume conservation since $\lambda=0$. However for cases where volume conservation is required, it is important to keep their product explicitly in the discretized equations so that the equations balance out. Otherwise, the curvature diffusion term causes the interface to eventually disappear. Therefore, the discrete Lagrange multipliers are written as:
\begin{equation}
\label{eq:lag_mu_2}
 \mu = -\ddfrac{\int_\Omega \delta \big( \phi \big) \bigg \{ \alpha \nabla \cdot \bigg (d_p \big( |\nabla \phi| \big) \nabla \phi \bigg) + \delta \big( \phi \big) \bigg[  \tau \, \nabla \cdot \bigg( \frac{\nabla \phi}{|\nabla \phi|} \bigg)  + \nabla \tau \cdot \frac{\nabla \phi}{|\nabla \phi|} + \Delta p H(-\psi) - \lambda H(-\psi)  \bigg ] \, \bigg\} H\big(\psi\big) \, \mathrm{d} \mathbf{x}} {\int_\Omega \delta \big(\phi \big) |\nabla \phi| H \big(\psi\big) \, \mathrm{d} \mathbf{x}} 
\end{equation}
and
\begin{equation}
\label{eq:lag_lambda_2}
 \lambda = \ddfrac{\int_\Omega \delta \big(\phi\big) \bigg \{ \alpha \nabla \cdot \bigg (d_p \big( |\nabla \phi| \big) \nabla \phi \bigg) + \delta \big(\phi\big) \bigg [ \tau \, \nabla \cdot \bigg( \frac{\nabla \phi}{|\nabla \phi|} \bigg) + \nabla \tau \cdot \frac{\nabla \phi}{|\nabla \phi|} + \Delta p H(-\psi) + \mu H(\psi) \bigg ] \, \bigg \} H\big(-\psi\big) \, \mathrm{d} \mathbf{x}} {\int_\Omega \delta^2 \big(\phi \big) H^2 \big(-\psi\big) \, \mathrm{d} \mathbf{x}} \, .
\end{equation}
The terms are integrated using a trapezoidal rule.
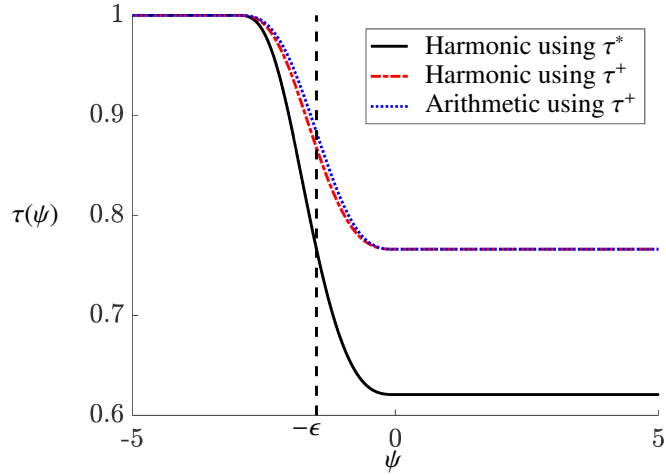
\begin{figure}
\begin{center}
  \includegraphics[scale=0.3]{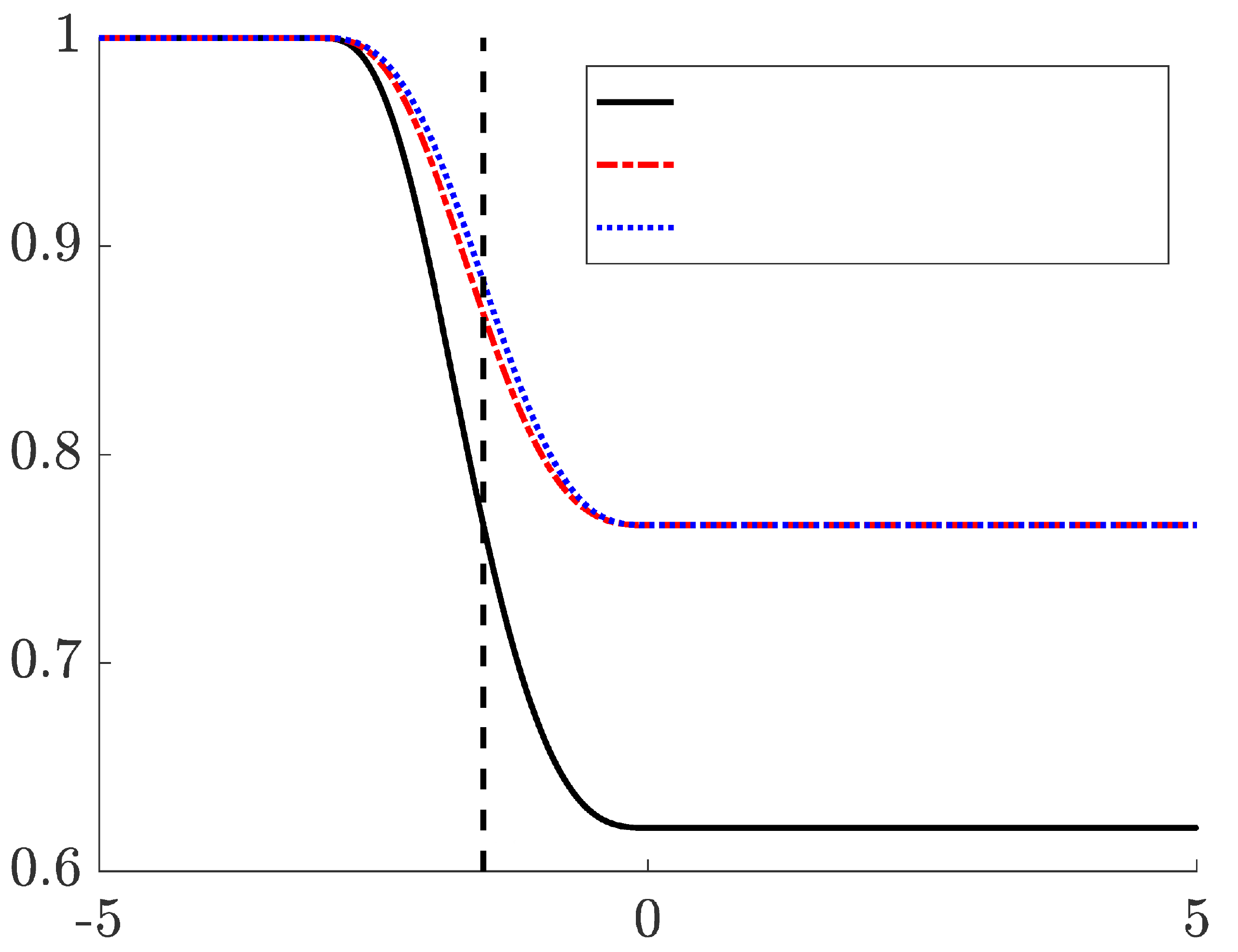}
  \put(-250,87){$\tau(\psi)$}
   \put(-110,-5){$\psi$}
   \put(-145,7){$-\epsilon$}
   \put(-95,150){Harmonic using $\tau^*$}
   \put(-95,139){Harmonic using $\tau^+$}
   \put(-95,128){Arithmetic using $\tau^+$}
\end{center}
\caption{Surface tension distribution using the traditional arithmetic mean with $\tau^+$ (dashed blue line), harmonic mean with $\tau^+$ (dash-dot red line) and harmonic mean with $\tau^*$ (solid black line). The solid interface location is denoted by the vertical dashed line at $\psi=-\epsilon$. The illustrated example uses $\tau^-=1$ and $\tau^+=|\tau^-\cos\theta_Y|$ where $\theta_Y=140^{\circ}$.}
\label{fig:surften_dist}
\end{figure}

The surface tension values are different for the LG interface, and the LS interface. Prescribing a binary distribution for $\tau(\mathbf{x})$ given by Eq. (\ref{eq:binary_tau}) leads to a stair-casing effect. Similar to \cite{zhao1998capturing}, surface tension is given as a distribution with a slight shift in the argument of the Heaviside function to move the stiffest change away from the solid boundary,
\begin{equation}
\label{eq:contin_tau}
 \tau(\mathbf{x}) = \tau^+ + (\tau^- - \tau^+) H(-\psi - \epsilon).
\end{equation}
In the paper by \cite{zhao1998capturing}, the bubble shape at equilibrium was not compared to a analytic solution and the contact angle at the triple-line junction was not reported. Qualitatively, the results were promising and demonstrated the effectiveness of the method. However, in our experience, the surface tension distribution caused the interface location to have a mismatch when compared to analytic solution, with errors ranging between $2\%$ to $8\%$. The reason for this error is due to the smoothing of the Heaviside function. At equilibrium, the Heaviside function at the interface location has a value of $H(\phi)=0.5$, this leads to an under predicted value of $\tau^+$ at the solid interface causing a mismatch in the contact angle and in turn the equilibrium position. To remedy this effect, we found that a temporary surface tension value $\tau^*$ can be defined such that the correct value of surface tension within the solid is obtained at the interface location. A harmonic mean is used in Eq. (\ref{eq:contin_tau}) to further reduce any errors that stem from the lower value of the dummy variable. The surface tension distribution becomes as follows
\begin{equation}
\label{eq:tau_geom}
 \tau(\mathbf{x}) = \bigg [ \frac{H(-\psi - \epsilon)}{\tau^-} + \frac{1-H(-\psi - \epsilon)}{\tau^*} \bigg ]^{-1},
\end{equation}
where 
\begin{equation}
\label{eq:tau_temp}
 \tau^* = \frac{\tau^+ \tau^-}{2\tau^- - \tau^+}.
\end{equation}
Note that substituting Eq. (\ref{eq:tau_temp}) in Eq. (\ref{eq:tau_geom}) gives the desired value of $\tau(\mathbf{x})=\tau^-$ at the interface where $H(\phi)=0.5$. Although the value inside the solid is incorrect, the value remains inconsequential since the no-penetration constraint and the formulation are only relevant outside the solid. Figure \ref{fig:surften_dist} presents a comparison between the different surface tension models. The harmonic mean using $\tau^*$ achieves the desired value of $\tau^+$ at a faster rate for a given interface location. The pseudo-code is summarized in Algorithm \ref{algo:drlsec}:
\begin{algorithm}
\caption{Pseudo-code using Heun's method}
\label{algo:drlsec}
\begin{algorithmic}[1]
\STATE Initialize $\Phi^0:=\{\phi^0_{ijk}\}_{i,j,k=1}^{n_x,n_y,n_z}$. Choose $\tau^+$, $\tau^-$, $\alpha$, $dt$ and $T$.
\STATE Set the solid boundary $\psi$. 
\STATE Calculate surface tension distribution using Harmonic mean with $\tau^*$
\STATE Compute $H_{\psi}:=H_{\epsilon}(\psi)$, $\tau_{\psi}:=\tau(\psi)$ and $H_{-\psi}:=H_{\epsilon}(-\psi)$.
\FOR{$n = 1$ to $T$}
\STATE Define $div(\cdot)$ = $\nabla \cdot ()$ using central difference.
\STATE Compute the discrete $\delta_{\phi}:=\delta_{\epsilon}(\phi)$ and $H_{\phi}:=H_{\epsilon}(\phi)$.
\STATE Compute gradients $\nabla \phi = \partial \phi / \partial x_i$ using central difference.
\STATE Compute magnitude of the gradient $|\nabla \phi|=\sqrt{(\partial \phi / \partial x)^2+(\partial \phi / \partial y)^2+(\partial \phi / \partial z)^2}$.
\STATE Compute normals $\mathbf{N}=\nabla \phi / |\nabla \phi|$.
\STATE Compute curvature term $\kappa(\phi):= div(\mathbf{N})$.
\STATE Compute the penalty term $\mathcal{P}(\phi):=\alpha \, div(d_p(|\nabla \phi|) \nabla \phi)$.
\STATE Compute gradient of surface tension distribution $\nabla \tau$ using central difference.
\STATE Compute Lagrange multipliers $\lambda(\phi)$, $\mu(\phi)$ using trapezoidal rule.
\STATE Set $f=\mathcal{P}(\phi)+\delta_{\phi}[\tau \kappa(\phi)+\nabla \tau \cdot \mathbf{N} +H_{-\psi}\Delta p+\mu(\phi)H_{\psi}-\lambda(\phi) H_{-\psi}]$
\STATE Define  $f$ = \textbf{L} ($\phi$)
\FOR{$k = 1$ to $n_z$}
\FOR{$j = 1$ to $n_y$}
\FOR{$i = 1$ to $n_x$}
\STATE $\bar{\phi}_{ijk}^{n+1}=\phi_{ijk}^{n}+dt$ \textbf{L}$[\phi_{ijk}^{n}]$
\STATE $\phi_{ijk}^{n+1}=\frac{1}{2}\phi_{ijk}^n+\frac{1}{2}\bar{\phi}_{ijk}^{n+1}+\frac{1}{2} dt$\textbf{L}$[\bar{\phi}_{ijk}^{n+1}]$
\ENDFOR
\ENDFOR
\ENDFOR
  \STATE Update next time level $\phi_{ijk}^n=\phi_{ijk}^{n+1}$
\
\ENDFOR
\STATE \textbf{return}
\end{algorithmic}
\end{algorithm}

\begin{remark}
In the formulation by \cite{zhao1998capturing}, the traditional level set is employed; the evolution of $\phi$ given by Eq. (\ref{eq:levset_evol_a}) without the penalty term and $\delta(\phi)$ is replaced by $|\nabla \phi|$. Note that the operator L[$\phi$] in \S\ref{subsec:traditional_levset} describes a convection-diffusion equation given in Eq. (\ref{eq:conv_diff}). The system of LSE contains Hamilton-Jacobi equations coupled to curvature and stiff source terms. Singularities may develop in the solution and the numerical implementation requires much of the modern level set technology to obtain a stable solution \citep{zhao1996variational}. The crucial ingredients are: \\
(i) Break up the parabolic and hyperbolic terms and address each one with the appropriate numerical discretization.
(ii) High-order accurate essentially non-oscillatory schemes (originating in the study of hyperbolic conservation laws) developed for Hamilton-Jacobi equations.\\
(iii) Reinitialization of each LSF to be an SDF using Eq. (\ref{eq:reinit}).\\
(iv) The curvature definition uses the explicit term involving the gradients of $\phi$ given by: 
\begin{equation}
\label{eq:trad_curv}
 \kappa = (\phi^2_x \phi_{yy} - 2 \phi_x \phi_y \phi_{xy} + \phi^2_y \phi_{xx} + \phi^2_x \phi_{zz} - 2 \phi_x \phi_z \phi_{xz} + \phi^2_z \phi_{xx} + \phi^2_y - 2 \phi_y \phi_z \phi_{yz} + \phi^2_z \phi_{yy})/|\nabla \phi|^{3/2} \, . 
\end{equation}
However, the Lagrange multiplier requires the use of a distance function to define curvature on or near the front using a separate definition given by:
\begin{equation}
 \kappa = \text{trace}[[I-\phi D^2 \phi]^{-1}D^2\phi],
\end{equation}
  where $D^2 \phi$ is the Hessian of $\phi$. This formula yields a constant value of $\kappa$ normal to the front and gives the correct value on the front. All of which is not necessary in the current formulation.  
\end{remark}
Note that in traditional level set methodologies, the reinitialization step enforces a local volume conservation \citep{sussman1999efficient}. The numerical integration uses a $9$-point quadrature formula in two spatial dimension. As mentioned earlier, the constraints in the current algorithm are enforced in a global sense. In order to test the local integration method, a $27$-point quadrature formula is employed for the general three spatial dimensions given by
\begin{equation}
  \int_{\Omega_{ijk}} f_{ijk} \, \mathrm{d} \mathbf{x} \approx \frac{1}{78} \bigg ( 52 f_{ijk} + \sum^{1}_{\substack{p,q,r=1 \\ (p,q,r)\ne(0,0,0)}} f_{i+p,j+q,k+r}\bigg ) \, \mathrm{d}\mathbf{x} \, .
\end{equation}
The local volume conservation could not be satisfied. It is worth noting that the local integration was also tested on the no-penetration constraint using the original formulation of $\mu$ using Eq. (\ref{eq:lag_mu_1}) instead of Eq. (\ref{eq:lag_mu_2}) with the standard surface tension model given by Eq. (\ref{eq:contin_tau}) instead of Eq. (\ref{eq:tau_geom}). The algorithm gave correct results for the interface contact points and radius of curvature over a cosine surface. However, for the locations where there is a liquid-solid overlap, the interface leaks about $3$-cells deep into the solid (for regions of high curvature) and $1$-cell deep (for shallow regions). It is not evident if the interface will leak more in regions of extremely large curvature (e.g. sharp peaks), therefore the method where the constraints are enforced globally will be used since it is more robust.

\section{Validation}
\label{sec:validation}
We begin with the effect of distance regularization as a method to satisfy $|\nabla \phi|=1$, next we build up in complexity of the evolution equation where we make sure that the method can solve traditional level set problems such as motion in the normal direction and curvature-driven flows. We validate our energy minimization model of the Gibbs free energy with the analytic solution of \cite{carbone2005}. We finally present some results over general geometries to demonstrate the effectiveness and flexibility of the proposed algorithm. 
\subsection{Distance Regularization as Reinitialization}
\begin{figure}
 \begin{center}
  \includegraphics[scale=0.23]{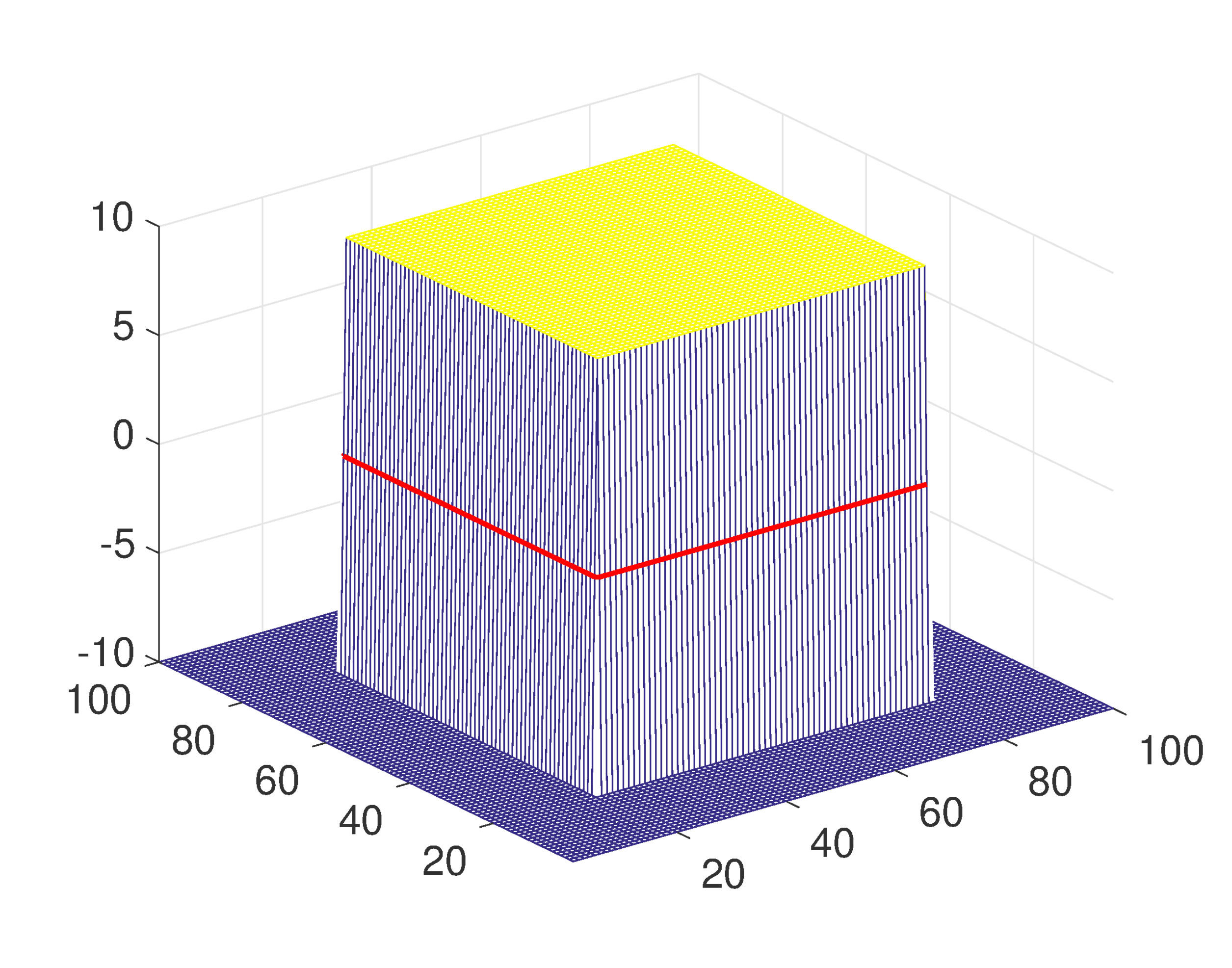}
  \put(-155,110){$(a)$}
  \\
  \includegraphics[scale=0.23]{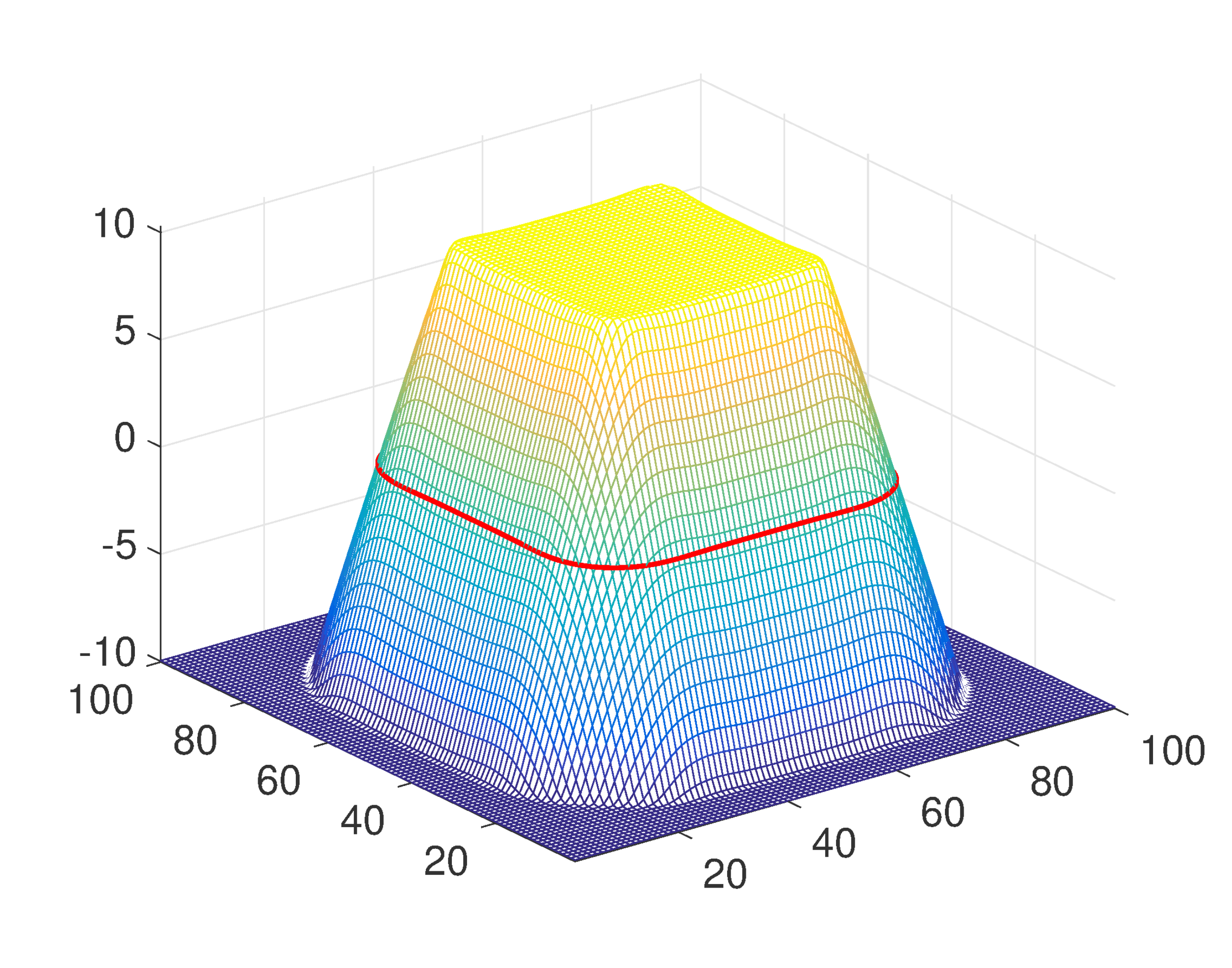} 
  \put(-155,110){$(b)$}
  \includegraphics[scale=0.2]{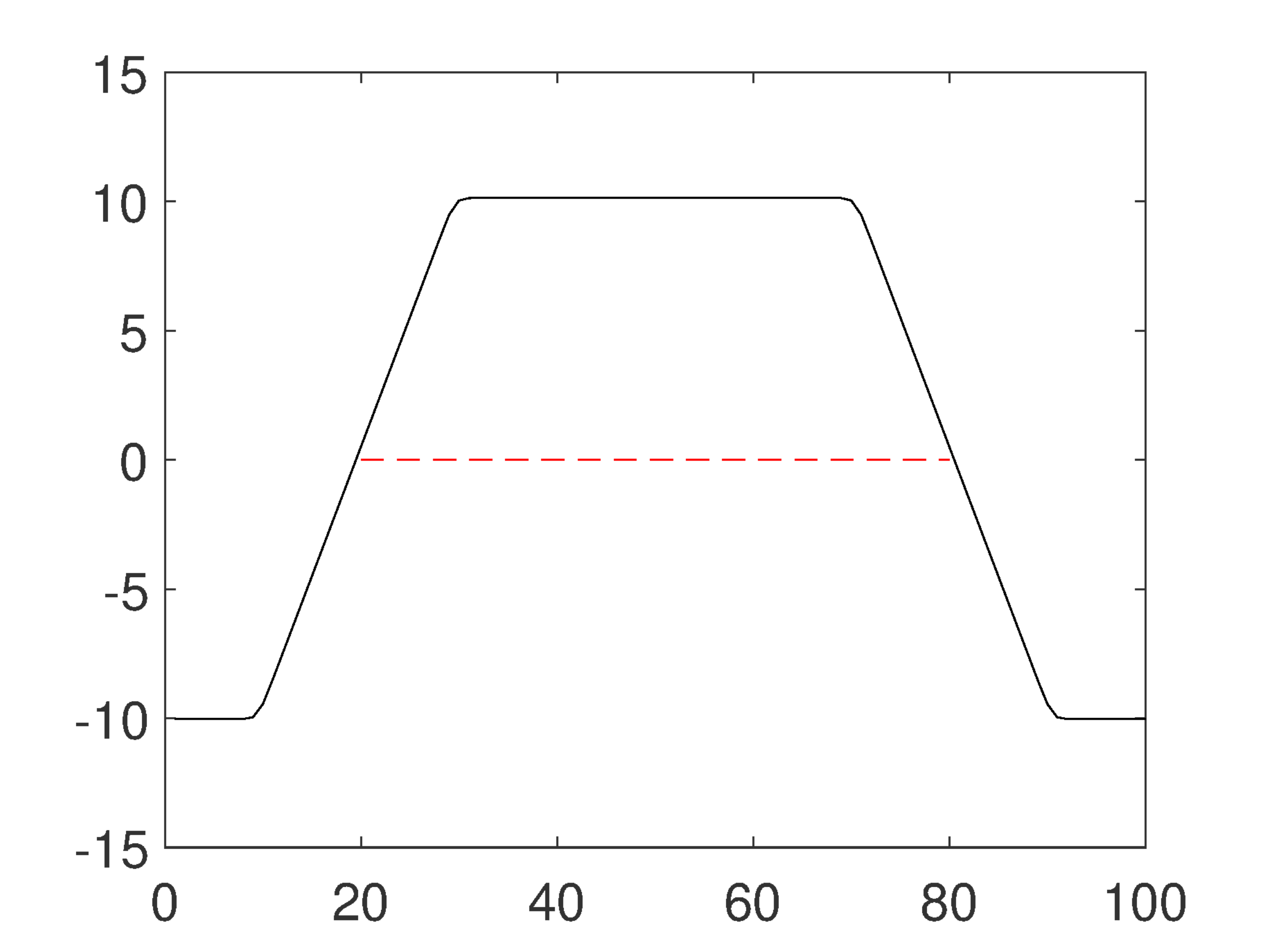}
  \put(-155,110){$(c)$}
  \includegraphics[scale=0.2]{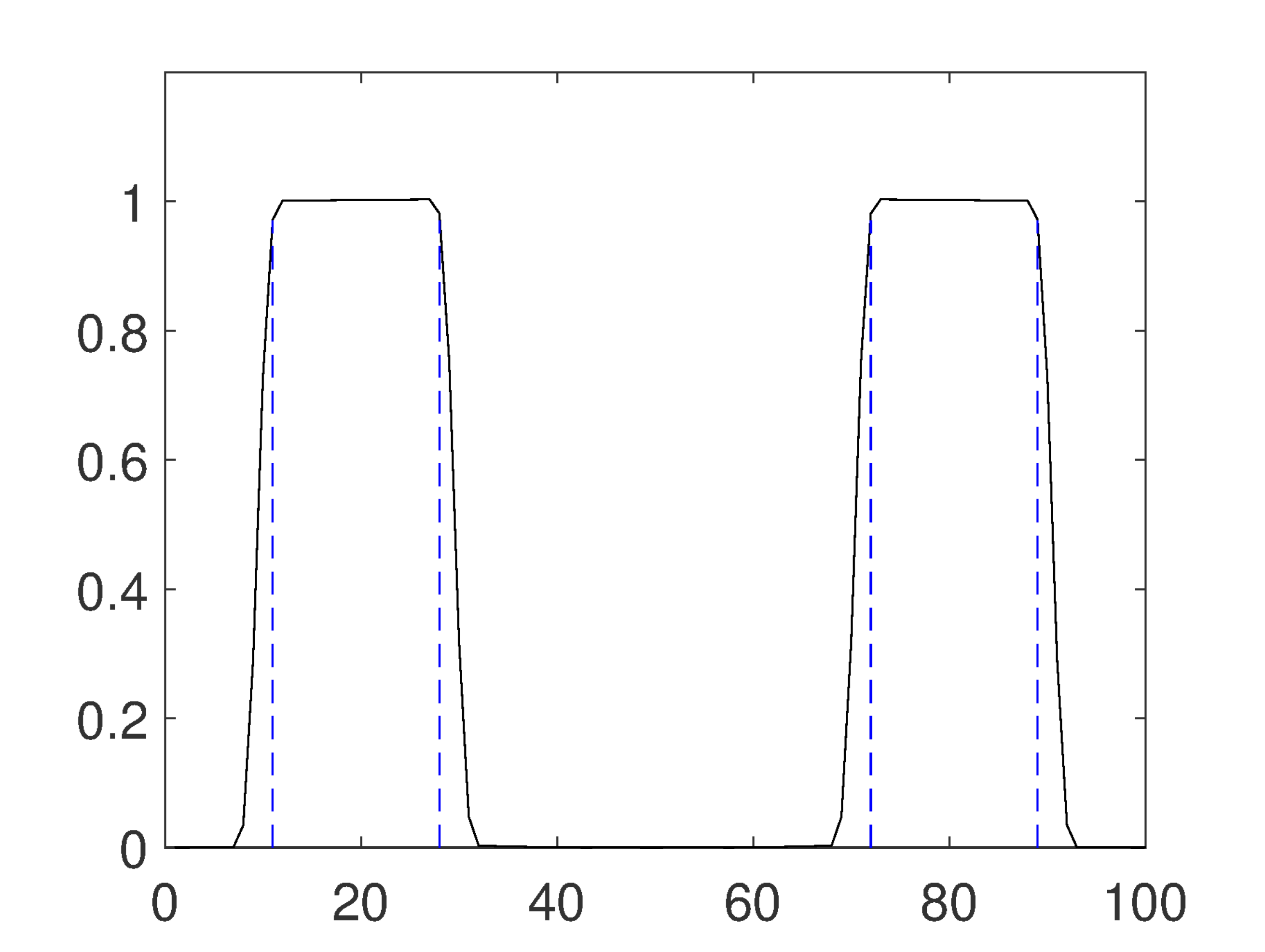}
  \put(-155,110){$(d)$}
  \\
  \includegraphics[scale=0.22]{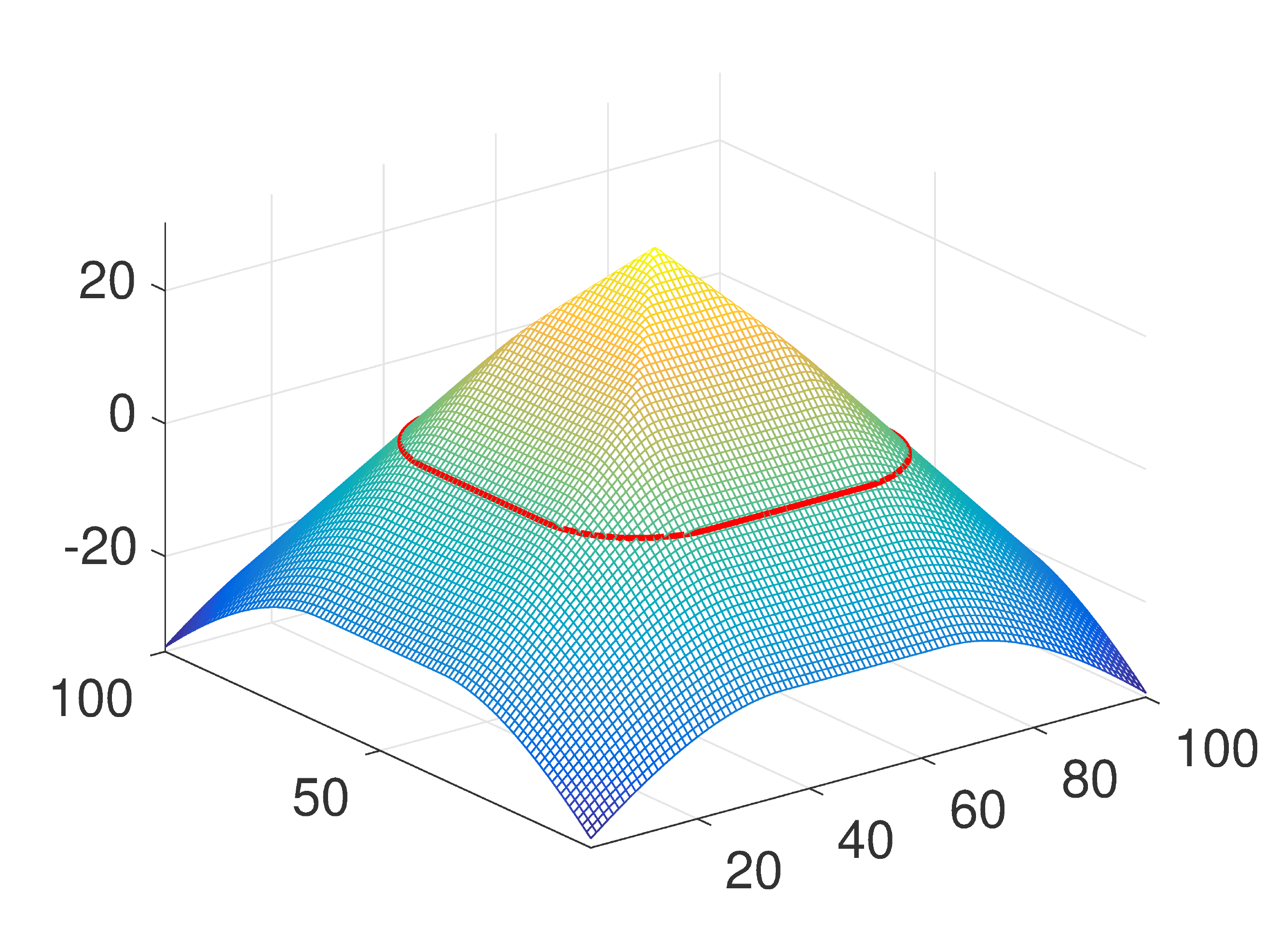}
  \put(-155,110){$(e)$}
  \includegraphics[scale=0.2]{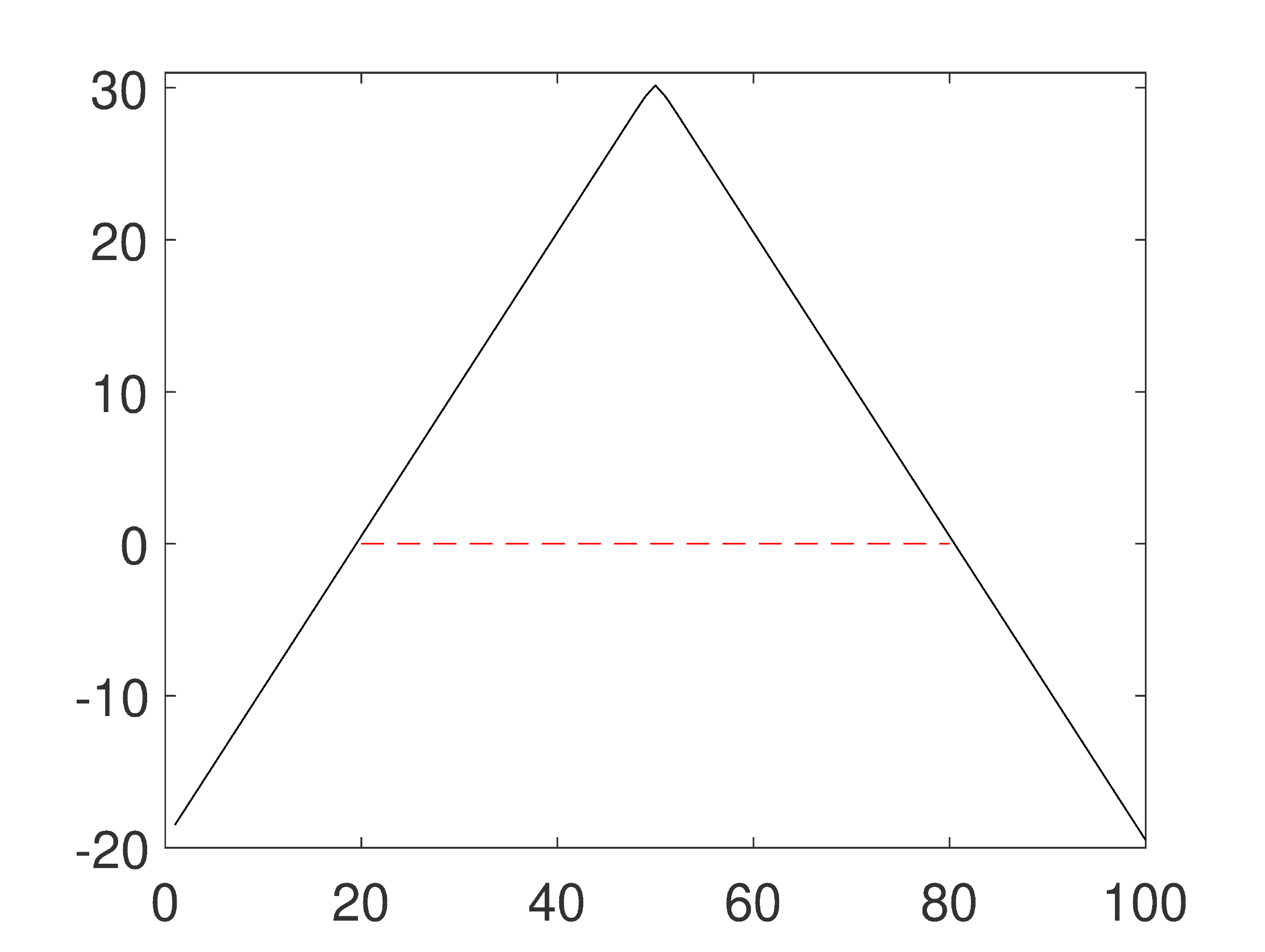}
  \put(-155,110){$(f)$}
  \includegraphics[scale=0.2]{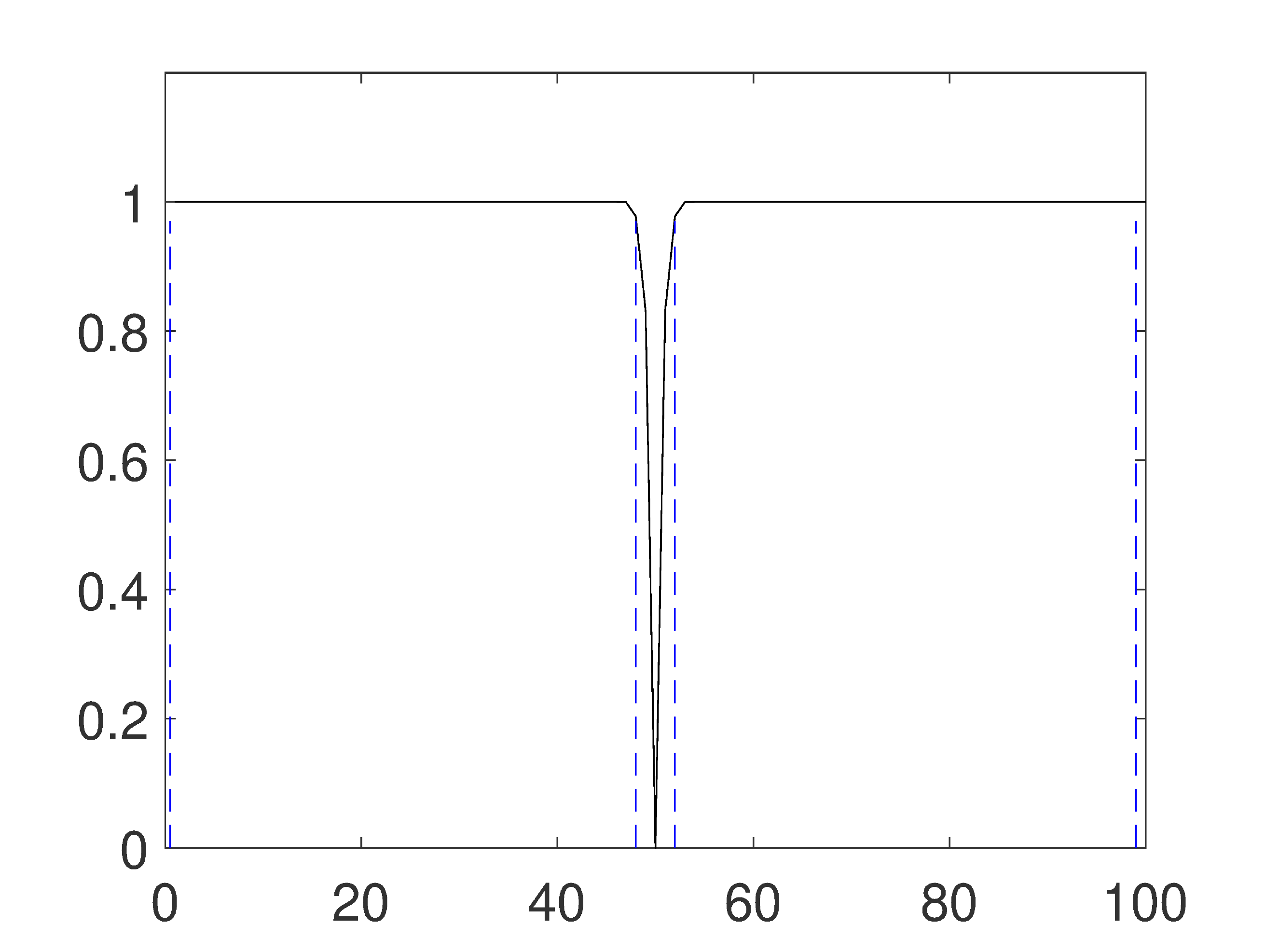}
  \put(-155,110){$(g)$}
 \end{center}
  \caption{The evolution of a level set function under the effect of distance regularization. $(a)$ The initial level set $\phi_o$ where the red solid line indicates $\phi_o=0$. $(b)$ The final level set after diffusion where the solid red line indicates $\phi=0$. $(c)$ A cross-section of the final level set function where the red dashed line indicates $\phi=0$. $(d)$ A cross-section of $|\nabla \phi|$ where the blue dashed lines indicate the width of the signed distance band. $(e)$ The final level set after standard reinitialization. $(f)$ A cross-section of the final level set function after standard reinitialization. $(g)$ A cross-section of $|\nabla \phi|$ after standard reinitialization.}
 \label{fig:binary_test}
\end{figure}

The effect of distance regularization as a method of reinitialization is demonstrated by simulating the forward and backward (FAB) diffusion problem with a binary LSF. Define the initial binary level set function $\phi_o$ by 
\begin{equation}
 \phi_o(\mathbf{x}) = \begin{cases}
		      -c_o, & \text{for $\mathbf{x} \in \Omega_o$} \\
			      0,  & \text{for $\mathbf{x} \in \partial \Omega_o$} \\
		       c_o,  & \text{for $\mathbf{x} \in \Omega - \Omega_o$} \, , 	      
		      \end{cases}
\end{equation}
where $c_o$ is a positive constant. Note that $\phi_o$ does not need to be initialized as an SDF like traditional LSM; this allows for flexibility in initialization due to its simplicity and practicality. The region $\Omega_o$ is taken to be a square of side length $60$ embedded in a square domain on a $100 \times 100$ grid, the time-step is taken to be $dt=0.5$ and $\alpha=0.4$. Define $\phi_o$ with a constant of $c_o=10$ which creates a steep jump of width $2c_o$ as shown in figure \ref{fig:binary_test}$(a)$, the solid red line highlights the zero crossing of $\phi_o$. The evolution of $\phi$ follows from the description of the properties of double-well potential given in \S\ref{subsec:penalty_regularization}. We solve a simple FAB diffusion evolution equation given by 
\begin{equation}
  \pop{\phi}{t} = - \pop{\mathcal{J}}{\phi} = -\alpha \pop{\mathcal{R}_p}{\phi} = \alpha \nabla \cdot (d_p(|\nabla \phi|) \nabla \phi) \, .
\end{equation}
At $t=0$ the zero crossing of $\phi_o$ has a $|\nabla \phi_o| >> 1$ due to the steep jump in the binary initialization. As a result, the diffusion rate $\alpha d_p(|\nabla \phi|)$ is positive and drives a forward diffusion which reduces $|\nabla \phi|$ until it reaches unity; the forward diffusion stops once $|\nabla \phi|=1$. For the case  where $|\nabla \phi|$ becomes less than unity, $\alpha d_p(|\nabla \phi|)$ becomes negative for $(1/2)<|\nabla \phi|<1$, therefore the backward diffusion increases the value of $|\nabla \phi|$ to unity. Figure \ref{fig:binary_test}$(b)$ shows the final result of $\phi$ where the level set is an SDF in the vicinity of the zero level set crossing. Figure \ref{fig:binary_test}$(c)$ shows a cross-section of $\phi$ for $y=50$ where the dashed red line indicates the zero crossing. Note that $|\nabla \phi|=1$ is satisfied within $c_o$ and $-c_o$ about the zero crossing, which will be defined as the signed distance band (SDB). Figure \ref{fig:binary_test}$(d)$ shows a cross-section of $|\nabla \phi|$ where the SDB, of width approximately $2c_o$, is highlighted with the dashed blue lines. On either sides of the SDB the binary function is constant, therefore $|\nabla \phi|=0$ and $\alpha d_p(|\nabla \phi|)>0$ indicating a forward diffusion that keeps $\phi$ flat at its constant value since the double-well potential is at a minimum.  

\begin{remark}
The bottom row of figure \ref{fig:binary_test} uses the standard reinitialization method to provide a comparison to distance regularization. Figure \ref{fig:binary_test}$(e)$ shows the final level set where the step function no longer exists compared to figure \ref{fig:binary_test}$(b)$. The cross-section in figure \ref{fig:binary_test}$(f)$ gives a clearer picture of the difference; the standard reinitialization ensures that the entire level set is an SDF, hence the slope on both ends of the discontinuity at $x=50$ have a value of unity satisfying $|\nabla \phi|=1$. Figure \ref{fig:binary_test}$(g)$ shows the cross-section of $|\nabla \phi|$, note how its value of unity is satisfied everywhere in the domain with the exception of the discontinuity, whereas figure \ref{fig:binary_test}$(d)$ shows the $|\nabla \phi|$ being satisfied locally near the region of $\phi=0$ over a band width of $2c_o$. However, both methods round the corners of the zero level set square contour. Therefore, one of the main differences between the two methods is that distance regularization only cares about the region around the zero level set within a certain band of width $2c_o$, and everything else in the far field remains a constant. This makes the DRLSE method highly localized.
\end{remark}

\subsection{Curvature-Driven Flow}\label{subsec:motion_curvature}
Traditional LSM has been used extensively in the context of externally driven flows. This involves both an externally generated velocity field, or a self-generated field $\mathbf{u}$ that depends directly on $\phi$. Consider the motion by mean curvature where the interface moves in the normal direction to the interface with a velocity proportional to its curvature. The velocity field is defined by $\mathbf{u}=-\kappa \ \mathbf{N}$ where $\kappa = \nabla \cdot (\nabla \phi/ |\nabla \phi|)$ such that the evolution equation becomes 
\begin{equation}
 \pop{\phi}{t} - \kappa \mathbf{N} \cdot \nabla \phi = 0, 
\end{equation}
and since the normal $\mathbf{N}$ is along the gradient direction $\nabla \phi$, it is equivalent to 
\begin{equation}
 \pop{\phi}{t} - \kappa |\nabla \phi| = 0.
\end{equation}
The above results in a parabolic equation $\phi_t=\kappa |\nabla \phi|$ that can be solved using central differencing with a stringent time-step restriction for stability. In the context of variational level set methodology, we can show that a mean curvature flow can be obtained by minimizing an external energy defined by the surface area of the LSF given by 
\begin{equation}
 \mathcal{E}_{ext}(\phi) = L_0(\phi) = \int_{\Omega} \delta(\phi)|\nabla \phi| \, \mathrm{d} \mathbf{x}.
\end{equation}
Taking the Fr{\'e}chet derivative of $\mathcal{E}_{ext}$ with respect to the $L^2$ inner product gives 
\begin{equation}
 \pop{\mathcal{E}_{ext}}{\phi}=-\delta(\phi) \nabla \cdot \bigg( \frac{\nabla \phi}{|\nabla \phi|} \bigg).
\end{equation}
This is similar to the curvature-driven evolution equation with the exception that $\delta(\phi)$ is $|\nabla \phi|$ instead. \cite{zhao1998capturing} simply replaced the terms in their formulation and reasoned that using $|\nabla \phi|$ is less stiff. Although both of the evolution equations are curvature driven, it is worth pointing out that they are inherently different. The equation given by $\phi_t=\delta(\phi) \nabla \cdot (\nabla \phi / |\nabla \phi|)$ moves the zero level set only which is only active due to $\delta(\phi)$. However, the equation given by $\phi_t=|\nabla \phi| \nabla \cdot (\nabla \phi / |\nabla \phi|)$ moves all the level sets by their respective curvatures i.e. all the neighboring values of the level set evolve with the same law. \cite{esodoglu2008variational} argues that by definition, we can allow all the level sets to move by a perturbation of an energy functional that measures the total length of all level curves  
\begin{equation}
  \mathcal{E}_{ext}(\phi) = \int_{\beta} L_\beta(\phi) \, \mathrm{d} \beta = \int_{\Omega} |\nabla \phi| \, \mathrm{d} \mathbf{x},
\end{equation}
where $L_\beta(\phi)= \int_{\Omega} \delta(\phi-\beta)|\nabla \phi|\, \mathrm{d} \mathbf{x}$. Note that the variation of $\phi$ only affects $\nabla \phi$ but does not move the level sets. Since the level set is a signed distance function, we can define a new level set norm \citep{esodoglu2008variational} instead of the usual $L^2$ norm as
\begin{equation}
 \langle f, g \rangle_{\phi} = \int_{\Omega} f \frac{g}{|\nabla \phi|} \, \mathrm{d} \mathbf{x}.
\end{equation}
The Fr{\'e}chet derivative obtained from this new norm is 
\begin{equation}
 \lim_{\epsilon \to 0} \frac{1}{\epsilon} \bigg [ \mathcal{E}_{ext}(\phi+\epsilon \chi)-\mathcal{E}_{ext}(\phi) \bigg ] = \bigg \langle \pop{\mathcal{E}_{ext}}{\phi}, \chi \bigg \rangle_{\phi} = - \int_{\Omega} \nabla \cdot \bigg( \frac{\nabla \phi}{|\nabla \phi|} \bigg) \, |\nabla \phi|  \, \frac{\chi}{|\nabla \phi|} \, \mathrm{d} \mathbf{x}.
\end{equation}
This results in 
\begin{equation}
 \pop{\mathcal{E}_{ext}}{\phi}=-\nabla \cdot \bigg( \frac{\nabla \phi}{|\nabla \phi|} \bigg) \, |\nabla \phi| \, ,
\end{equation}
which lends itself to the traditional curvature-driven flow discussed earlier in this section. The evolution equation solved is given by
\begin{equation}
 \pop{\phi}{t} = \alpha \nabla \cdot (d_p(|\nabla \phi|) \nabla \phi) + \nabla \cdot \bigg( \frac{\nabla \phi}{|\nabla \phi|} \bigg) \, |\nabla \phi| \,.
\end{equation}
Note that as mentioned earlier we can advect all the level curves in the context of energy minimization such that we can replicate the motion of a curvature-driven flow. Isolating the zero level set curve gives an evolution equation that does not look like the typical curvature-driven flow, but since the current method is only concerned with the region around the interface, it is able to handle both equations without running into stability problems. This was tested for both formulations but we will only present the traditional cases for brevity. The only difference observed between the two formulations is the rate at which the curves reach the desired final solution.

\subsubsection{Seven-Pointed Star}
\label{subsubsec:sps_curv}

\begin{figure}
 \begin{center}
  \includegraphics[scale=0.2]{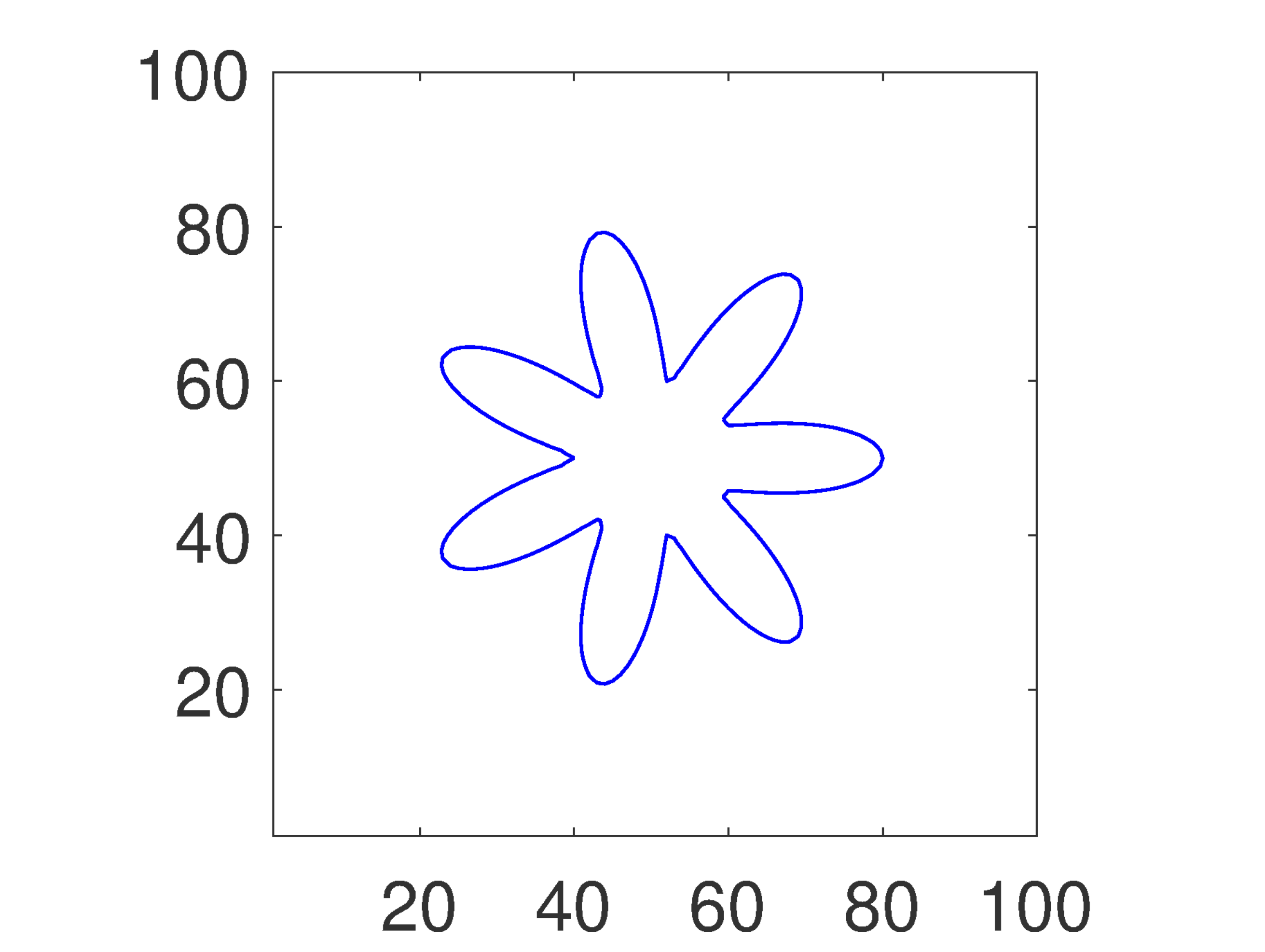}
  \includegraphics[scale=0.2]{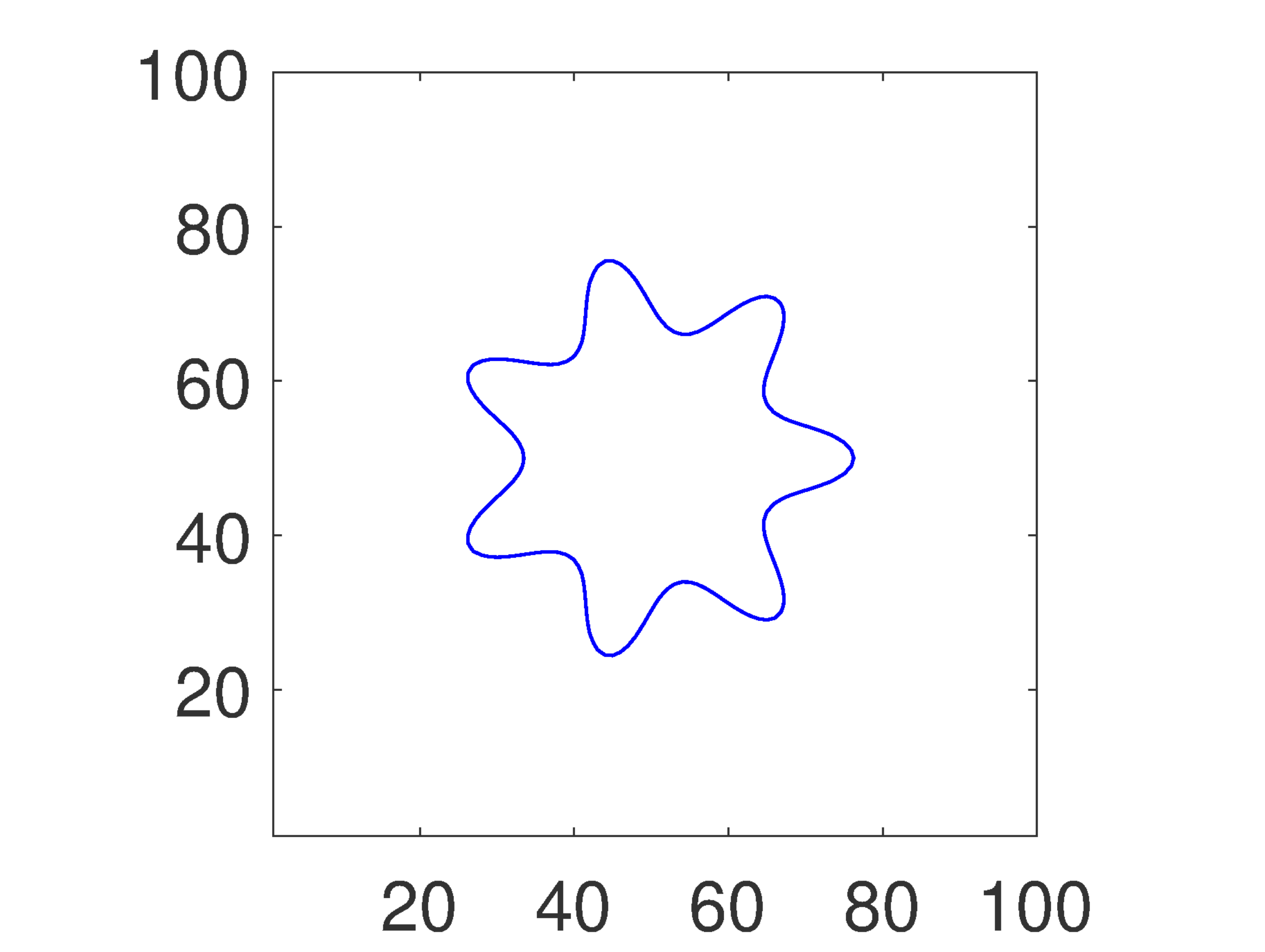}
  \includegraphics[scale=0.2]{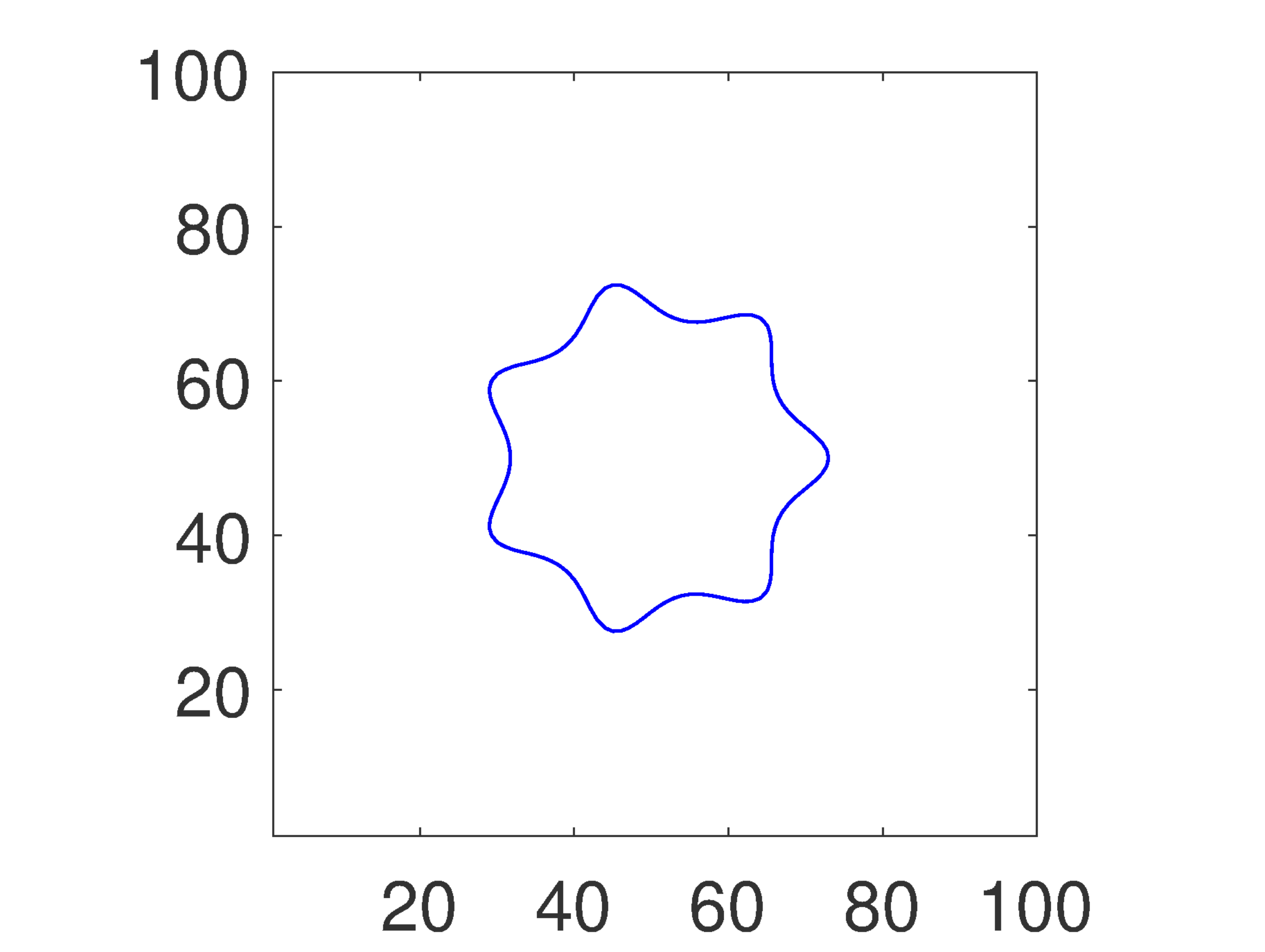}
  
  \includegraphics[scale=0.2]{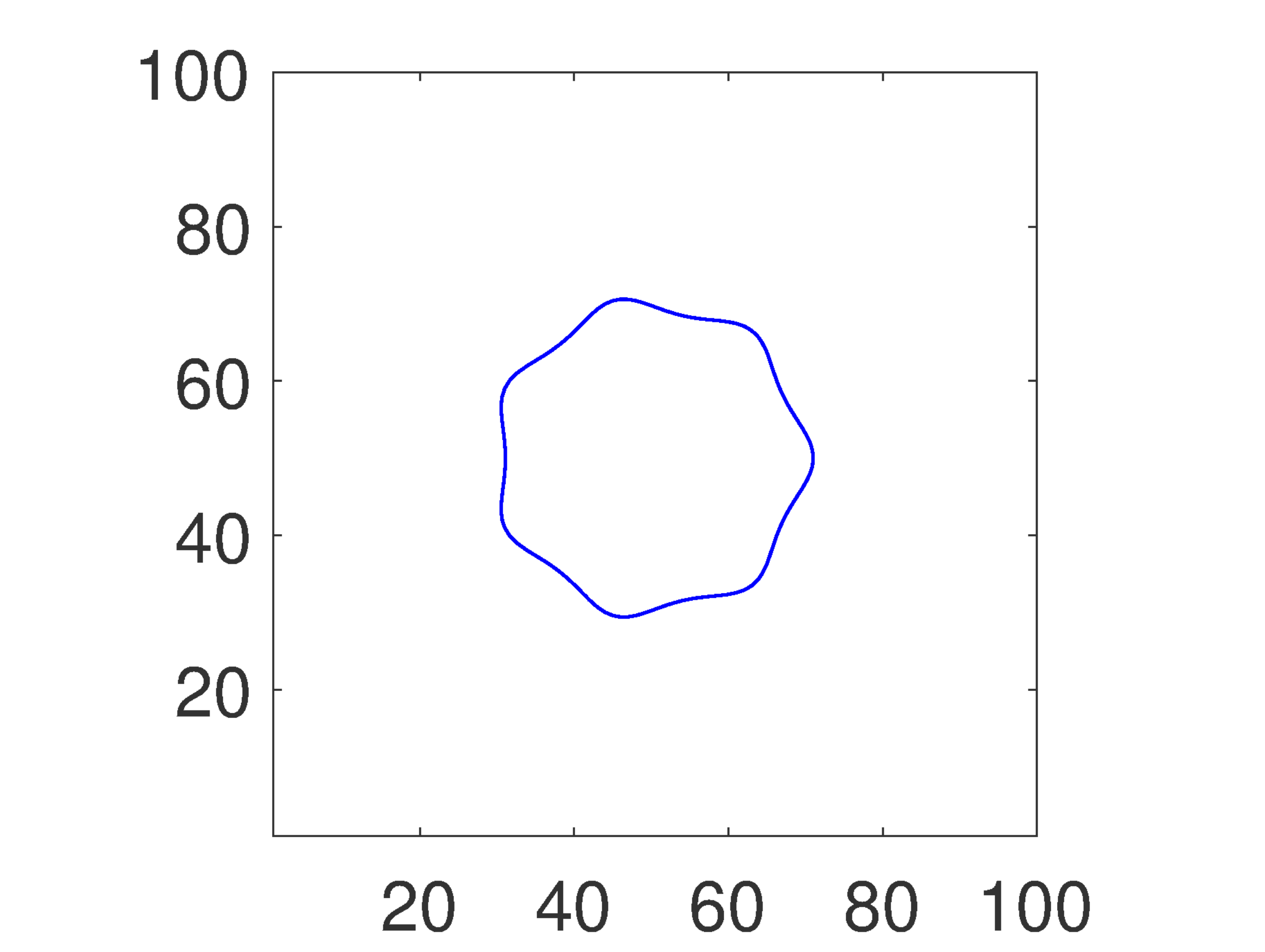}
  \includegraphics[scale=0.2]{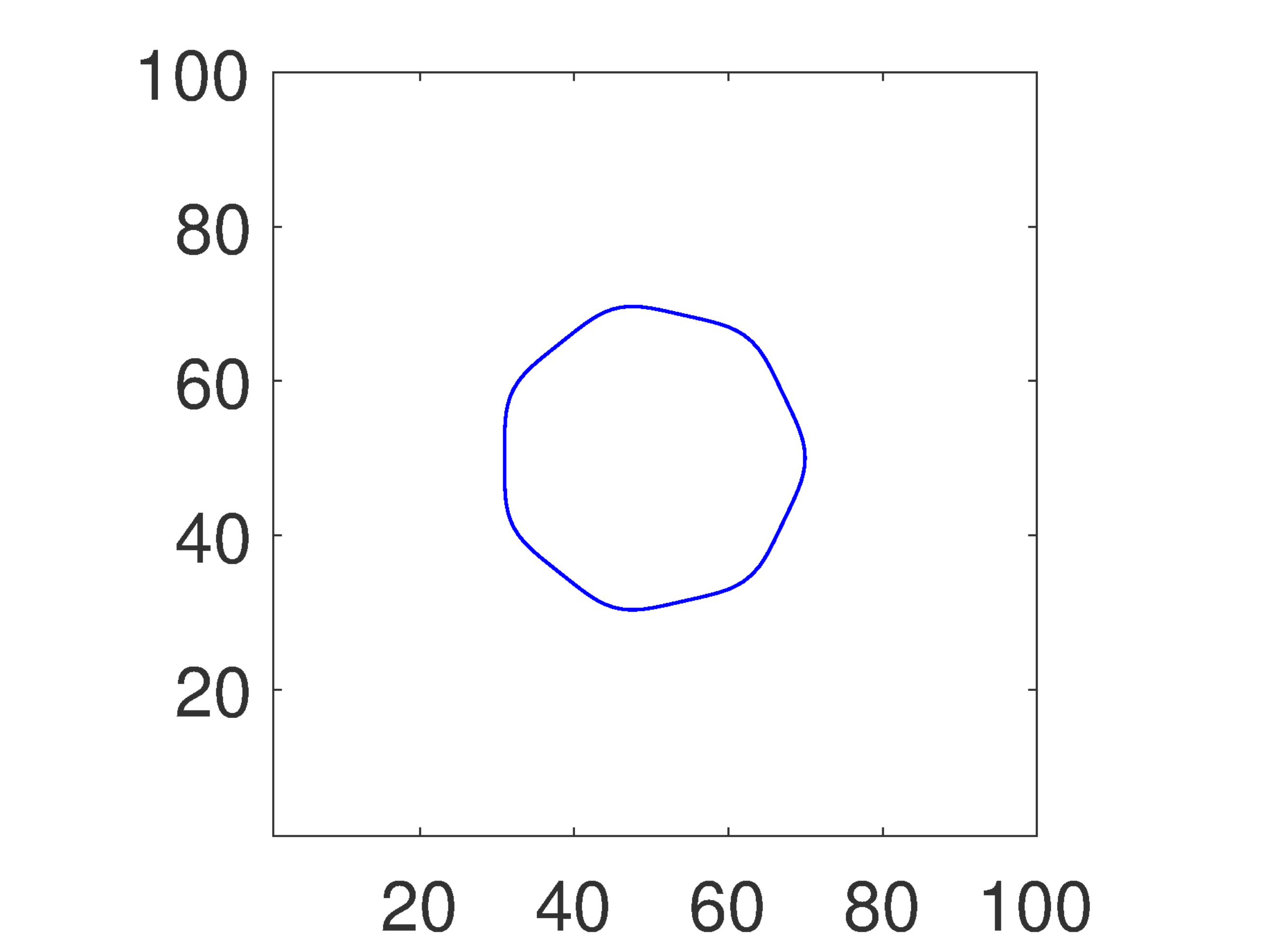}
  \includegraphics[scale=0.2]{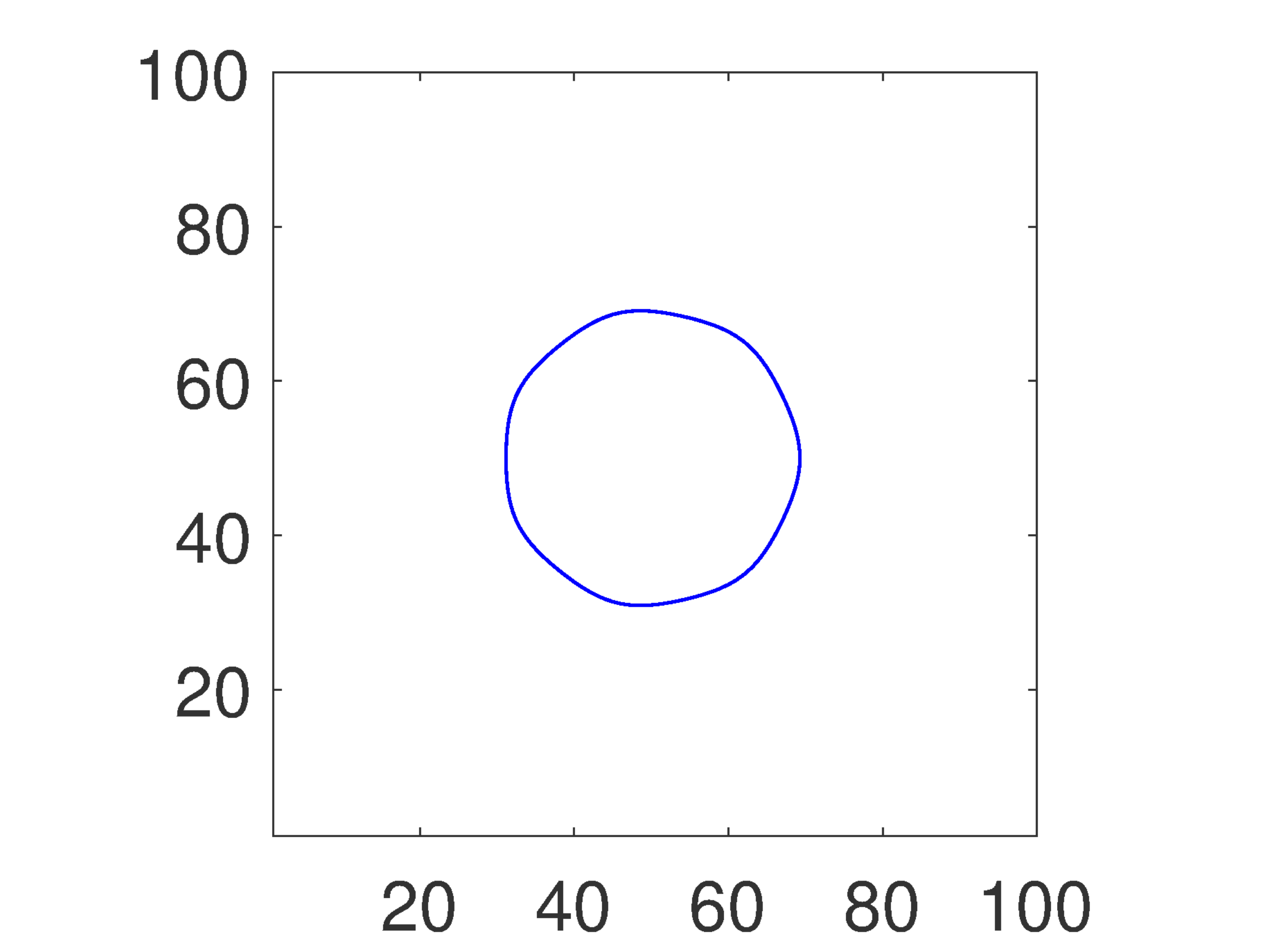}
 
  \includegraphics[scale=0.2]{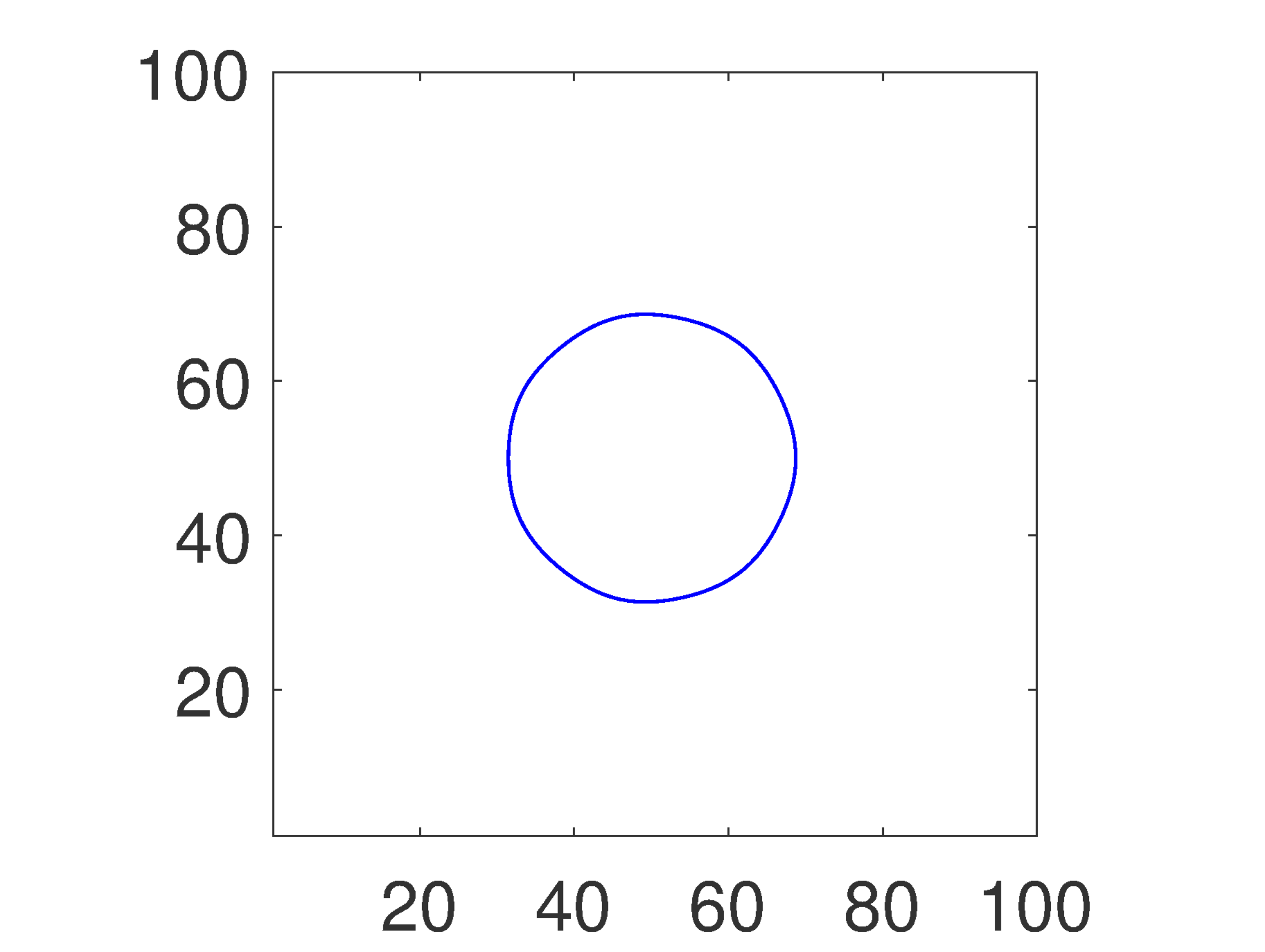}
  \includegraphics[scale=0.2]{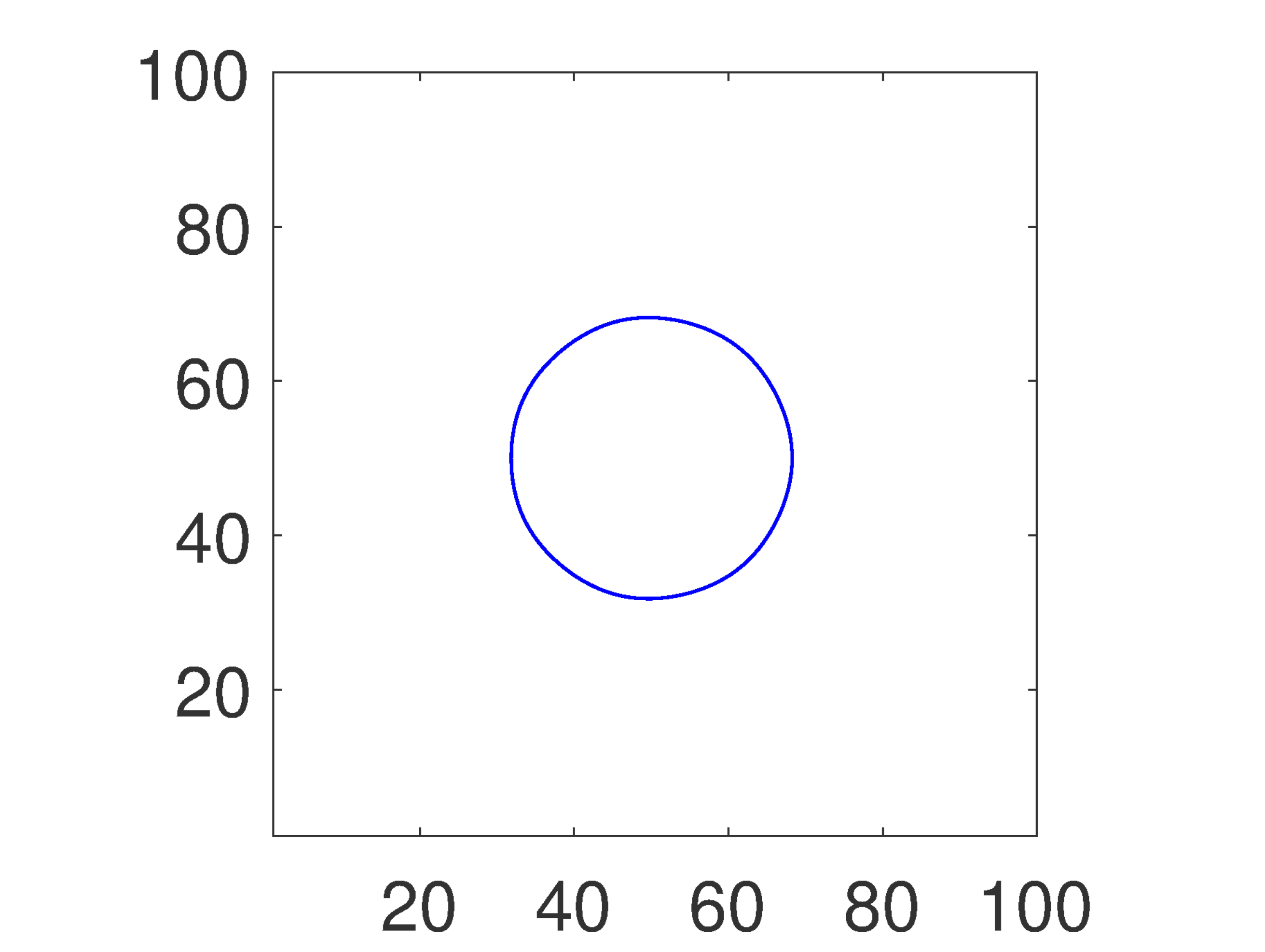}
  \includegraphics[scale=0.2]{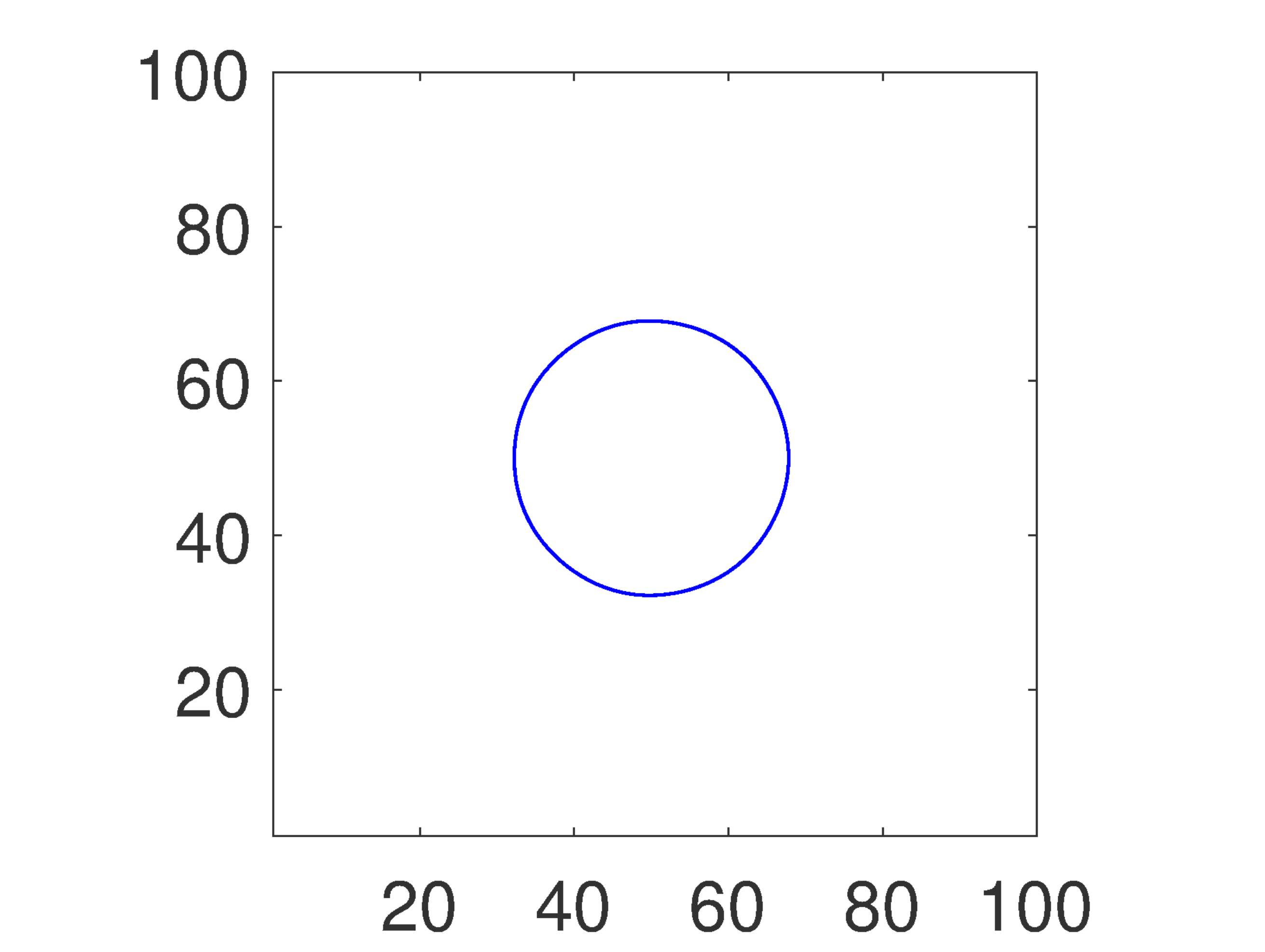}
 \end{center}
 \caption{ Evolution of the star-shape under a curvature-driven flow. The images are snapshots in time from left to right, top to bottom where the star collapses unto a circle for $t = 0.0, 7.5, 15.0, 22.5, 30.0, 37.5, 45.0, 52.5, 60.0$.}
 \label{fig:curv_star}
\end{figure}

%
%
  Consider a seven pointed star given by the following parameterized curve
\begin{equation}
 \gamma(s) = \big[20 + 10 \cos(7 \cdot 2 \pi s)\big] \big[\cos(2\pi s), \sin(2\pi s)\big] \quad \textrm{for} \quad s \in [0,1],
\end{equation}
such that the zero level set is defined from  polar coordinates is given by
\begin{equation}
 \phi(\mathbf{x},0) = ||\mathbf{x-x_c}|| -  \bigg\{20 +  10 \cos\bigg[7 \cdot \arctan\big((y-y_c)/(x-x_c)\big)\bigg]\bigg\},
\end{equation}
where $\mathbf{x_c}$ is the offset parameter. We shift the petal shapes of the star from its center by $20$ units of length which effectively increases the relative size of the petal shape compared to the main body, a scaling factor of $10$ for the star size and $7$ points for the number of petals in the star. The computational domain is a square with side length of $100$ units and a grid size of $100\times100$. The time-step is $dt=0.1$ and $\alpha=0.4$. The star-shaped level curve is centered at $(50,50)$. The simulation is run for $600$ iterations. Figure \ref{fig:curv_star} shows the progression of the front evolving under the curvature-driven flow for $t = 0.0, 7.5, 15, 22.5, 30.0, 37.5, 45.0, 52.5, 60.0$. The star-shaped curve collapses to a circle under this motion where the region of positive curvature (peaks) collapses inwards and the regions of negative curvature (troughs) propagate outward.  

\subsubsection{Wound Spiral}
\begin{figure}
 \begin{center}
  \includegraphics[scale=0.2]{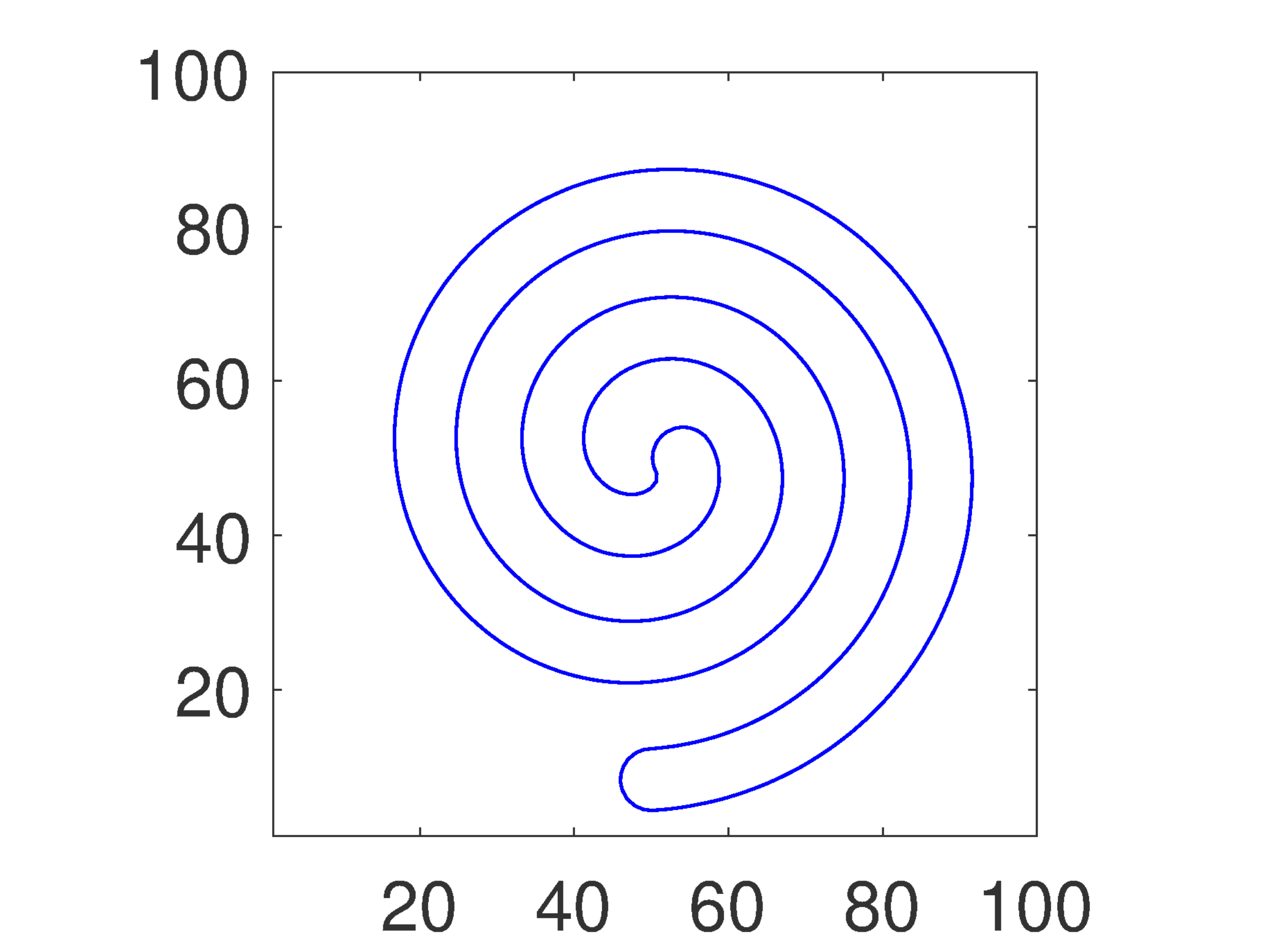}
  \includegraphics[scale=0.2]{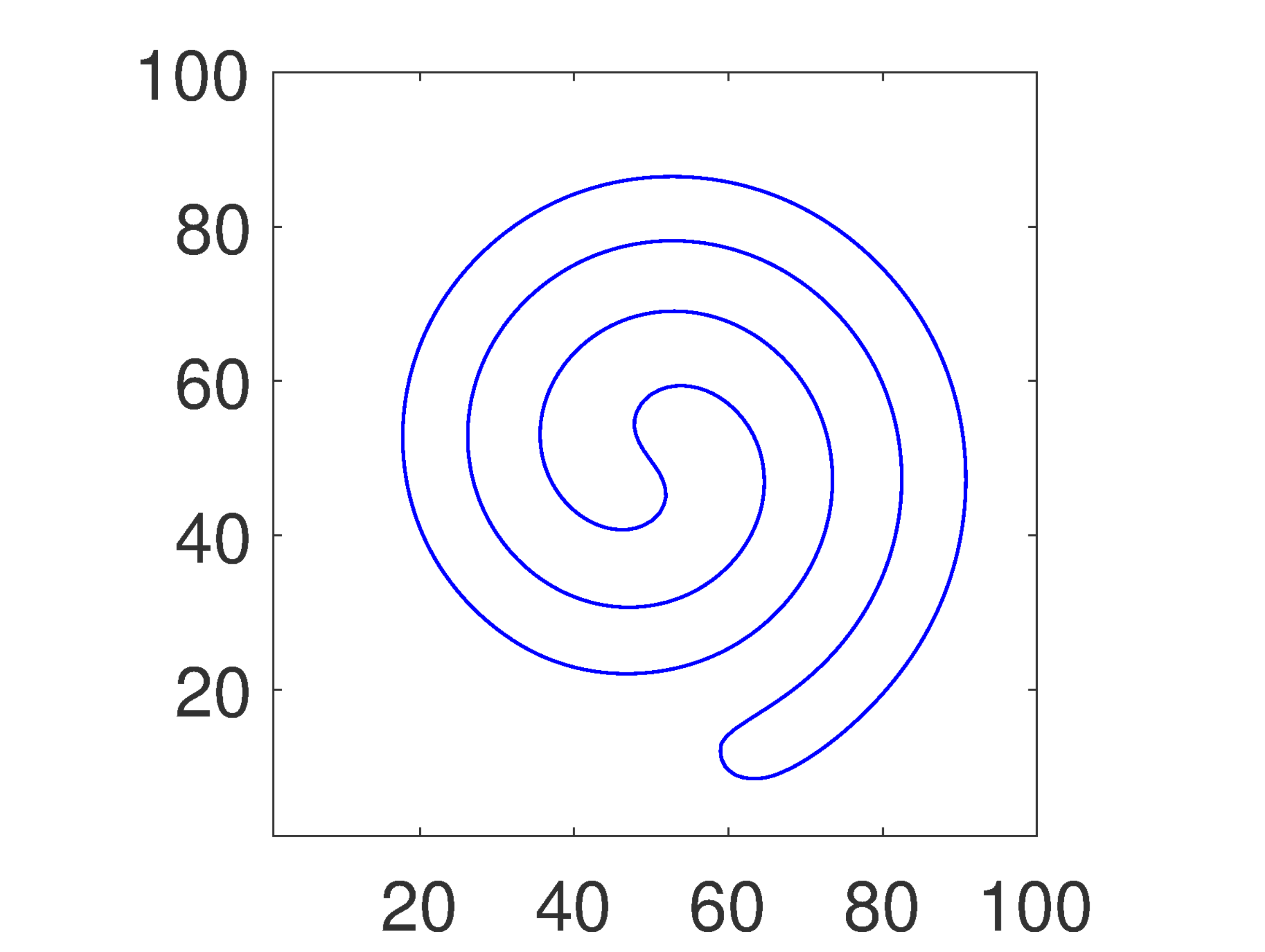}
  \includegraphics[scale=0.2]{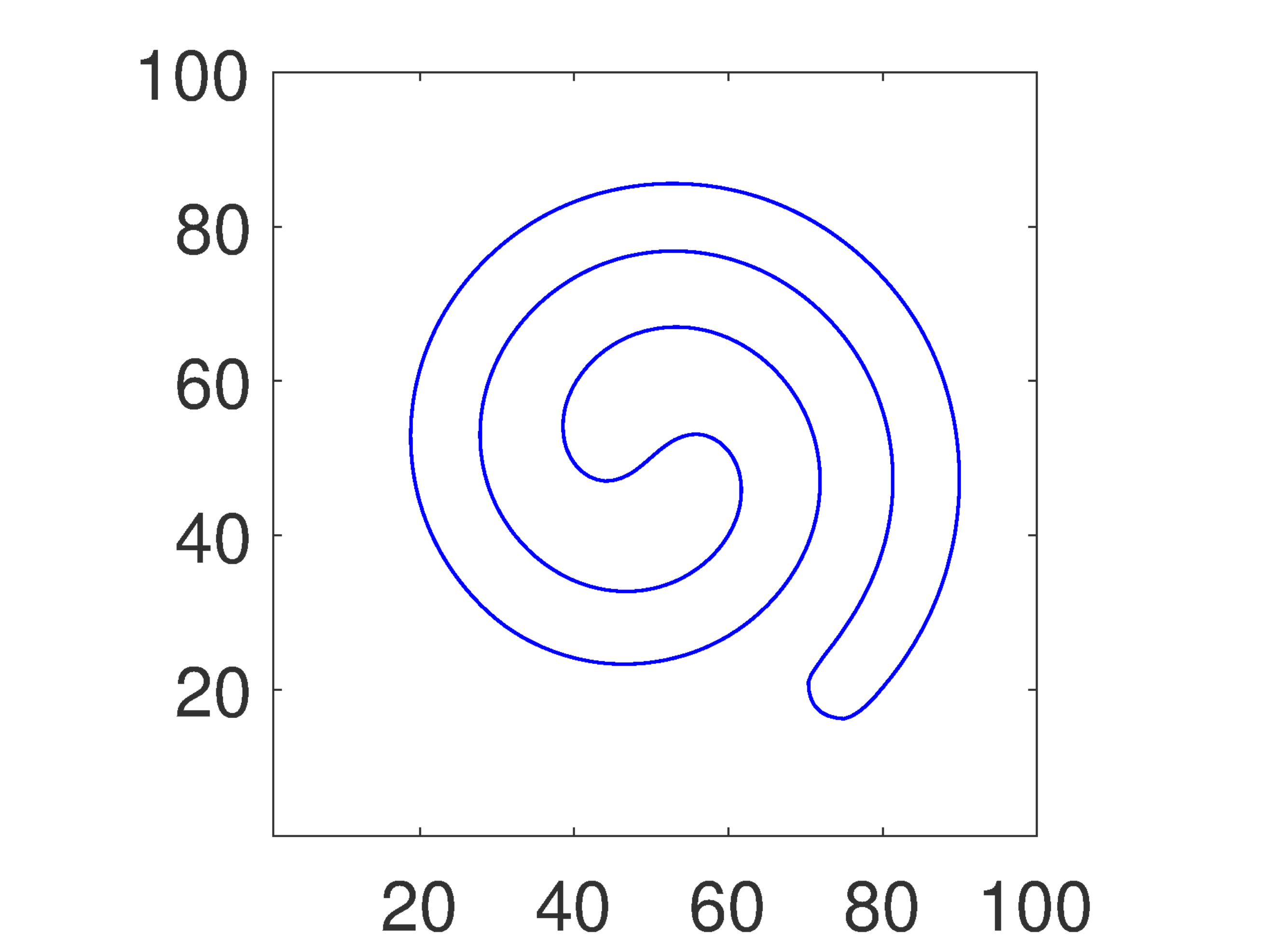}
  
  \includegraphics[scale=0.2]{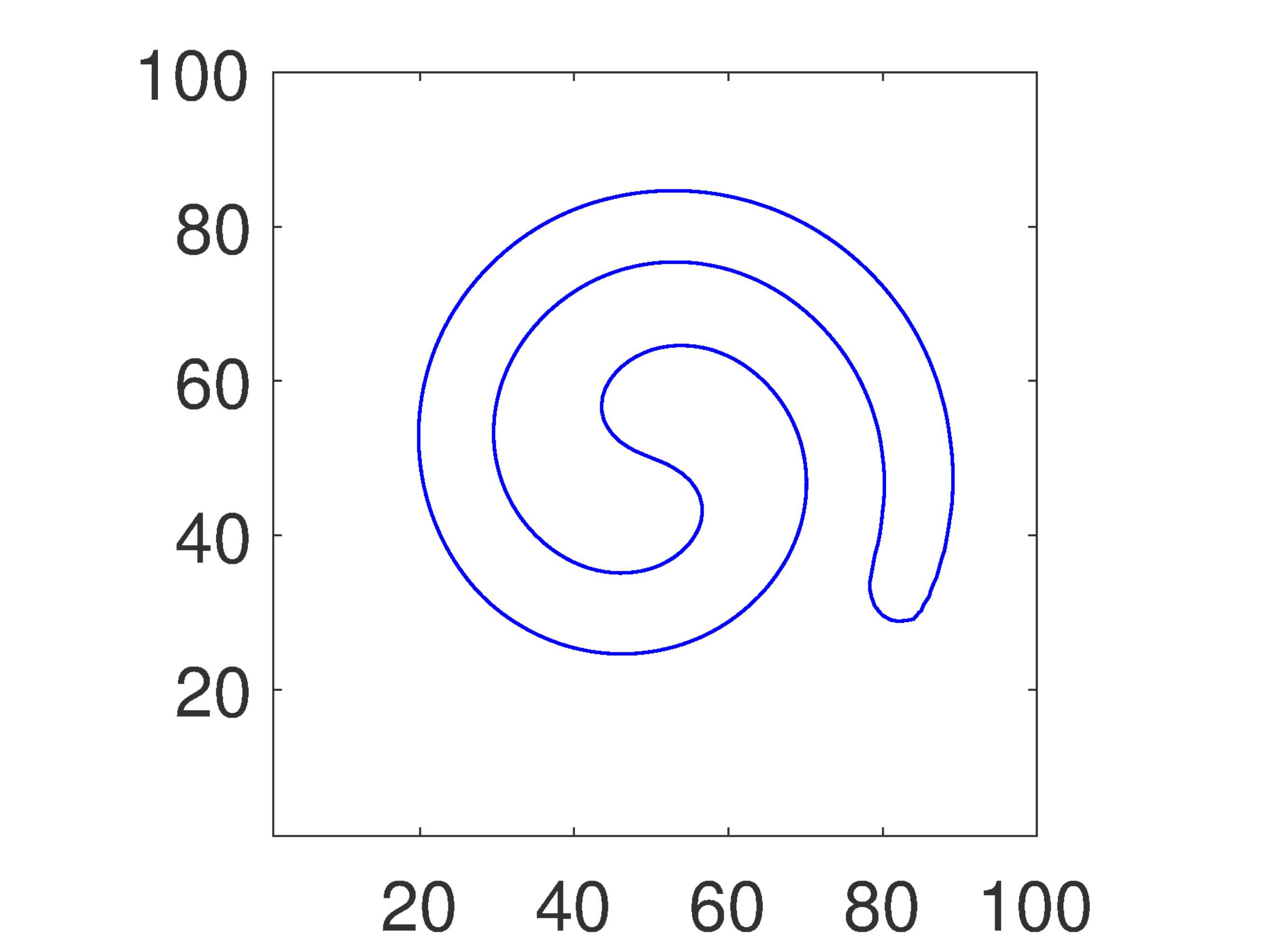}
  \includegraphics[scale=0.2]{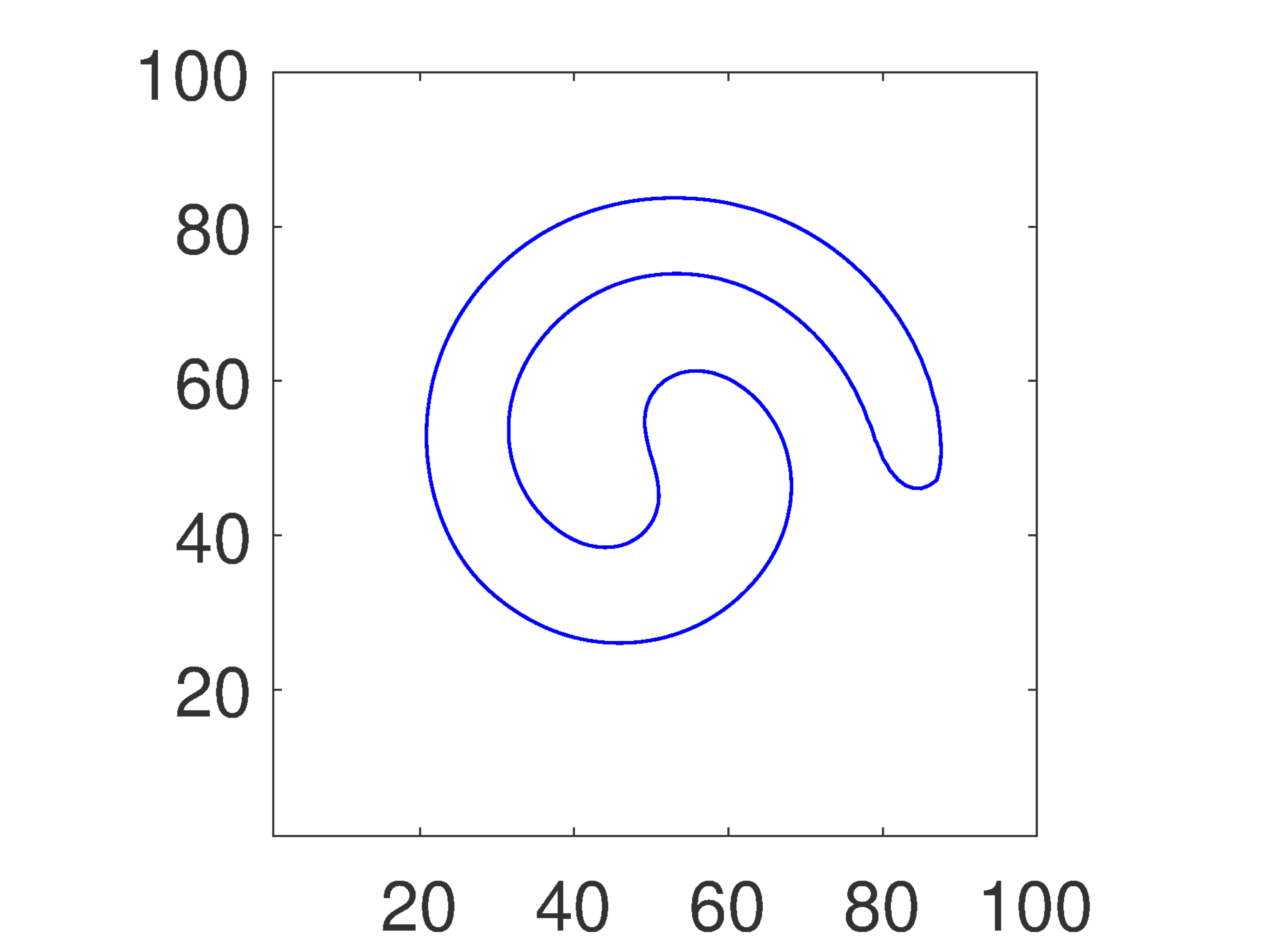}
  \includegraphics[scale=0.2]{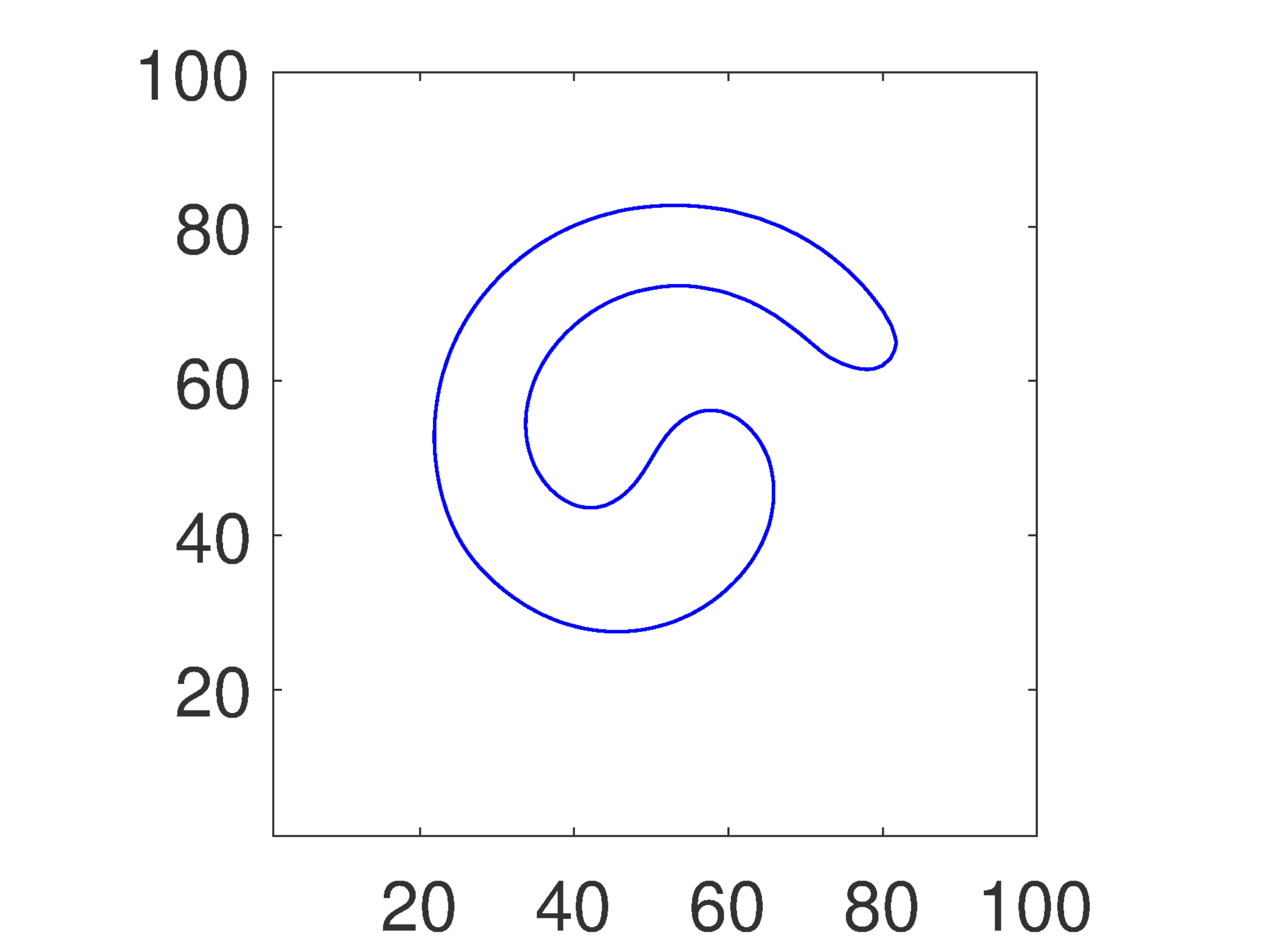}
 
  \includegraphics[scale=0.2]{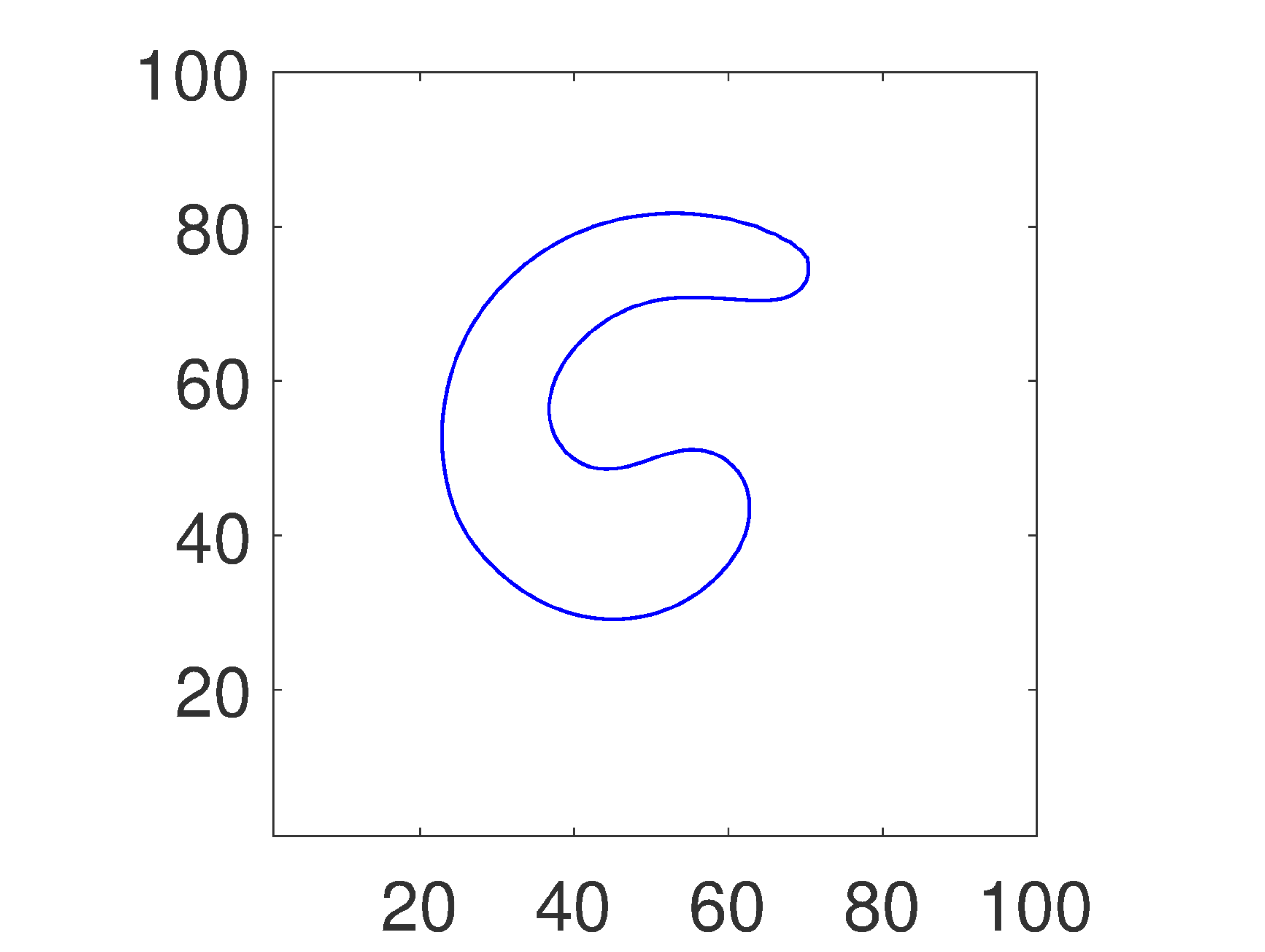}
  \includegraphics[scale=0.2]{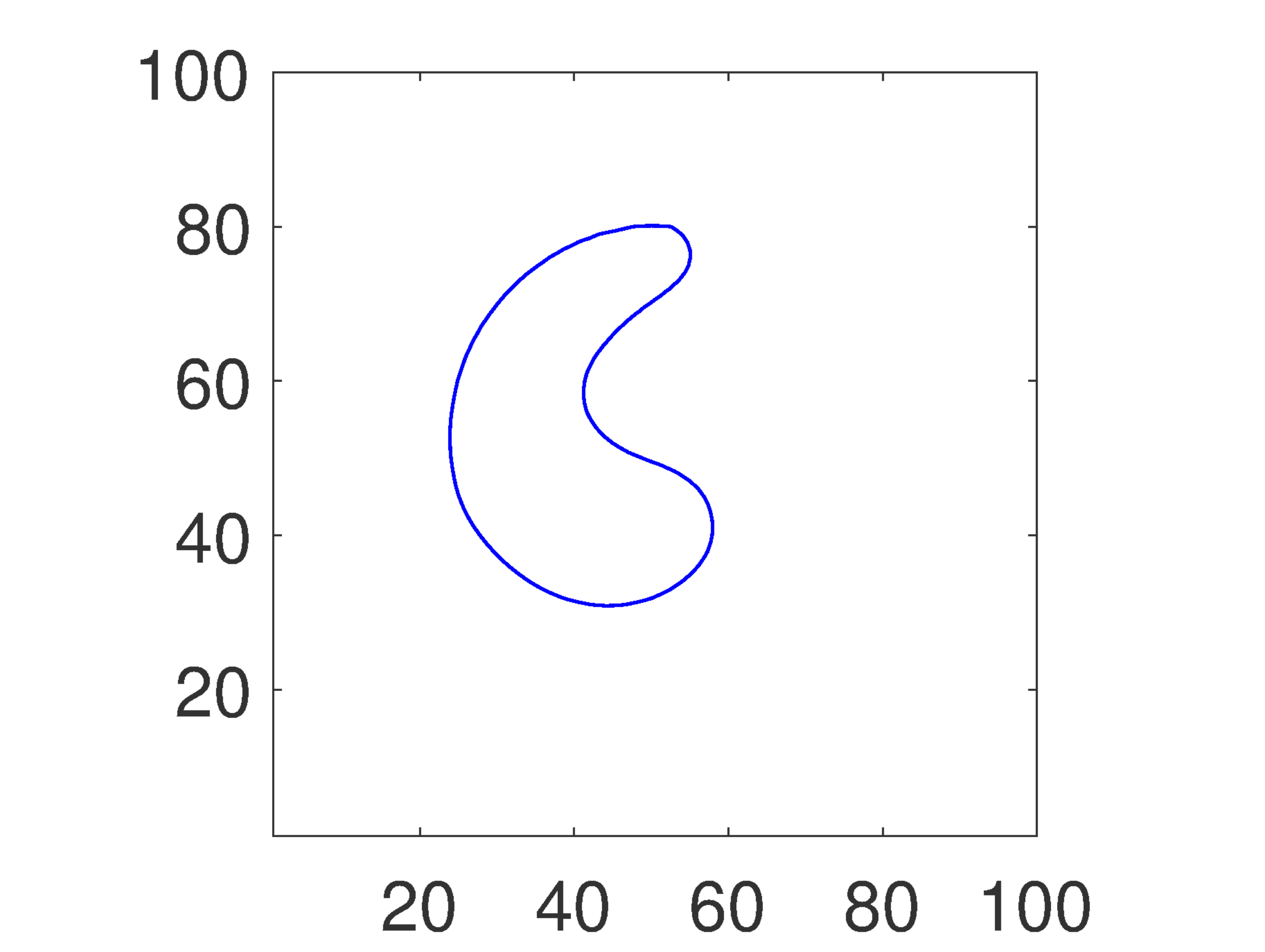}
  \includegraphics[scale=0.2]{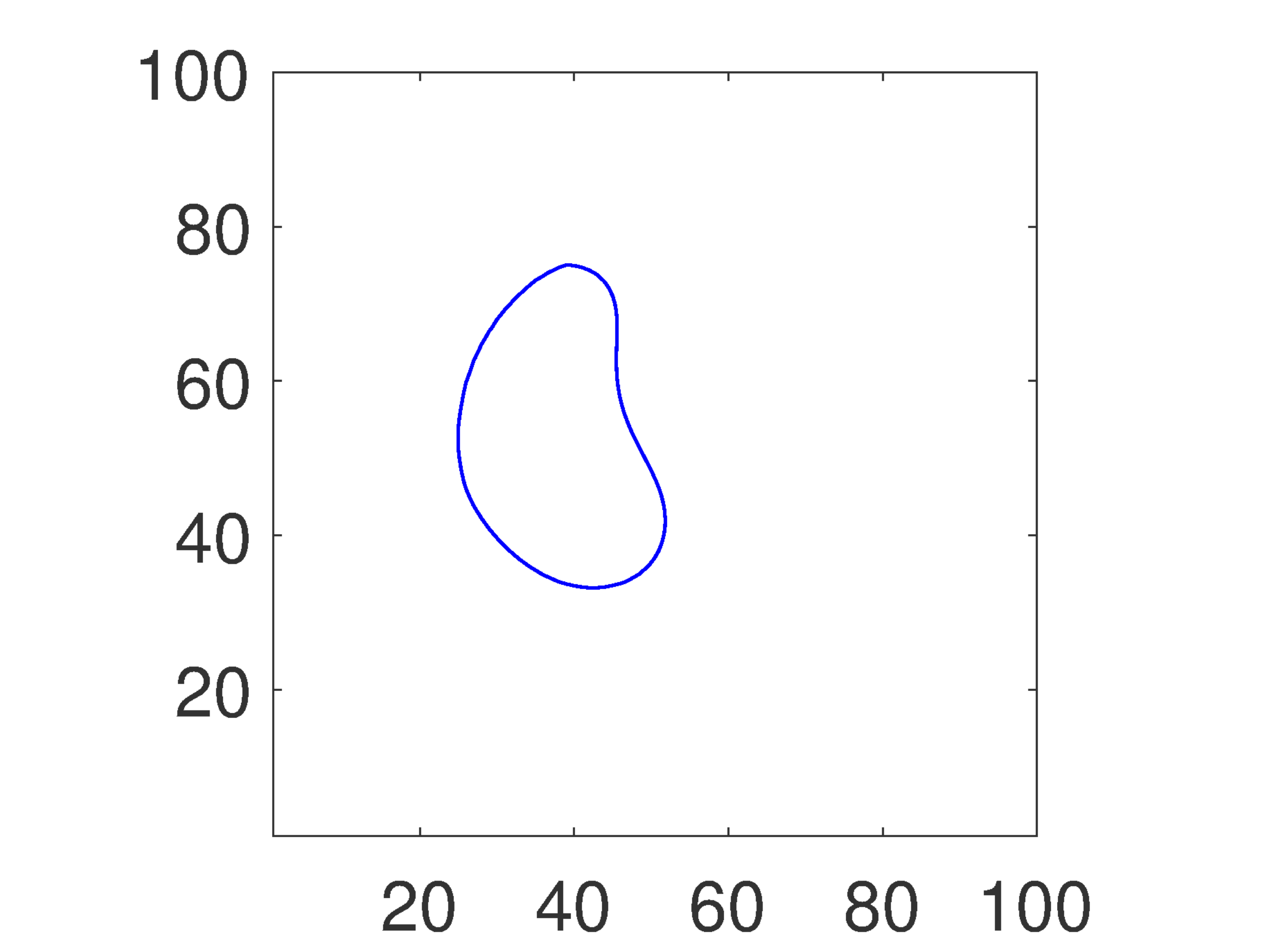}
 \end{center}
 \caption{ Evolution of the spiral under a curvature-driven flow. The images are snapshots in time from left to right, top to bottom where the ends of the spiral unwind faster than the interior region for $t = 0.0, 37.5, 75.0, 112.5, 150.0, 187.5, 225.0, 262.5, 300.0$.}
 \label{fig:curv_spiral}
\end{figure}
  Figure \ref{fig:curv_spiral} shows results of the same curvature-driven motion for a wound spiral given by the following parameterization
\begin{equation}
 \theta = 2 \pi D \sqrt{s} \quad \textrm{for} \quad s \in [0,1], 
\end{equation}
such that $s = (k+a)/(np+a)$ where $k$ is an integer that loops over the total number of points $np$, and the value of $a$ determines the shape of the spiral head in the center. The location of the points is defined by  
\begin{equation}
x_p = x_c + sf \, \big[D/(1+D) \sqrt{s} \cos (\theta)\big] \quad \textrm{and} \quad y_p = y_c + sf \, \big[D/(1+D) \sqrt{s} \sin(\theta)\big],
\end{equation}
where $\mathbf{x_c}$ is the center location of the domain, $sf$ the scaling factor for the spiral size and $D$ the number of spirals. Define a distance function $d(\mathbf{x})=||\mathbf{x}-\mathbf{x_p}||$ such that the zero level set comes out as 
\begin{equation}
 \phi(\mathbf{x},0) = d(\mathbf{x}) - w 
\end{equation}
where $w$ is the width of the spirals. The computational domain is a square with side length of $100$ units and a grid size of $100\times100$. The time-step is $dt=0.1$ and $\alpha=0.4$. The wound spiral level curve is centered at $\mathbf{x_c}=(50,50)$ with $np=400$, $D=2.5$, $a=3$ and $sf=50$. The simulation is run for $3000$ iterations. Figure \ref{fig:curv_spiral} shows the progression of the front evolving under a curvature-driven flow for $t = 0.0, 37.5, 75.0, 112.5, 150.0, 187.5, 225.0, 262.5, 300.0$. The wound spiral collapses to a circle under this motion where the region of positive curvature (spiral ends) collapse faster than the elongated regions. 
%
%

\subsubsection{Dumbbell}
\begin{figure}
 \begin{center}  
  \includegraphics[scale=0.2]{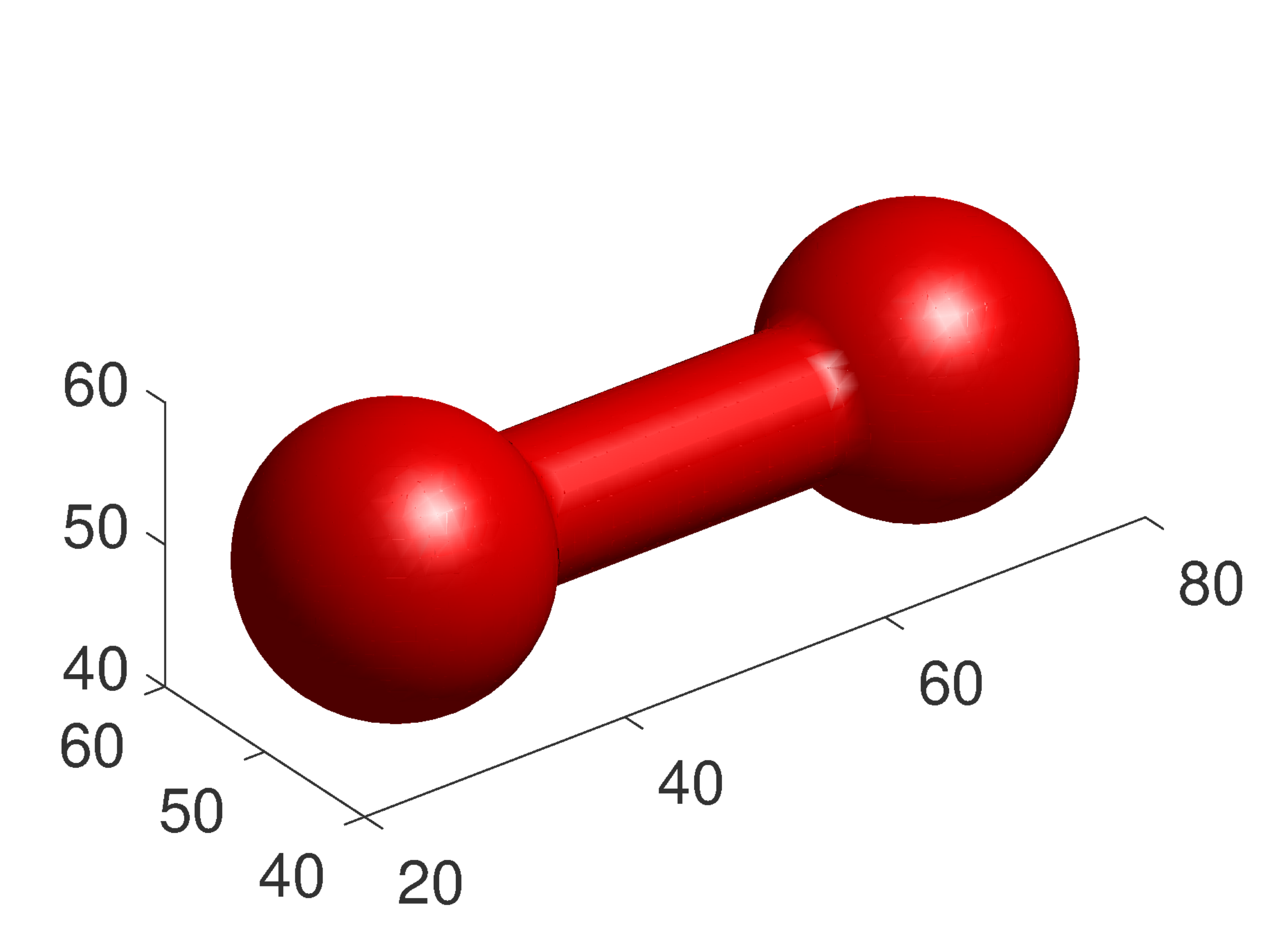}
  \includegraphics[scale=0.2]{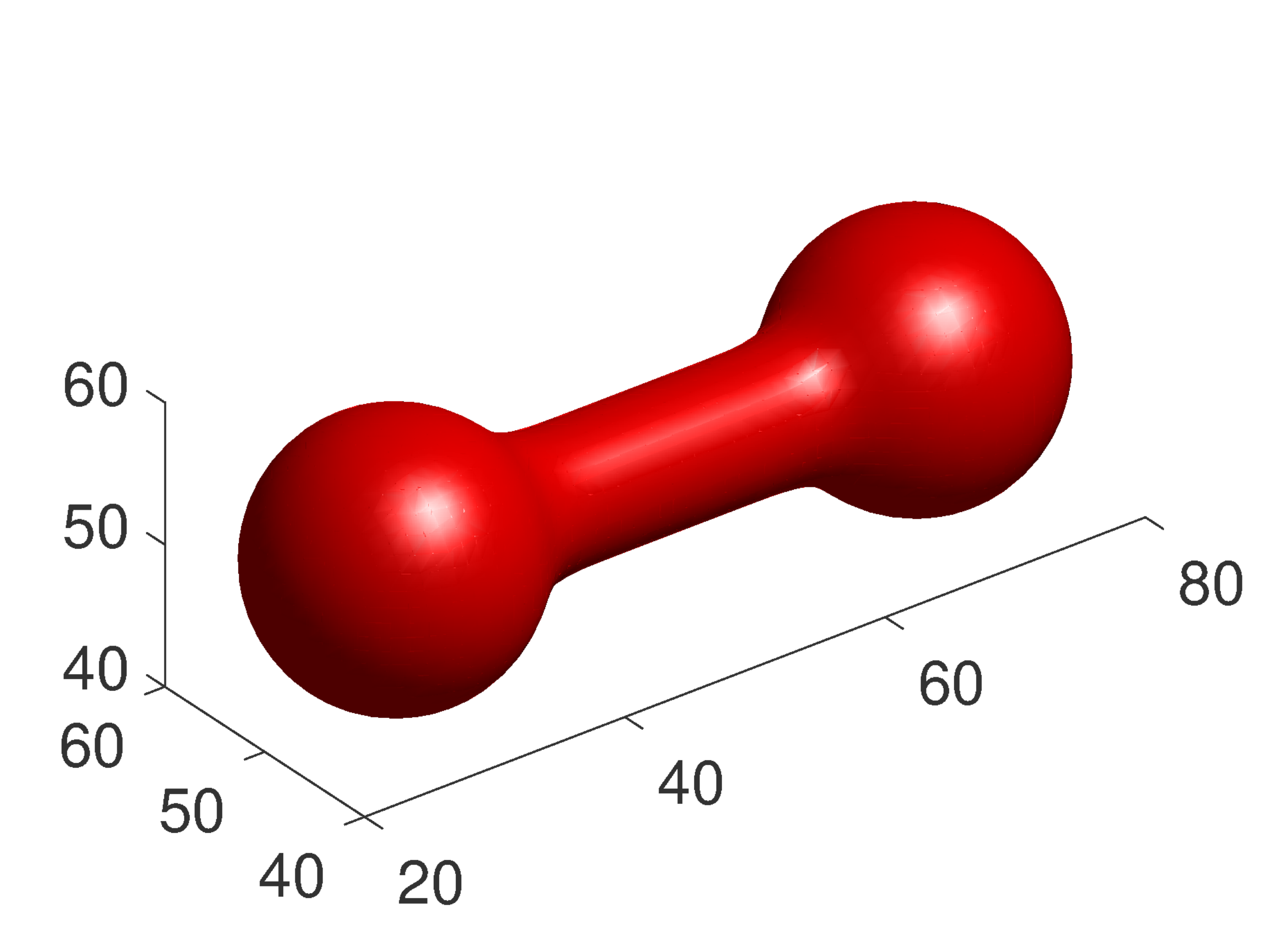}
  \includegraphics[scale=0.2]{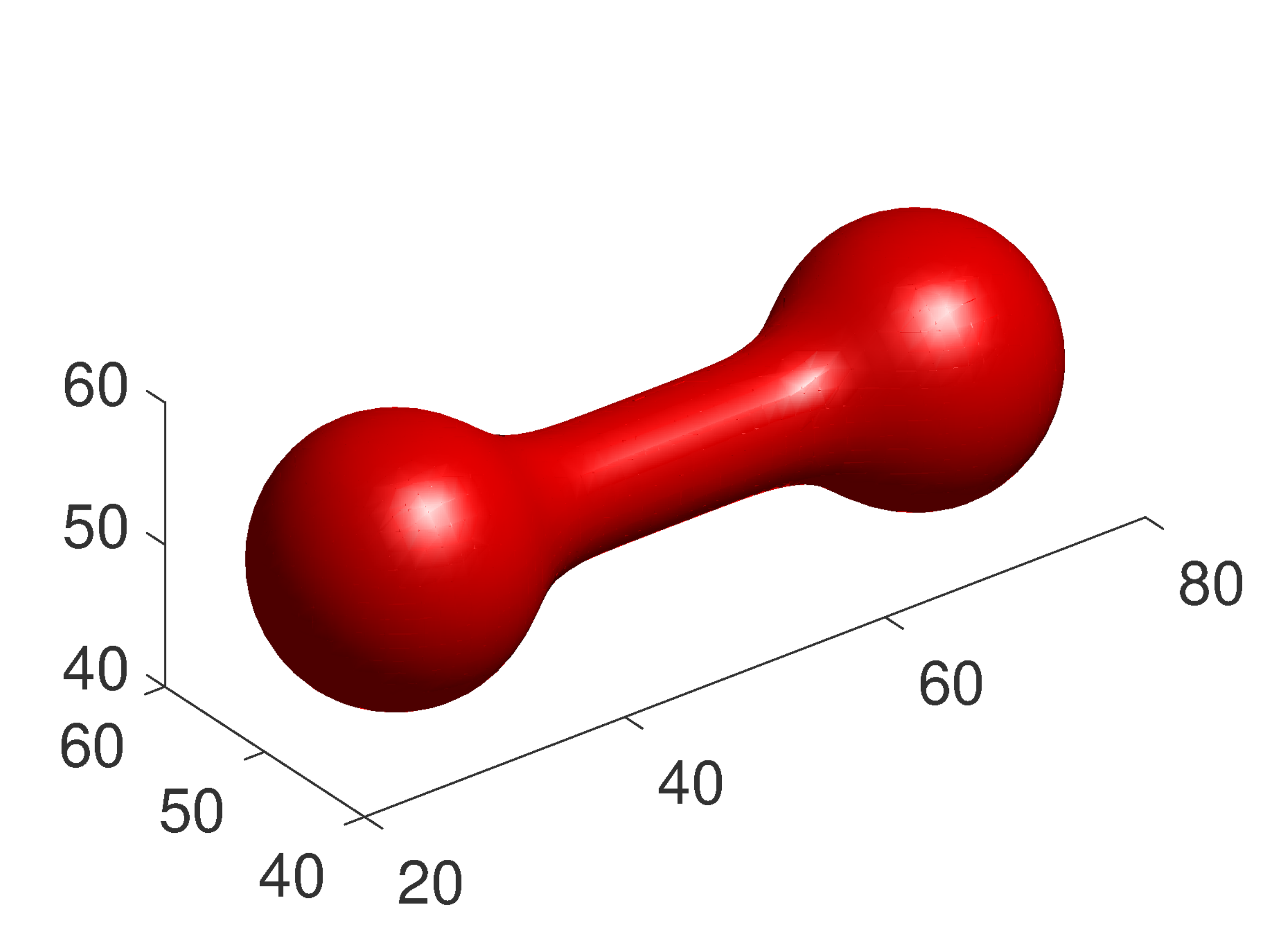}
  
  \includegraphics[scale=0.2]{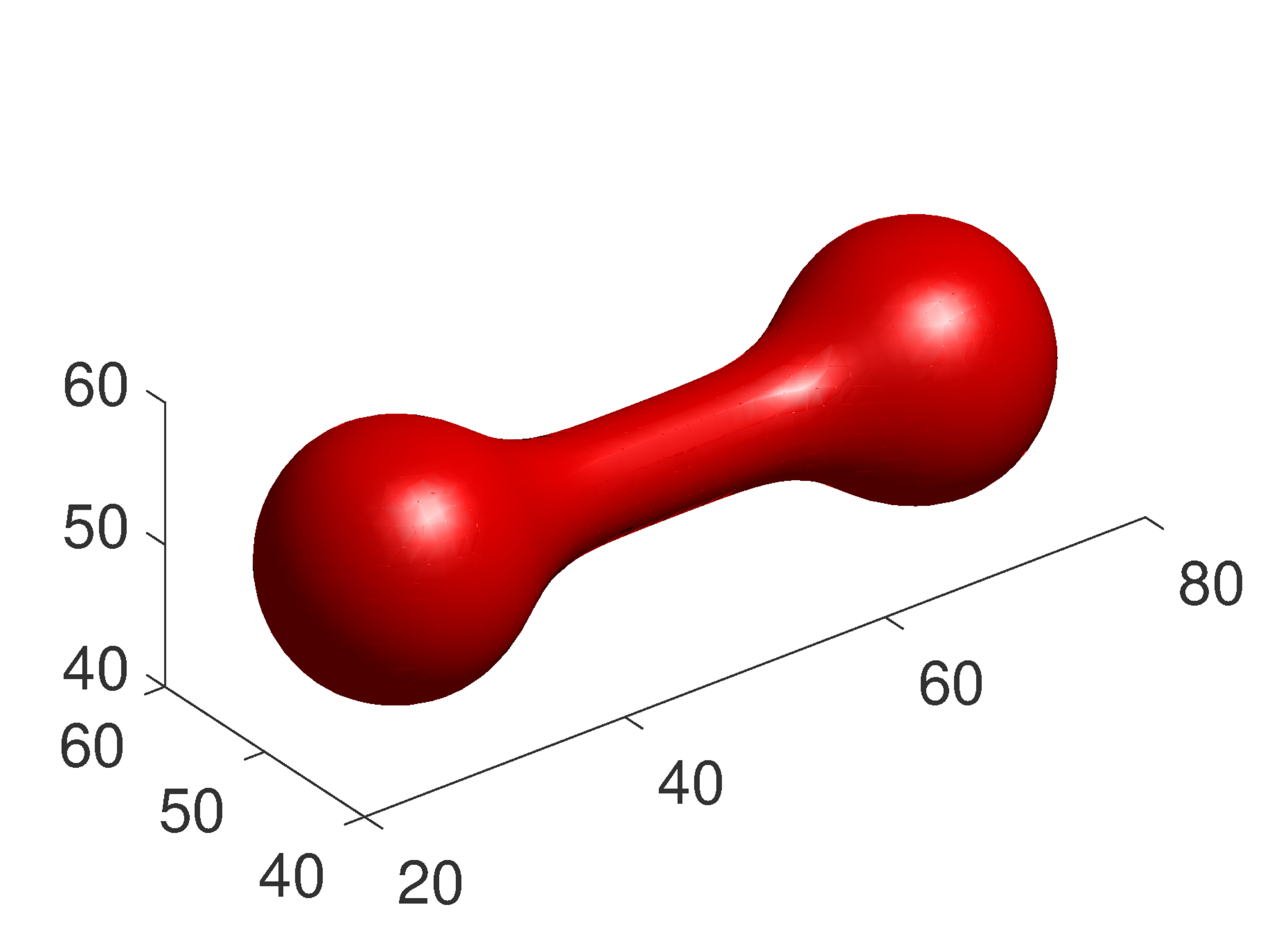}
  \includegraphics[scale=0.2]{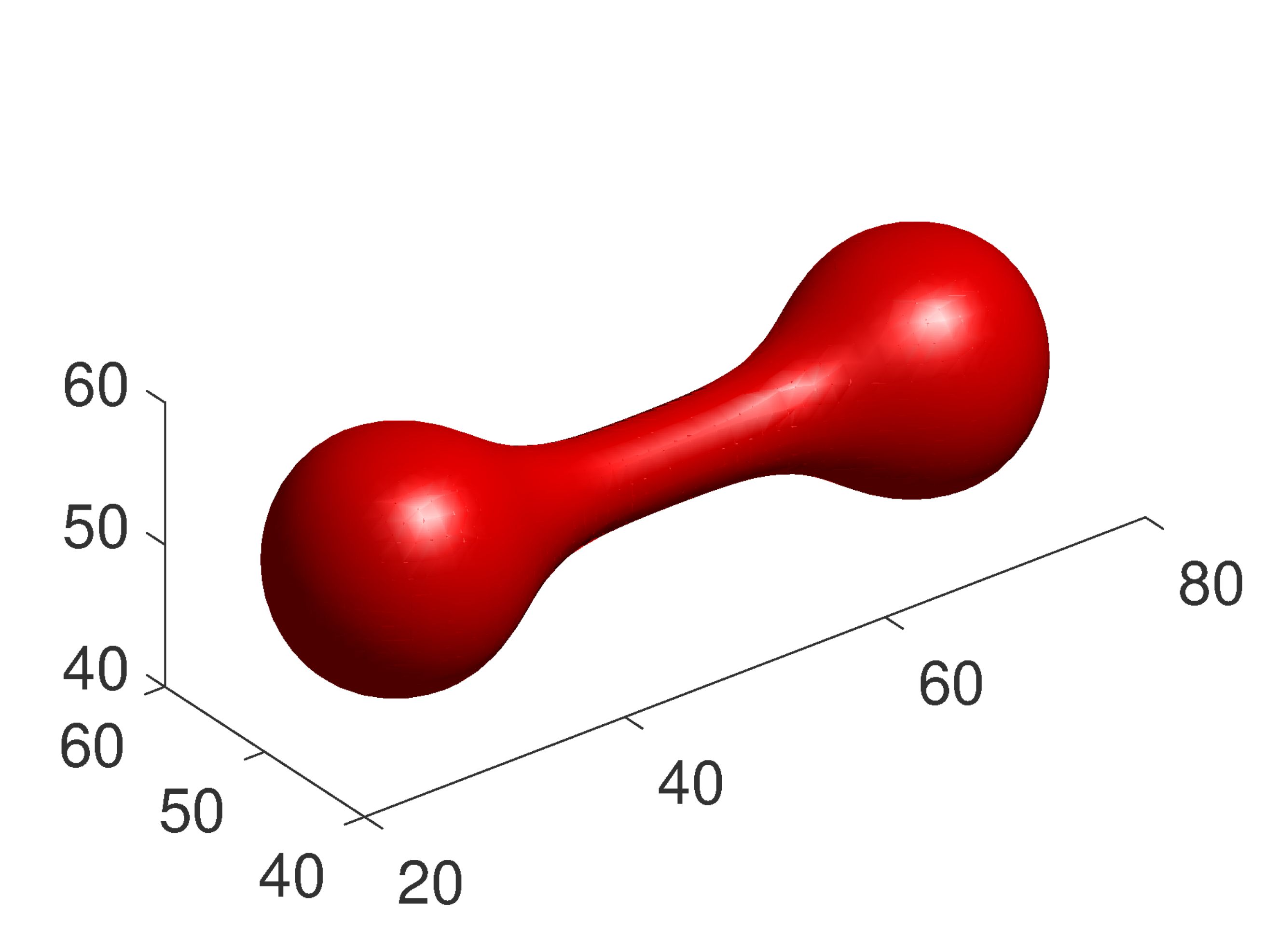}
  \includegraphics[scale=0.2]{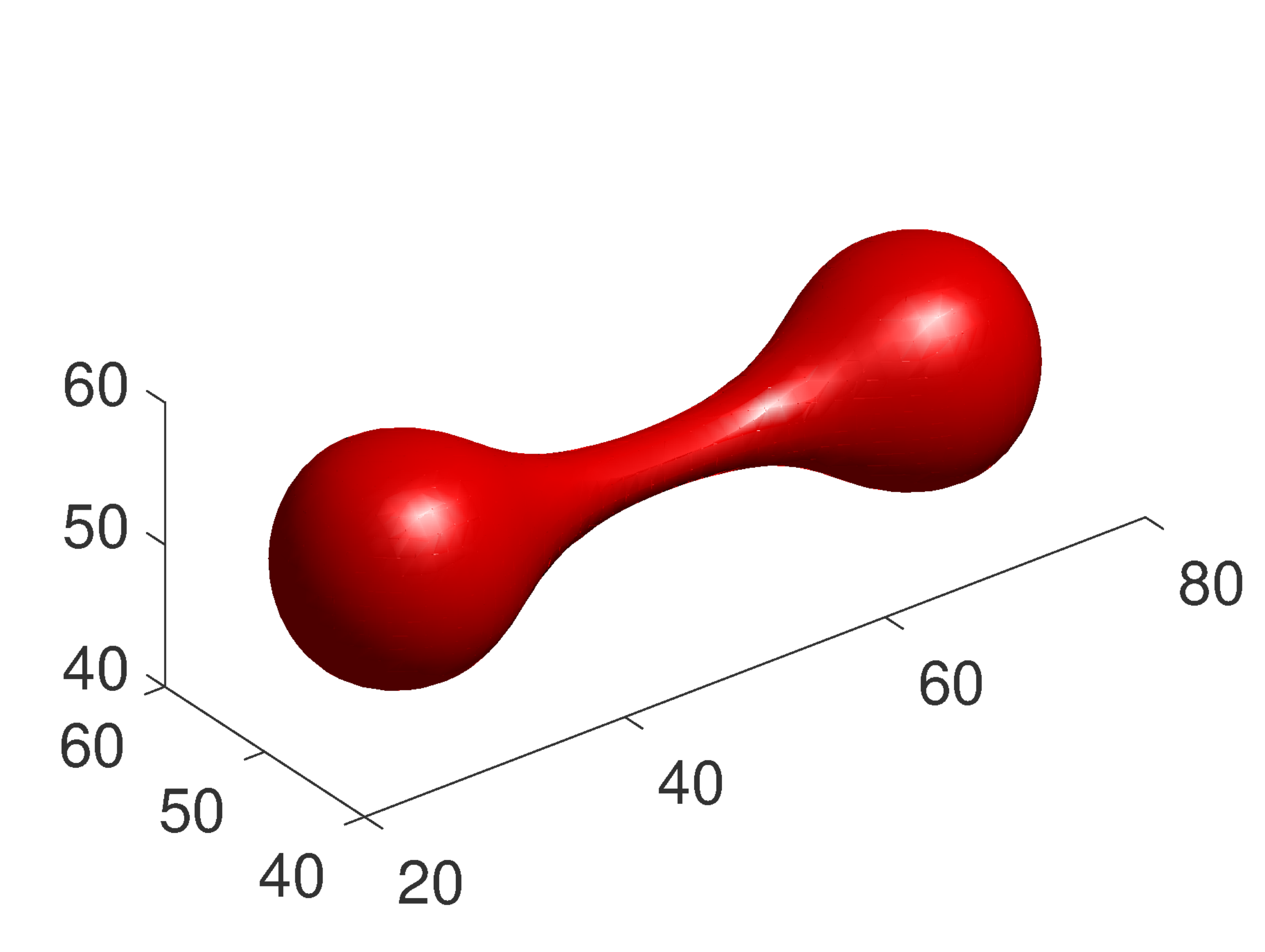}
 
  \includegraphics[scale=0.2]{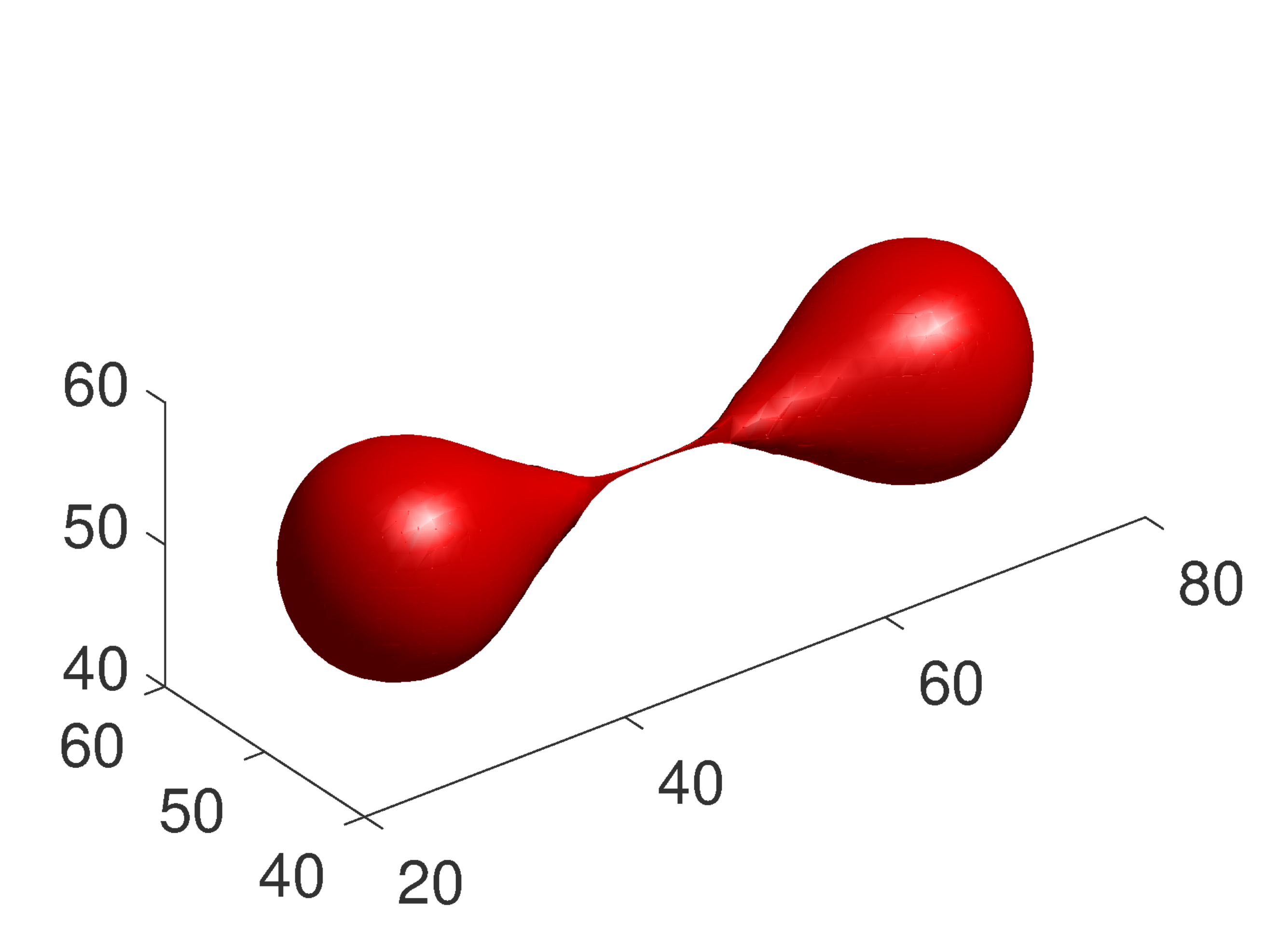}
  \includegraphics[scale=0.2]{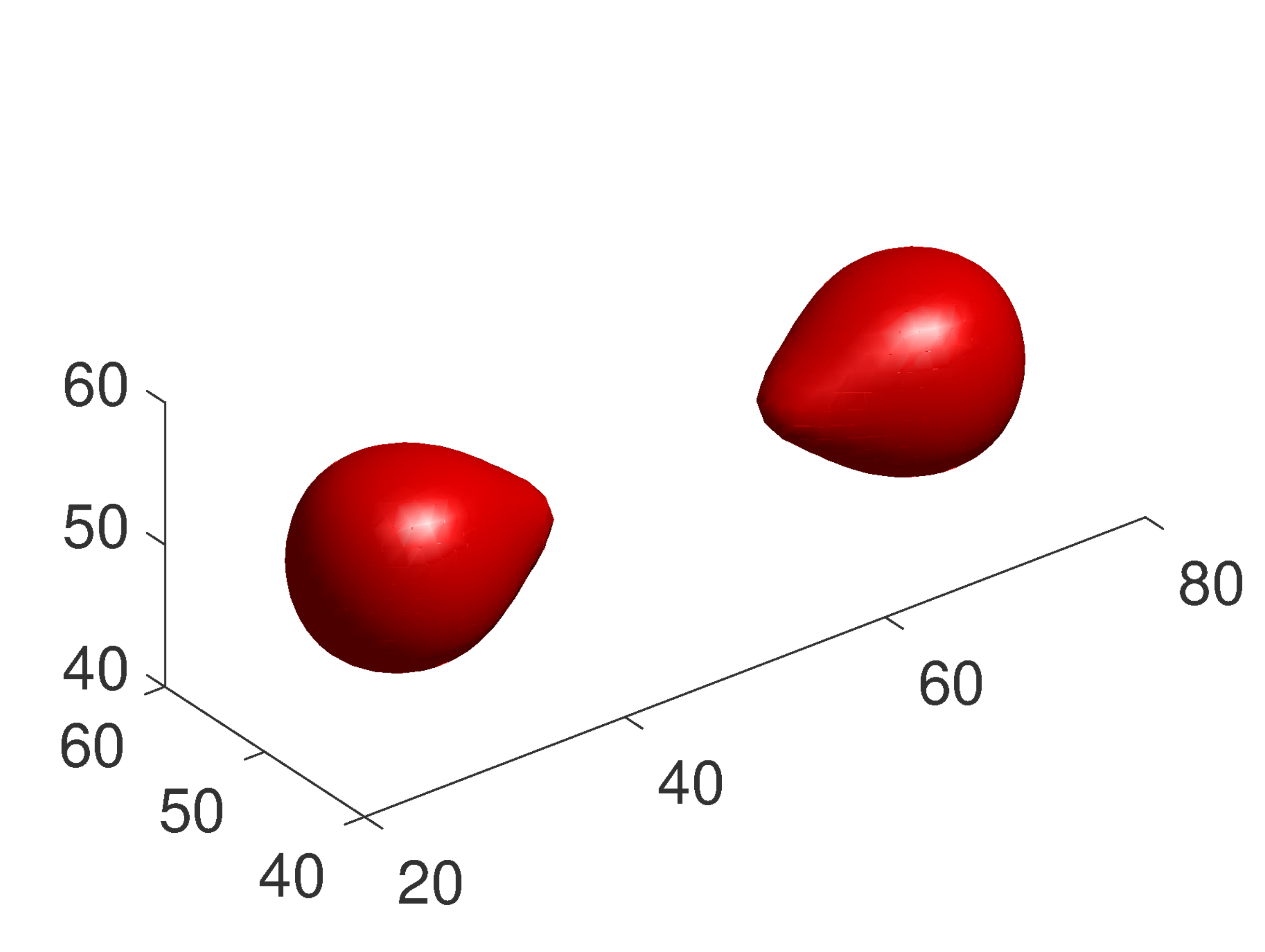}
  \includegraphics[scale=0.2]{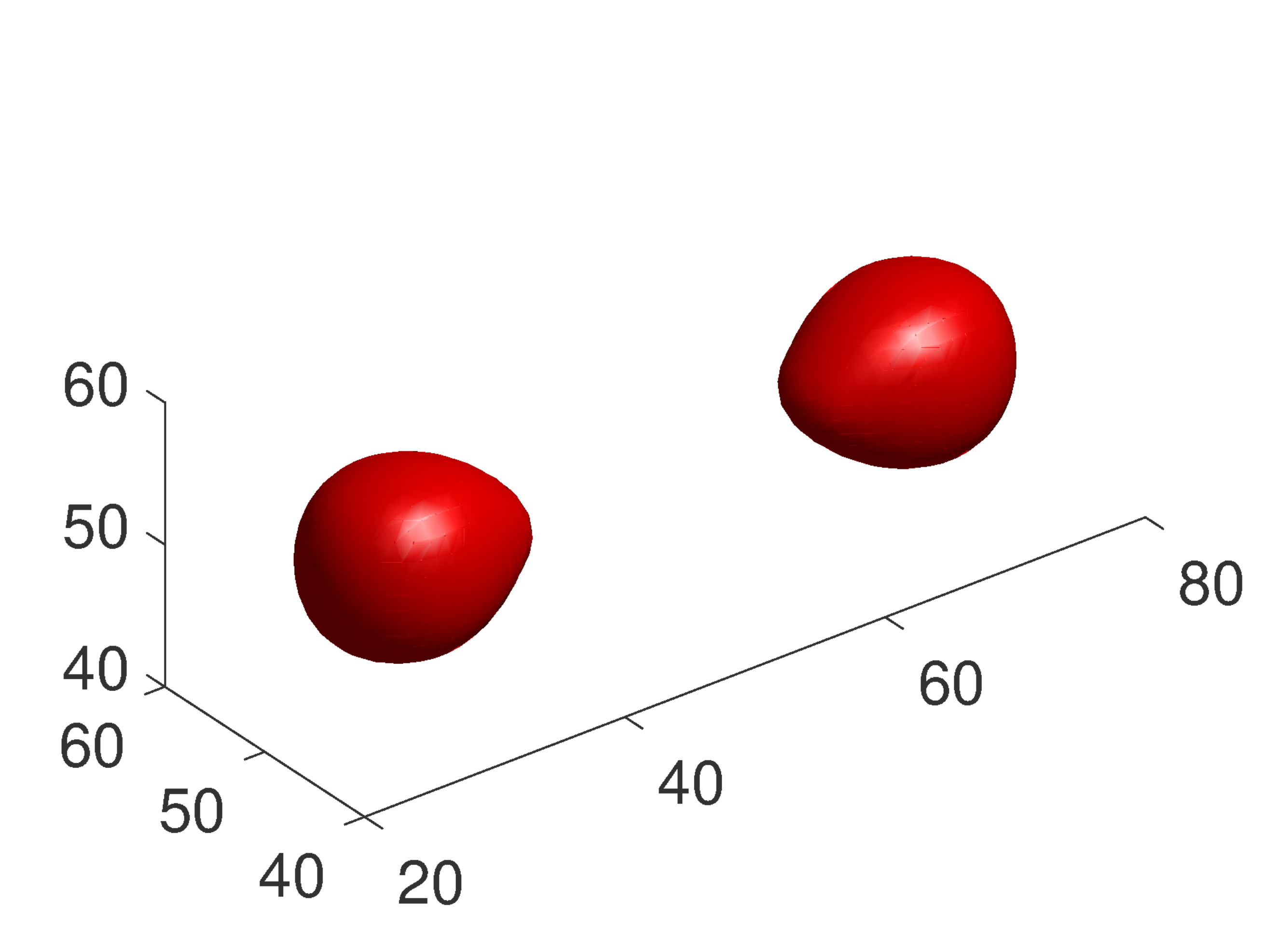}
 \end{center}
 \caption{ Evolution of a dumbbell shaped surface under a curvature-driven flow. The images are snapshots in time from left to right, top to bottom where the handle of the dumbbell shrinks faster than the spherical shells due to its high curvature leading to a pinch-off for $t = 0.0, 2.0, 4.0, 6.0, 8.0, 10.0, 12.0, 14.0, 16.0$.}
 \label{fig:curv_dumbbell}
\end{figure}

\begin{figure}
 \begin{center}
    \includegraphics[scale=0.4]{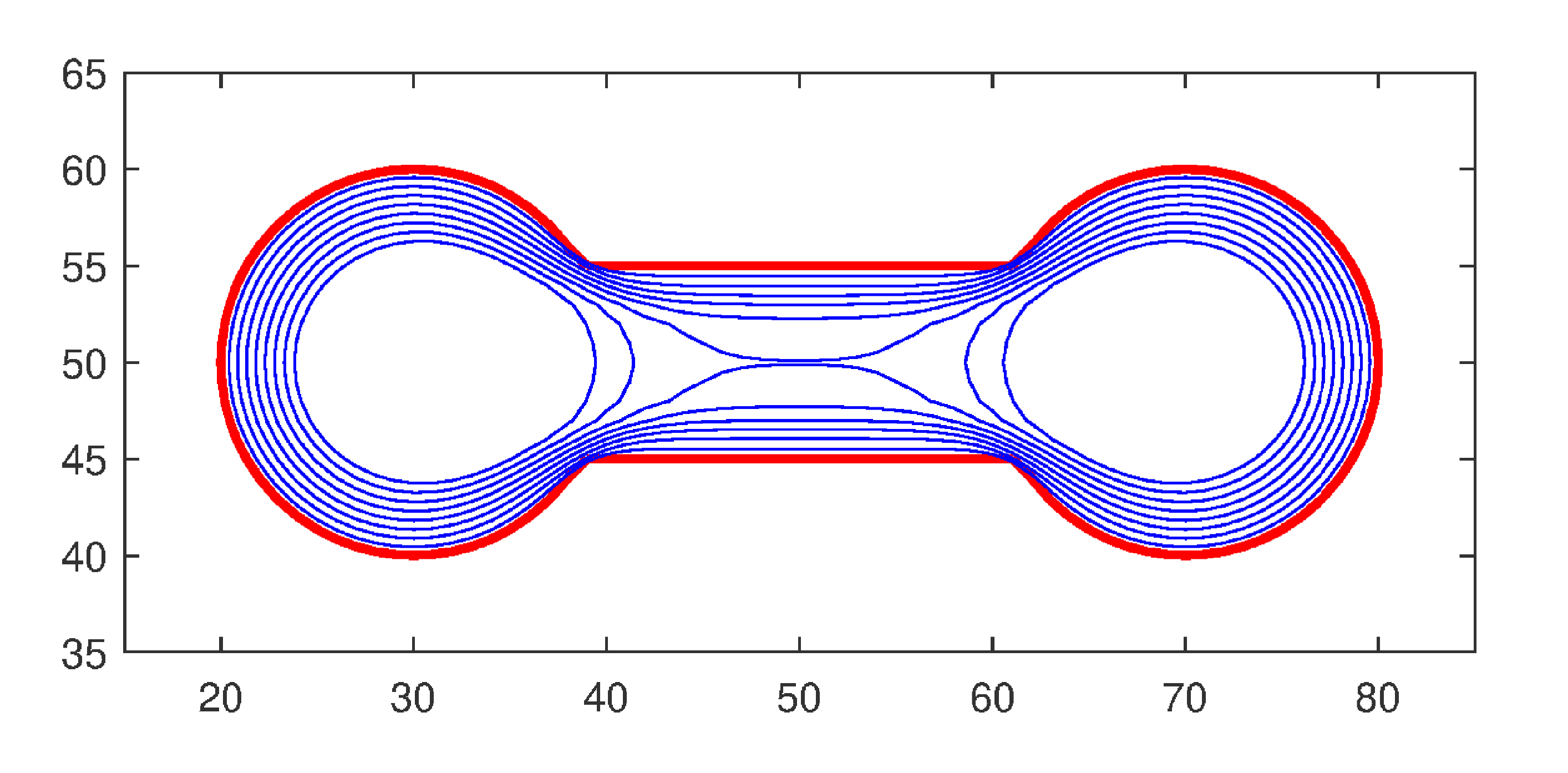}
 \end{center}
 \caption{ A $2$D slice in the middle of the dumbbell at $z=50$ for the zero level set contour evolving under curvature-driven motion for $t = 0.0, 2.0, 4.0, 6.0, 8.0, 10.0, 12.0, 14.0, 16.0$.}
 \label{fig:curv_dumbbell_cut}
\end{figure}

%
%
 The previous examples presented level curves in $2$D. Here, we show that the method is also applicable in $3$D where we examine the curvature-driven motion of a dumbbell-shaped level surface. The level surface is initialized by taking advantage of the simplicity of Boolean functions in the LSM. Consider the union of two spheres and a cylinder given by: 
 
 \begin{equation}
 \begin{split}
  \phi_{left}(\mathbf{x}) = \sqrt{(x-x_c+o)^2+(y-y_c)^2+(z-z_c)^2}-r, \\
  \phi_{right}(\mathbf{x}) = \sqrt{(x-x_c-o)^2+(y-y_c)^2+(z-z_c)^2}-r, \\
  \phi_{center}(\mathbf{x}) = \max \bigg [ \big(|x-x_c|-o\big),\bigg (\sqrt{(y-y_c)^2+(z-z_c)^2}-w \bigg)  \bigg ], \\
  \phi(\mathbf{x},0) = \min [\phi_{left},\phi_{right},\phi_{center}],
 \end{split}
 \end{equation}
 where $\mathbf{x_c}$ are the domain center coordinates, $o$ is the distance between the center of the spherical shells and the center of the cylinder, $r$ the radius of the spherical shells and $w$ the radius of the center cylinder. Figure \ref{fig:curv_dumbbell} demonstrates the ability of the level set methodology to handle pinching, merging and separation of surfaces without any special mathematical or algorithmic `surgery'. The initial dumbbell level set evolves under the motion of mean curvature, and the handle shrinks faster due to its higher curvature than the spherical shells on both ends. This leads to a pinch-off at the center where the implicit surface separates into two tear drop surfaces that also shrink due to curvature. The computational domain is a cube with side length of $100$ units and grid size of $100\times100\times100$. The time-step is $dt=0.1$ and $\alpha=0.4$. The dumbbell level surface is centered at $\mathbf{x_c}=(50,50,50)$ with $r=10$, $w=5$ and $o=20$. The case is run for $160$ iterations. Figure \ref{fig:curv_dumbbell} shows the progression of the front evolving under a curvature-driven flow for $t = 0.0, 2.0, 4.0, 6.0, 8.0, 10.0, 12.0, 14.0, 16.0$. Figure \ref{fig:curv_dumbbell_cut} shows a cut slice at $z=50$ where the evolution of the dumbbell leading up to pinching is more clear.

\subsection{Motion in the Normal Direction}\label{subsec:motion_normal}
Consider motion in the normal direction where the interface evolves under an internally generated velocity field. The velocity field is defined by $\mathbf{u}=a\mathbf{N}$ where $a$ is a constant that can be either positive or negative. The equations of motion for the level set equation are given by:
\begin{equation}
 \pop{\phi}{t} + a \mathbf{N} \cdot \nabla \phi = 0,
\end{equation}
which can be rewritten as 
\begin{equation}
 \pop{\phi}{t} + v_n |\nabla \phi| = 0,
\end{equation}
where $v_n=a$ is the constant velocity in the normal direction. 
This motion can also be represented as a case of energy minimization. Express the energy functional in terms of a volume enclosed by the surface $\phi<0$ such that 
\begin{equation}
  \mathcal{E}_{ext}(\phi) = v_n \int_{\Omega} H(-\phi) \, \mathrm{d} \mathbf{x}.
\end{equation}
Taking the Fr{\'e}chet derivative of $\mathcal{E_{ext}}$ with an $L^2$ norm gives
\begin{equation}
\pop{\mathcal{E}_{ext}}{\phi}=v_n \delta(-\phi)=-v_n \delta(\phi).
\end{equation}
A similar argument to \S\ref{subsec:motion_curvature} can be made. The Dirac delta functional singles out the zero level set. In order to obtain motion in the normal direction, all the level sets should minimize their values such that the volume enclosed under each respective level curve $V_{\beta}(\phi)= \int_{\Omega} H(-\phi+\beta) \, \mathrm{d} \mathbf{x}$ tends to zero. The definition of the energy functional becomes:
\begin{equation}
  \mathcal{E}_{ext}(\phi) =  v_n \int_{\beta} \int_{\Omega} H(-\phi+\beta) \, \mathrm{d} \mathbf{x} \, \mathrm{d} \beta \, .
\end{equation}
Therefore by definition, the Fr{\'e}chet derivative of $\mathcal{E}_{ext}$ using the new norm is defined as:
\begin{equation}
 \lim_{\epsilon \to 0} \frac{1}{\epsilon} \bigg [ \mathcal{E}_{ext}(\phi+\epsilon \chi)-\mathcal{E}_{ext}(\phi) \bigg ] = \bigg \langle \pop{\mathcal{E}_{ext}}{\phi}, \chi \bigg \rangle_{\phi} =  v_n \int_{\beta} \int_{\Omega} \delta(-\phi+\beta) \, |\nabla \phi| \frac{\chi}{|\nabla \phi|} \, \mathrm{d} \mathbf{x} \, \mathrm{d} \beta,
\end{equation}
where
\begin{equation}
 \int_{\beta} \int_{\Omega} \delta(-\phi+\beta) \, |\nabla \phi| \frac{\chi}{|\nabla \phi|} \, \mathrm{d} \mathbf{x} \, \mathrm{d} \beta = -\int_{\Omega} |\nabla \phi| \frac{\chi}{|\nabla \phi|} \, \mathrm{d} \mathbf{x}.
\end{equation}
Therefore, the expression can be written as  
 \begin{equation}
 \pop{\mathcal{E}_{ext}}{\phi}= -v_n|\nabla \phi| \, .
\end{equation}
The evolution equation becomes
\begin{equation}
 \pop{\phi}{t} = \alpha \nabla \cdot (d_p(|\nabla \phi|) \nabla \phi) + v_n|\nabla \phi|.
\end{equation}
The above equation leads to the traditional motion in the normal direction. We validate this type of motion for $v_n = -1$ and $v_n=1$, where the latter describes the motion by which all the level curves maximize the volume they enclose. Given the above, we compare the effect of reinitialization using DRLSE, traditional reinitialization and no reinitialization for several cases that will be discussed in the following sections. 

\subsubsection{Circle with $v_n=-1$}
\begin{figure}
 \begin{center}
  \includegraphics[scale=0.21]{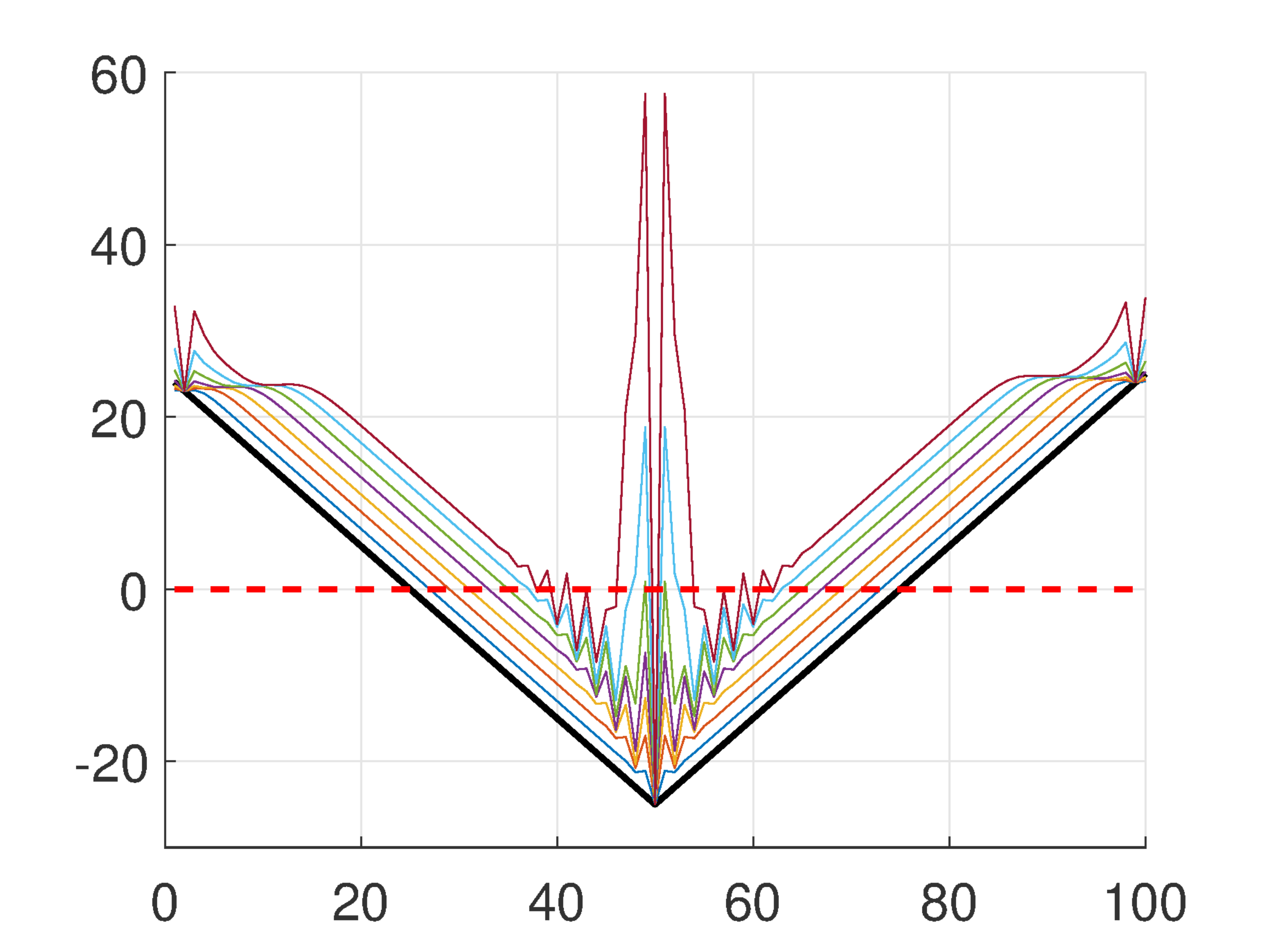}
  \put(-165,110){$(a)$}
  \includegraphics[scale=0.22]{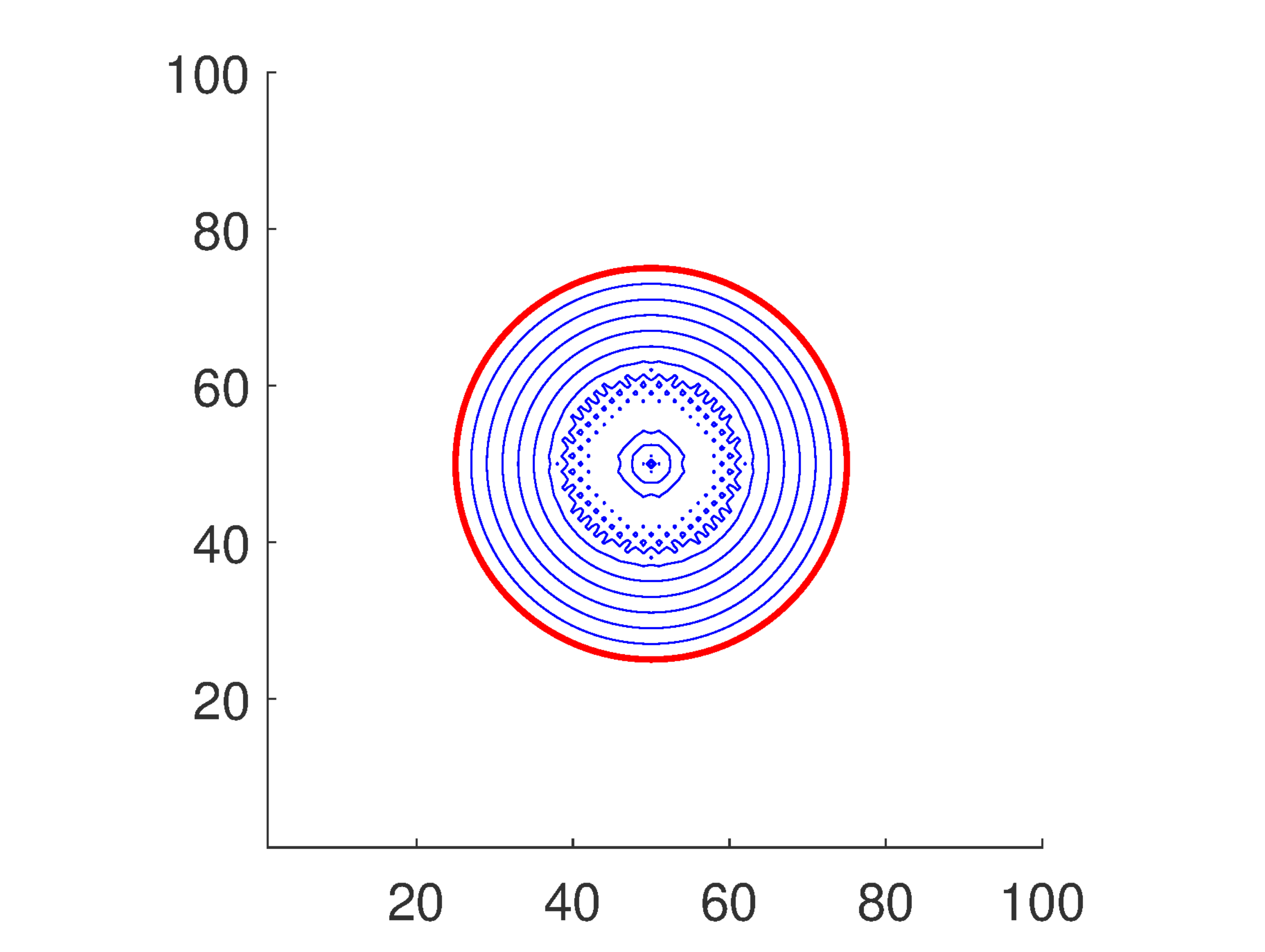}
  \put(-160,110){$(b)$}
  \includegraphics[scale=0.23]{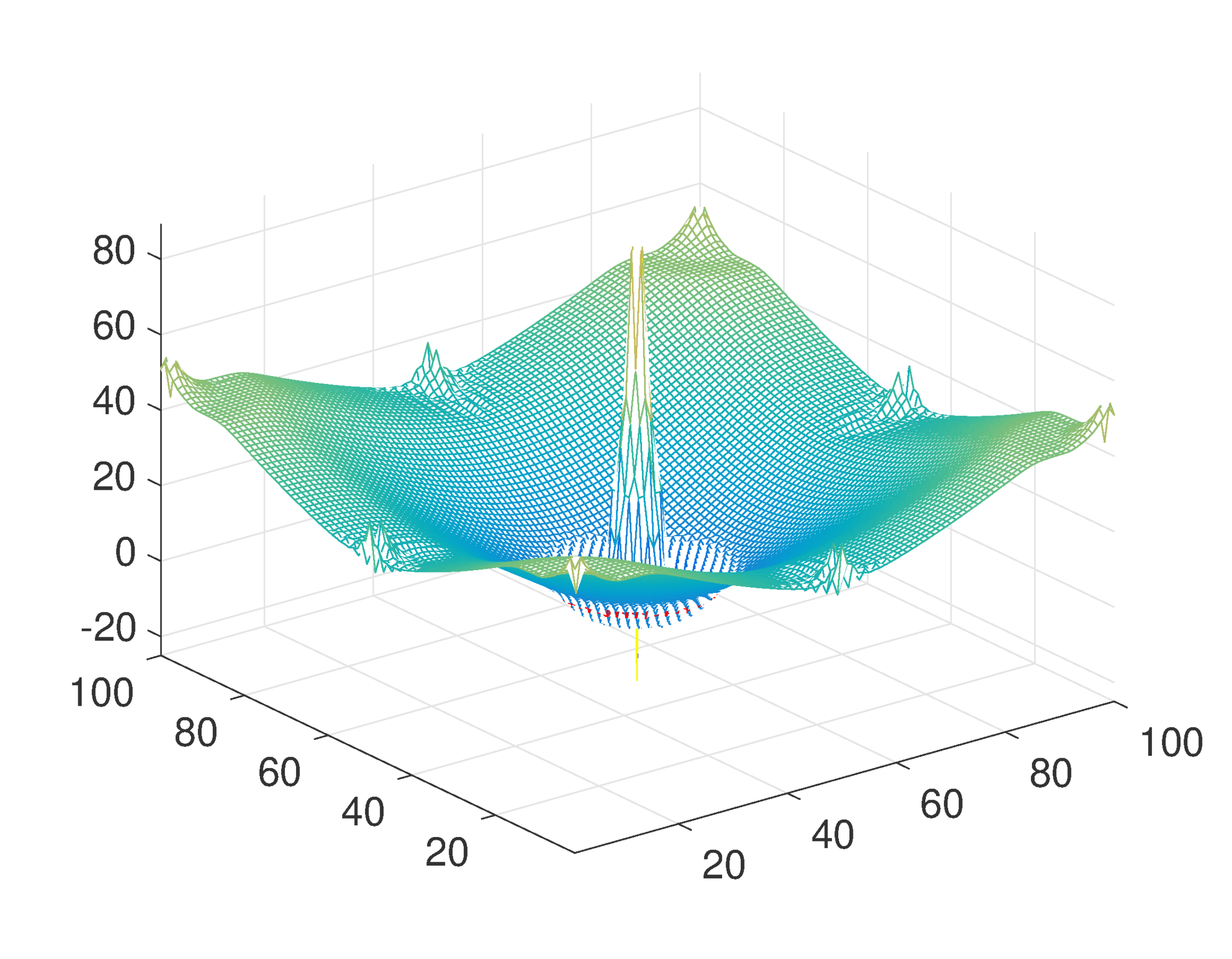}
  \put(-165,110){$(c)$}\\
  \includegraphics[scale=0.21]{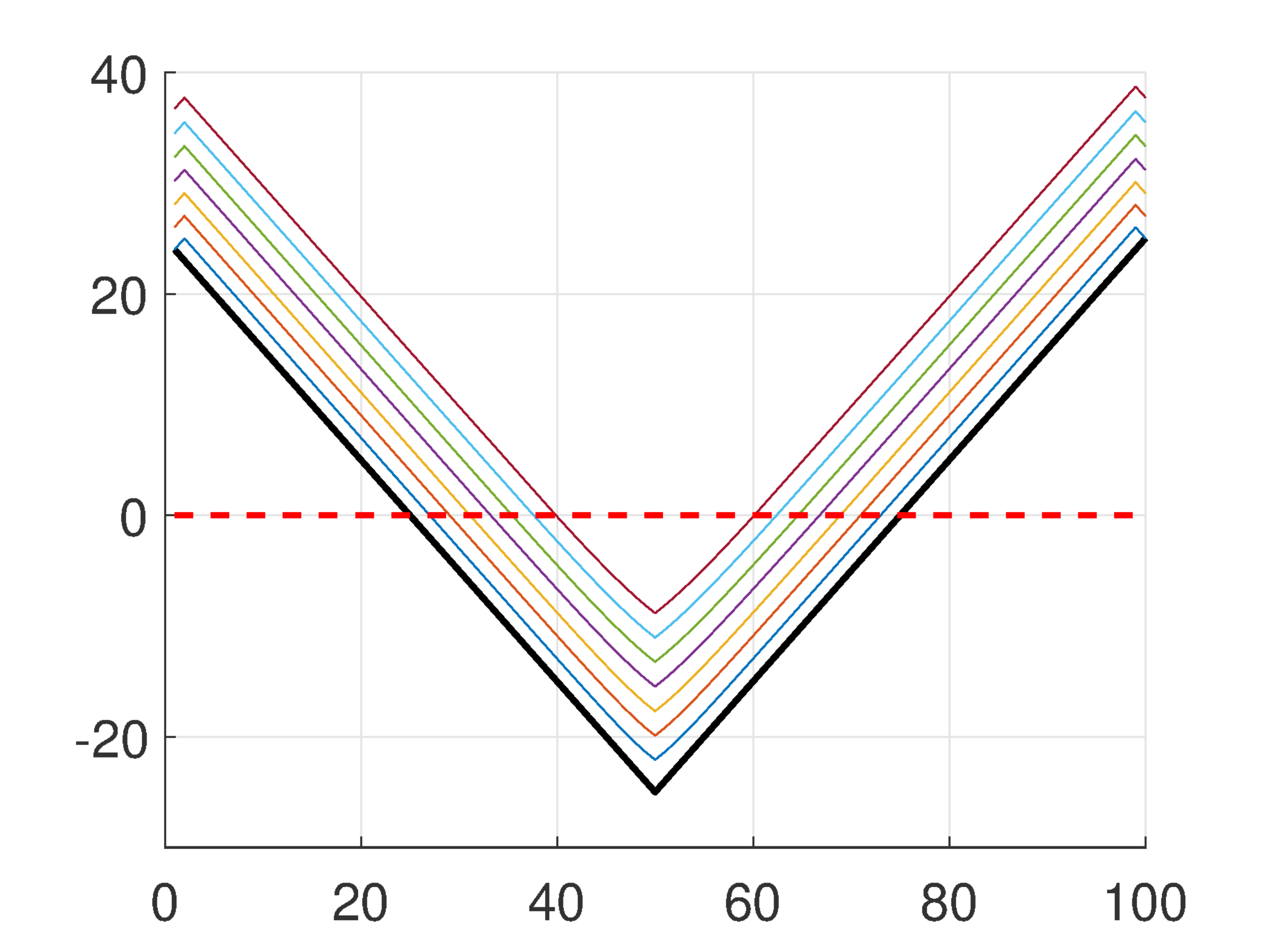}
  \put(-165,110){$(d)$}
  \includegraphics[scale=0.22]{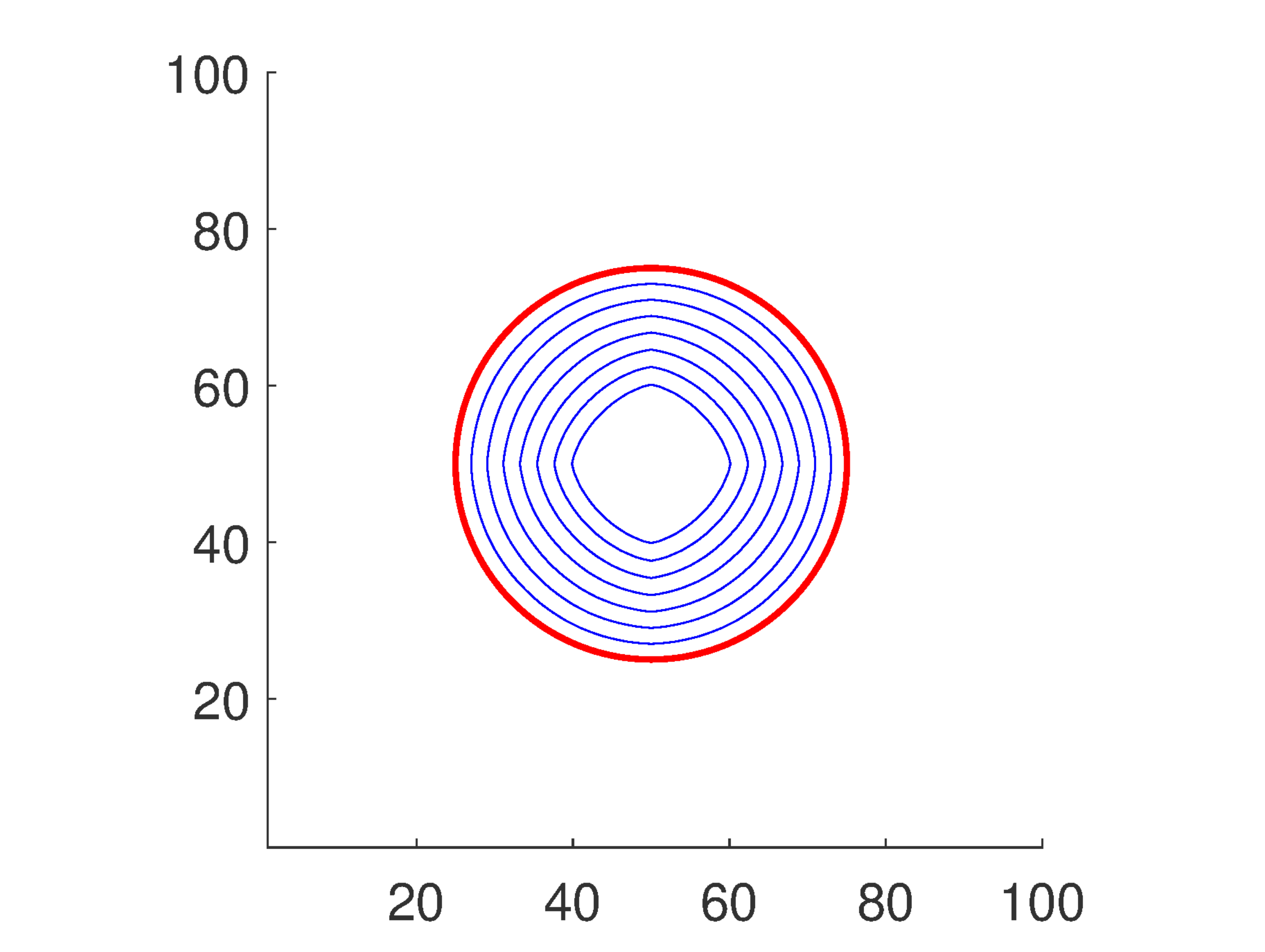}
  \put(-160,110){$(e)$}
  \includegraphics[scale=0.23]{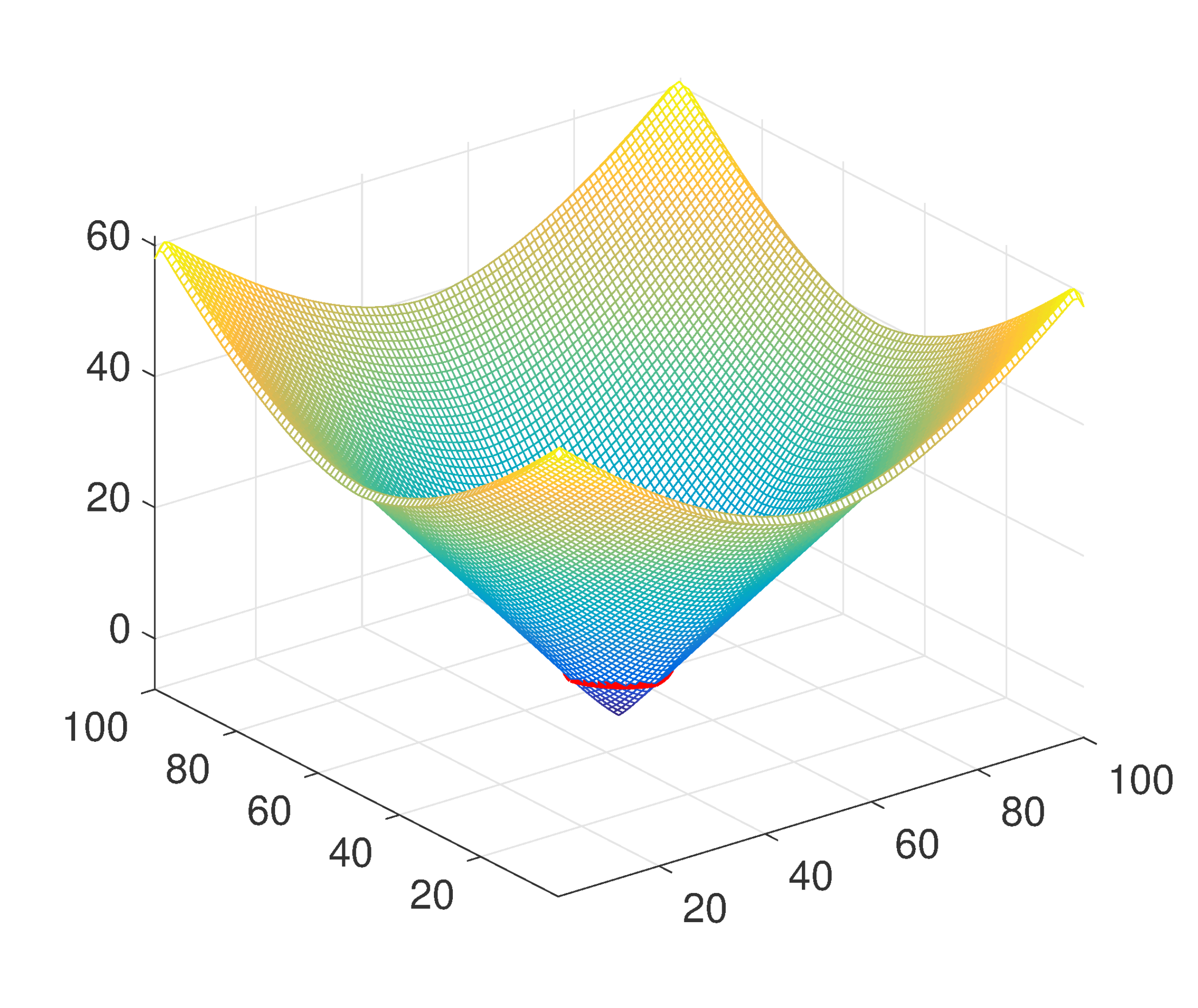}
  \put(-155,110){$(f)$}\\
  \includegraphics[scale=0.21]{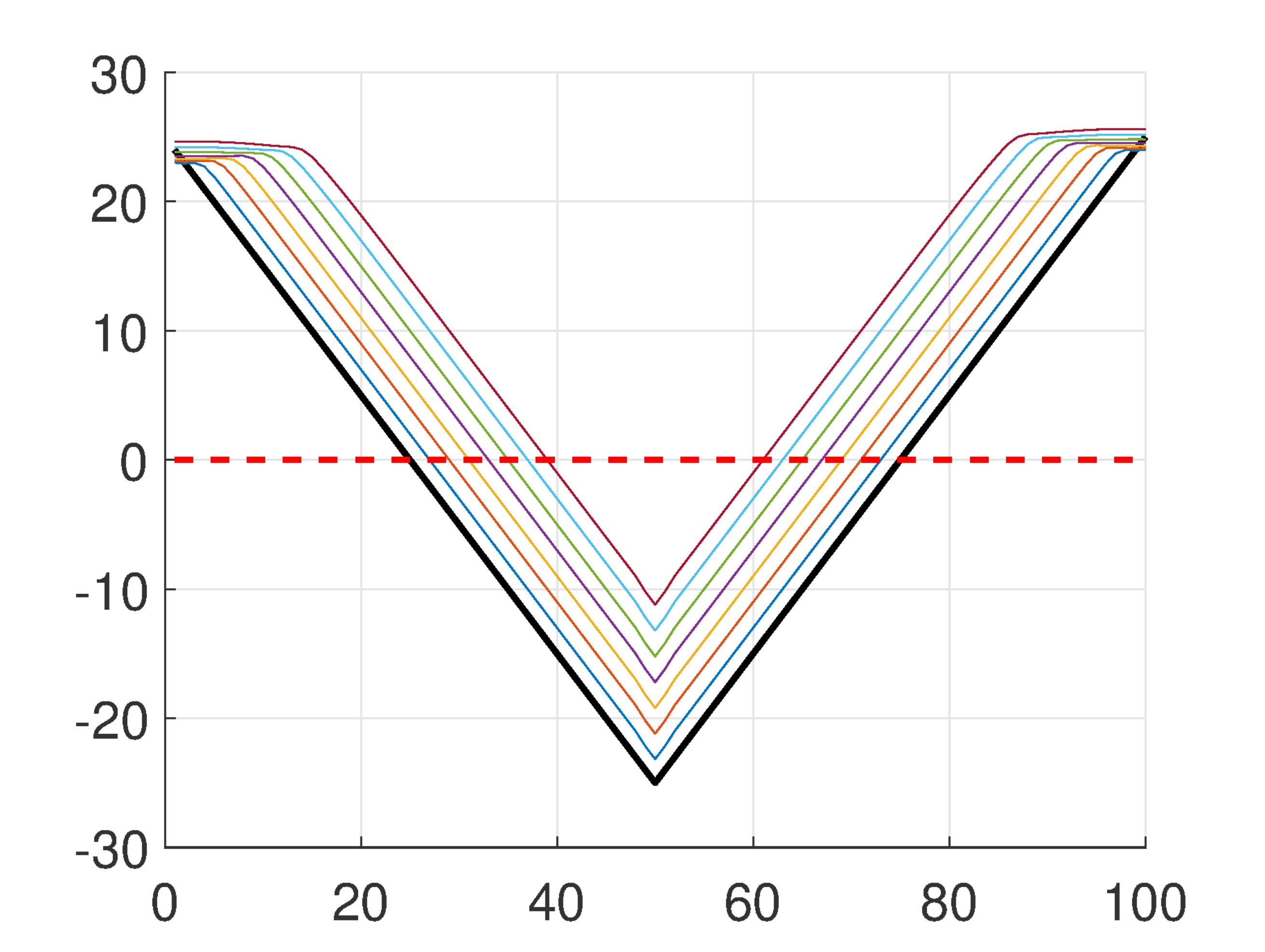}
  \put(-165,110){$(g)$}
  \includegraphics[scale=0.22]{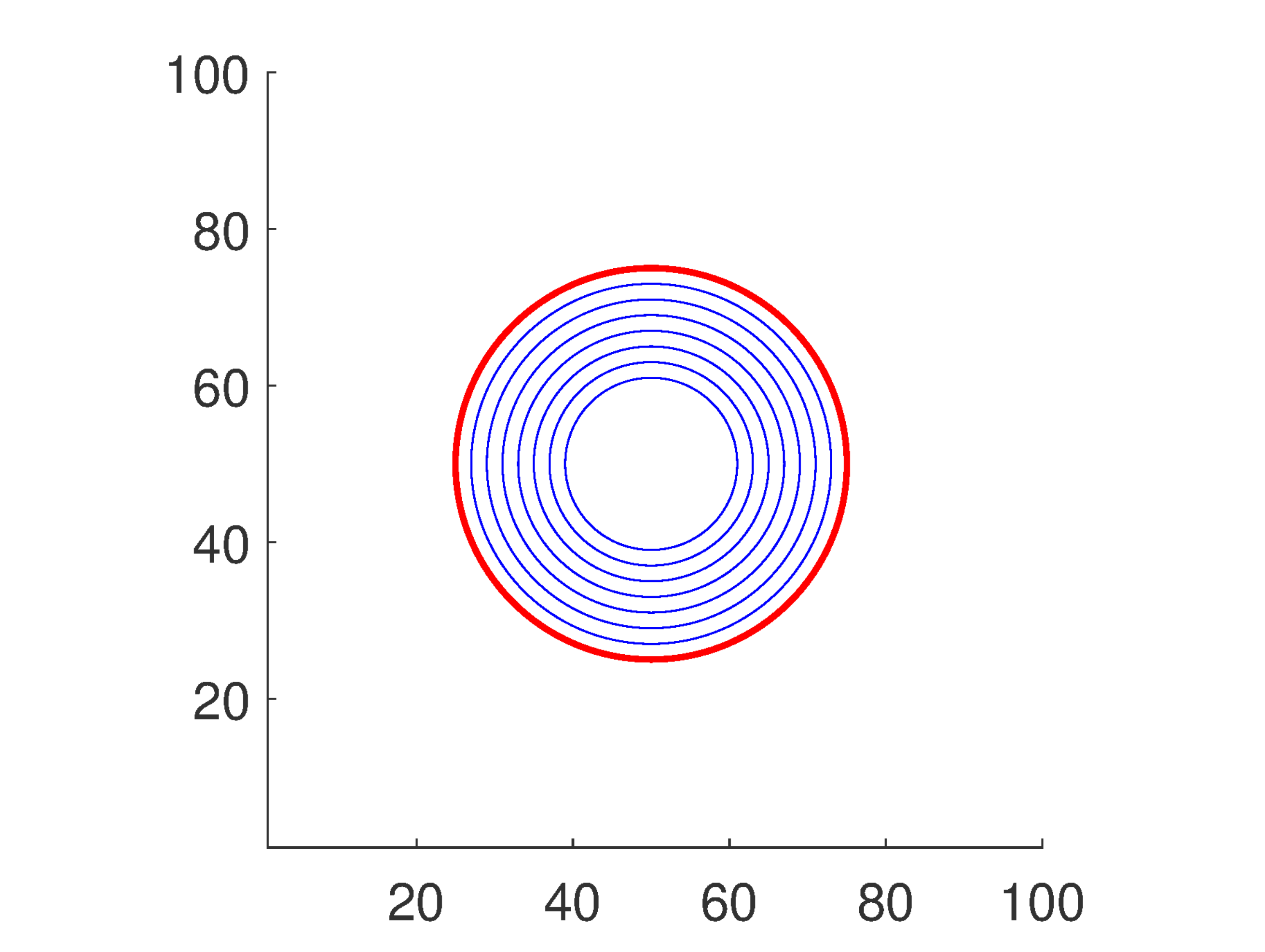}
  \put(-160,110){$(h)$}
  \includegraphics[scale=0.23]{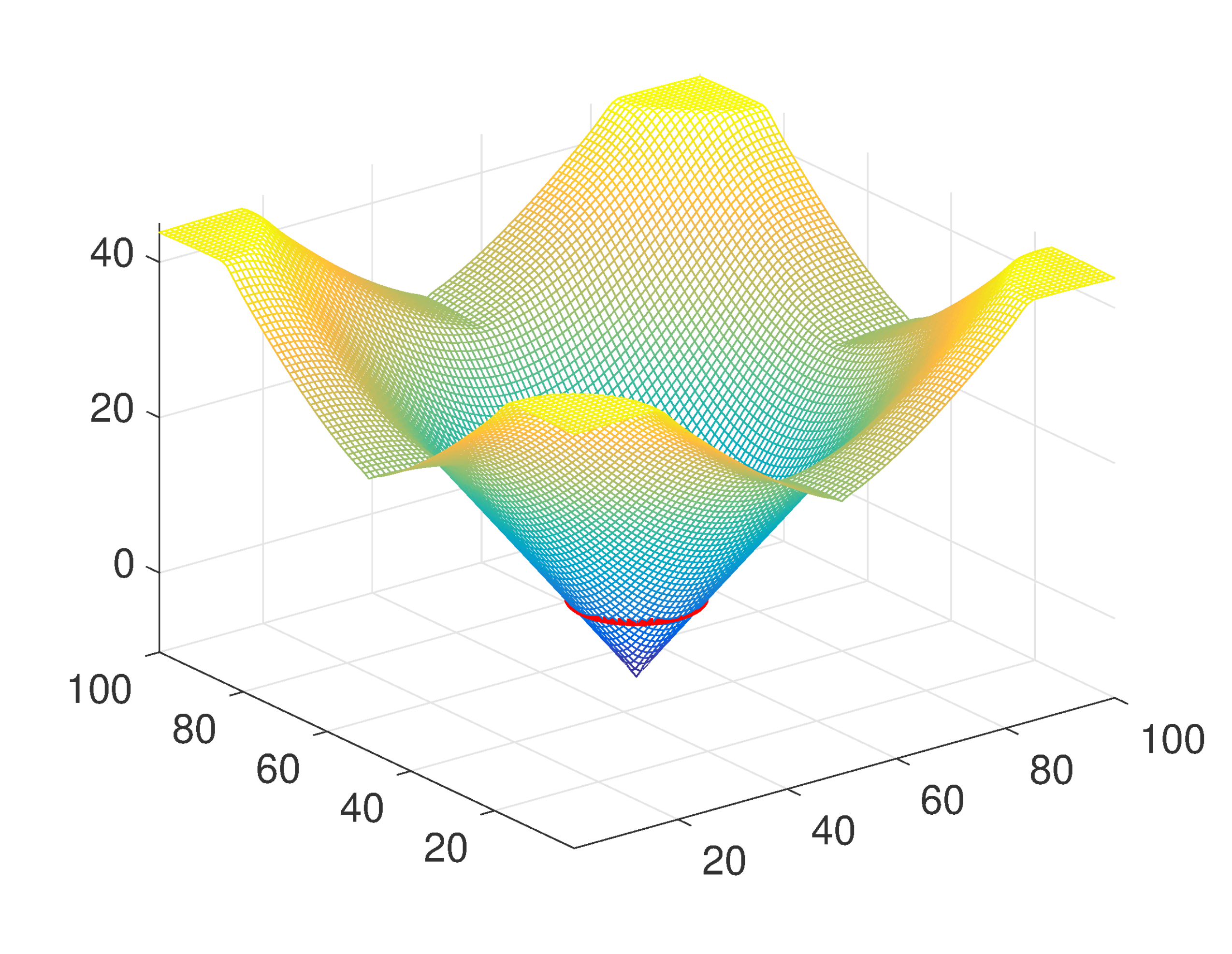}
  \put(-165,110){$(i)$}
 \end{center}
  \caption{A comparison of the inward motion in the normal direction of a circle for the level set methodology without reinitialization, traditional reinitialization and DRLSE from top to bottom respectively. Left column is a cut plane at $y=50$ for the evolution of the contour over time. The solid thick black line is the initial state, the colored lines are the subsequent time-steps and the red dashed line is the zero level set crossing. The center column is the evolution of the zero level set curves with time, solid red line is the initial condition and the solid blue lines represent the subsequent time-steps. The right most column shows the evolution of the level set surface with the solid red line highlighting the zero level set crossing.}
 \label{fig:inm_circle}
\end{figure}
Consider a level curve of a circle given by the following implicit equation:
\begin{equation}
 \phi(\mathbf{x},0) = ||\mathbf{x}-\mathbf{x_c}||-r.
\end{equation}
The domain is a square of size $100\times100$ units where $\mathbf{x_c}=(50,50)$ and $r = 25$, the time-step is $dt=0.1$ and $\alpha=0.4$. Figure \ref{fig:inm_circle} shows the evolution of a contracting circle under a constant velocity $v_n=-1$ which is a result of minimizing the volume enclosed by $\phi<0$. We solve the equation given by $\phi_t=|\nabla \phi|$ using DRLSE which is posed as a level set evolution for a reaction-diffusion type problem. For the traditional level set, the equation solved keeps the velocity term on the left hand side ($\phi_t-|\nabla \phi|=0$), this requires a class of high-order schemes for the advection where the level set is reinitialized after each time-step. 

The top row in figure \ref{fig:inm_circle} shows the results from the DRLSE, middle row for the traditional level set with reinitialization, and the bottom row for traditional level set without reinitialization. The left column (figures \ref{fig:inm_circle}$a$, $d$ and $g$) shows a cut plot in the plane of $y=50$, the solid thick black line is the initial level set location, the subsequent colored lines are the time evolved level sets. The dashed red line indicates the zero level set crossing. The center column (figures \ref{fig:inm_circle}$b$, $e$ and $h$) is a contour plot of the zero level set crossing shown in the thick solid red line, expanding zero level sets over time are shown in the solid blue line. The right column (figures \ref{fig:inm_circle}$c$, $f$ and $i$) shows the final level set function. 
The first striking feature of not reinitializing the level set is the ringing observed in figure \ref{fig:inm_circle}$(a)$. This is a well known issue; the level curves are no longer signed distance functions as they tend to drift away from their initialized value. The errors accumulate leading each level set curve to intersect and develop noisy features and steep gradients that further corrupt the numerical approximation of $|\nabla \phi|$. The advection equation becomes stable once the level set function is reinitialized. The regions where the level set pile up on each other (e.g. near the center of the domain where a kink exists), noisy features develop. This is also evident in figure \ref{fig:inm_circle}$(b)$ where we clearly observe the unstable level set contours near the center of the domain. In the regions where the level sets separate from each other (domain boundaries), we get the flattening of $\phi$ as shown in figure \ref{fig:inm_circle}$(a)$ in the region furthest from the kink. The steepening and flattening effects increase numerical errors as shown in figure \ref{fig:inm_circle}$(c)$ where numerical errors are the largest in the center. Thus in order to reduce these errors, one has to reset the location of the isocontours of each level curve via reinitialization. Figures \ref{fig:inm_circle}$(d)$, $(e)$ and $(f)$ show the results of the traditional level set advection with reinitialization. The level set functions are well-behaved, figure \ref{fig:inm_circle}$(d)$ shows the level set moving upwards while maintaining the kink at $x_c$ with a slight rounding of the edges. The signed distance function is maintained. Figure \ref{fig:inm_circle}$(e)$  shows the isocontours as the level set moving inwards, note the loss of symmetry in the regions where the normals have either of their components equal to zero. Higher order schemes, finer grids and lower time-steps can correct for the asymmetries. Figures \ref{fig:inm_circle}$(g)$, $(h)$ and $(i)$ show the results for the DRLSE. Note that as the level set curve shifts upwards in figure \ref{fig:inm_circle}$(g)$, the kink is maintained at $x_c$ without any rounding in the corner. The far field has level sets piling up but this has no effect on the solution since the level sets are moving away from the boundaries and towards the center of the circle. Accuracy is maintained as it nears the center, the level curves maintain their circular shape as they advect inwards. Figures \ref{fig:inm_circle}$(f)$ and $(i)$ show differences in the far field where the lobes are less uniform but are inconsequential to the final solution. Note that a similar test case was performed using $v_n=1$ for the circular shape (not shown here); similar behavior is observed.
\subsubsection{Seven-Pointed Star with $v_n=1$}
\begin{figure}
 \begin{center}
  \includegraphics[scale=0.21]{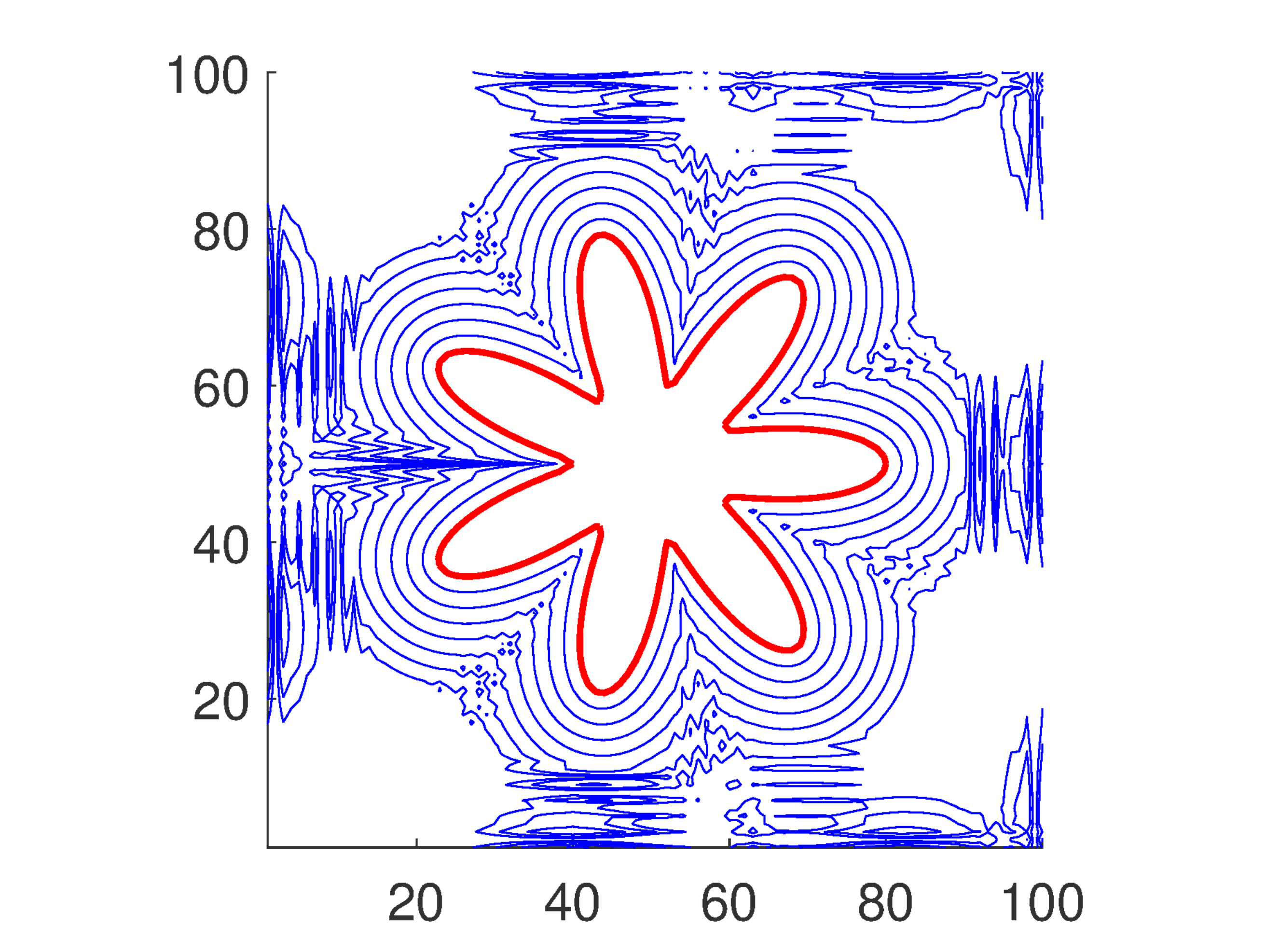}
  \put(-155,110){$(a)$}
  \includegraphics[scale=0.21]{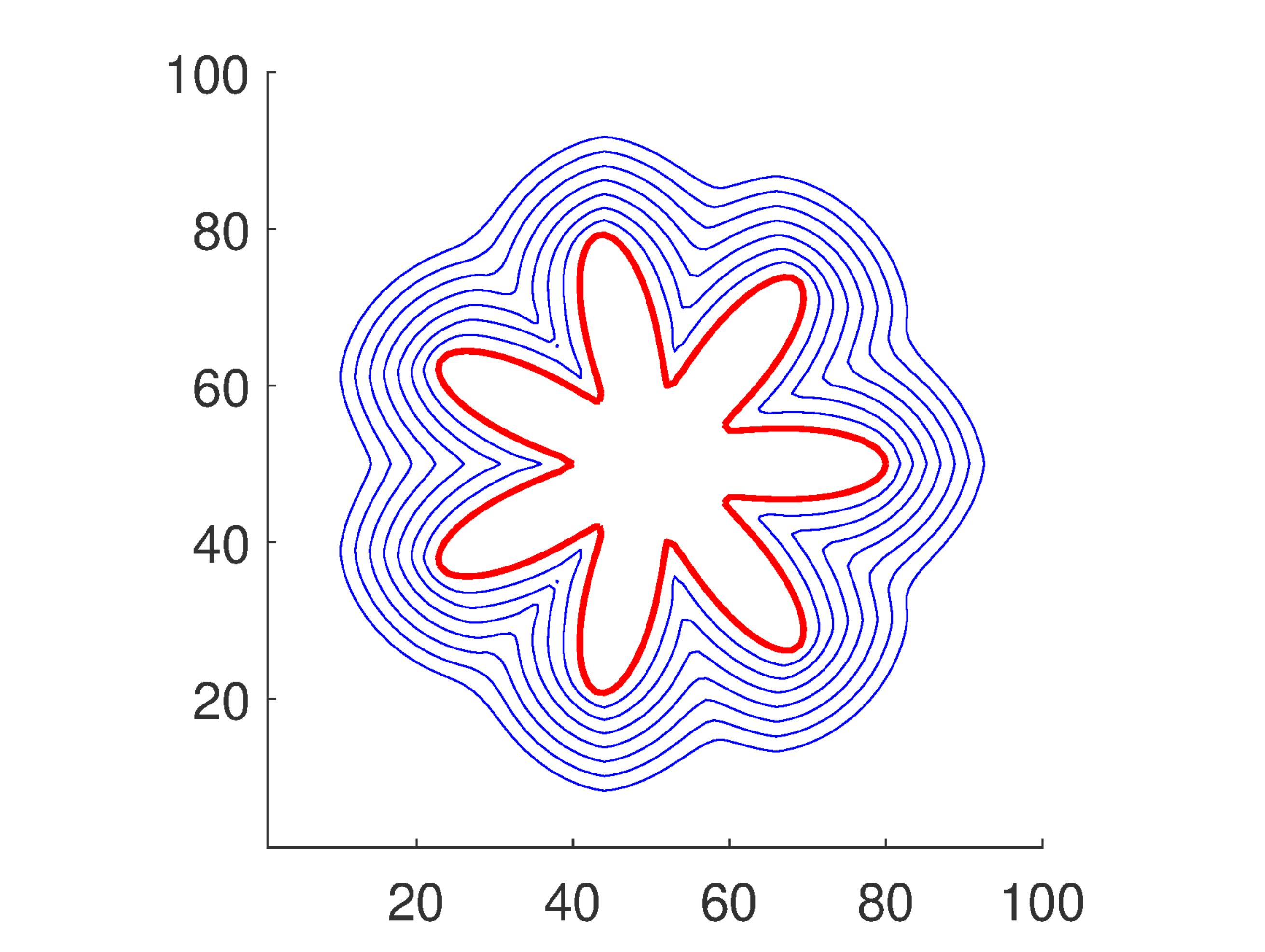}
  \put(-155,110){$(b)$}
  \includegraphics[scale=0.21]{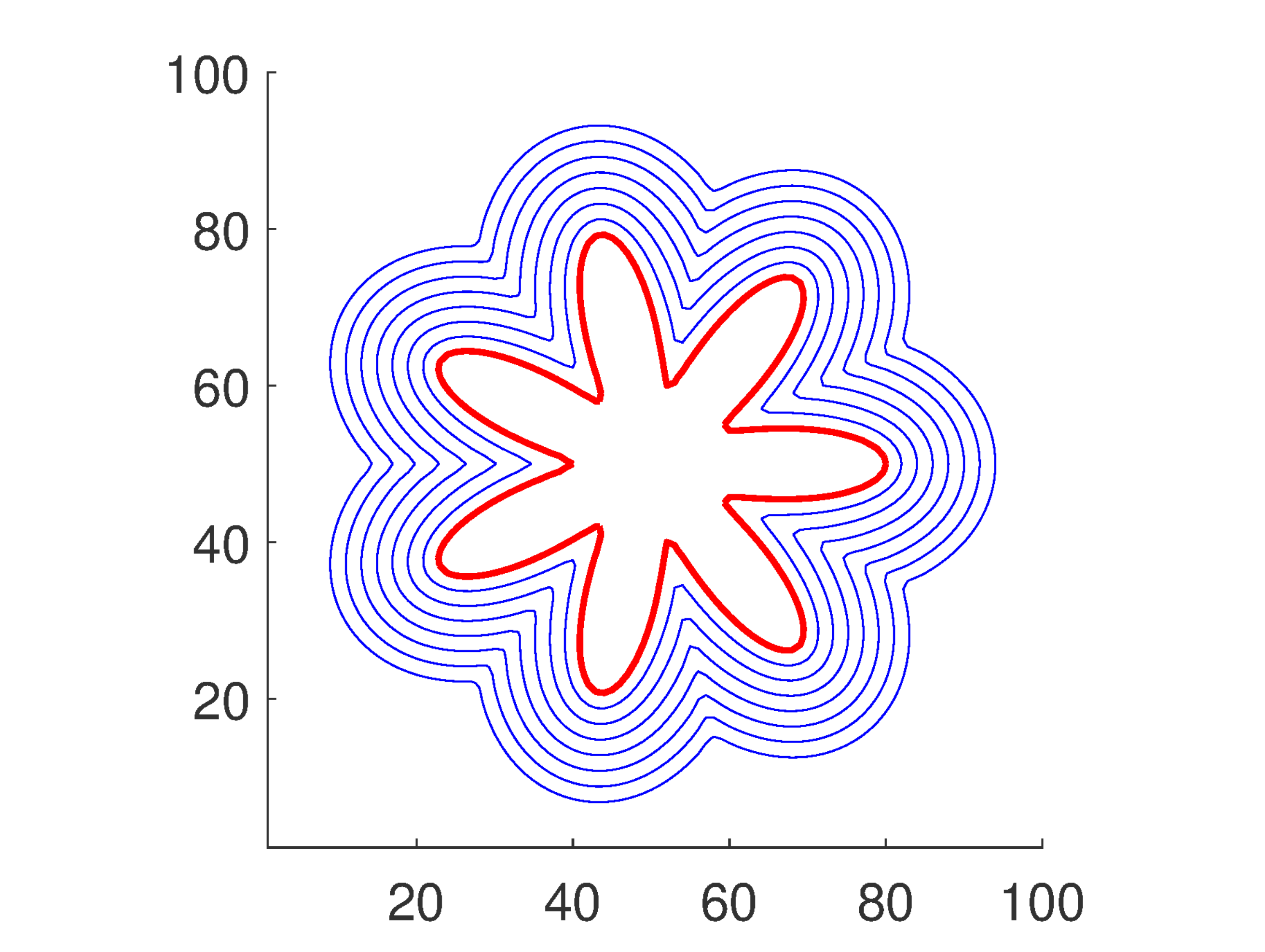}
  \put(-155,110){$(c)$}\\
  \includegraphics[scale=0.23]{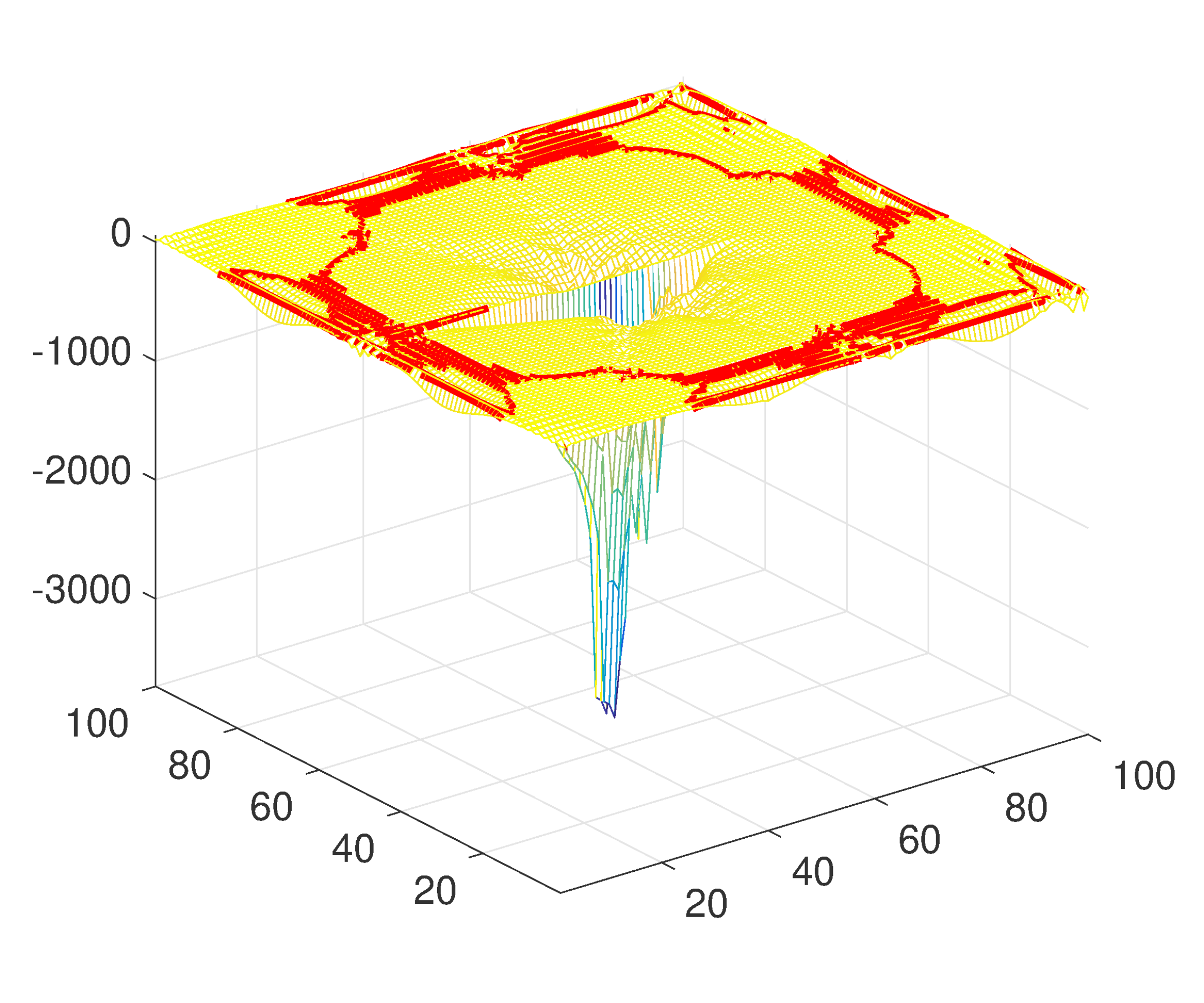}
  \put(-155,110){$(d)$}
  \includegraphics[scale=0.23]{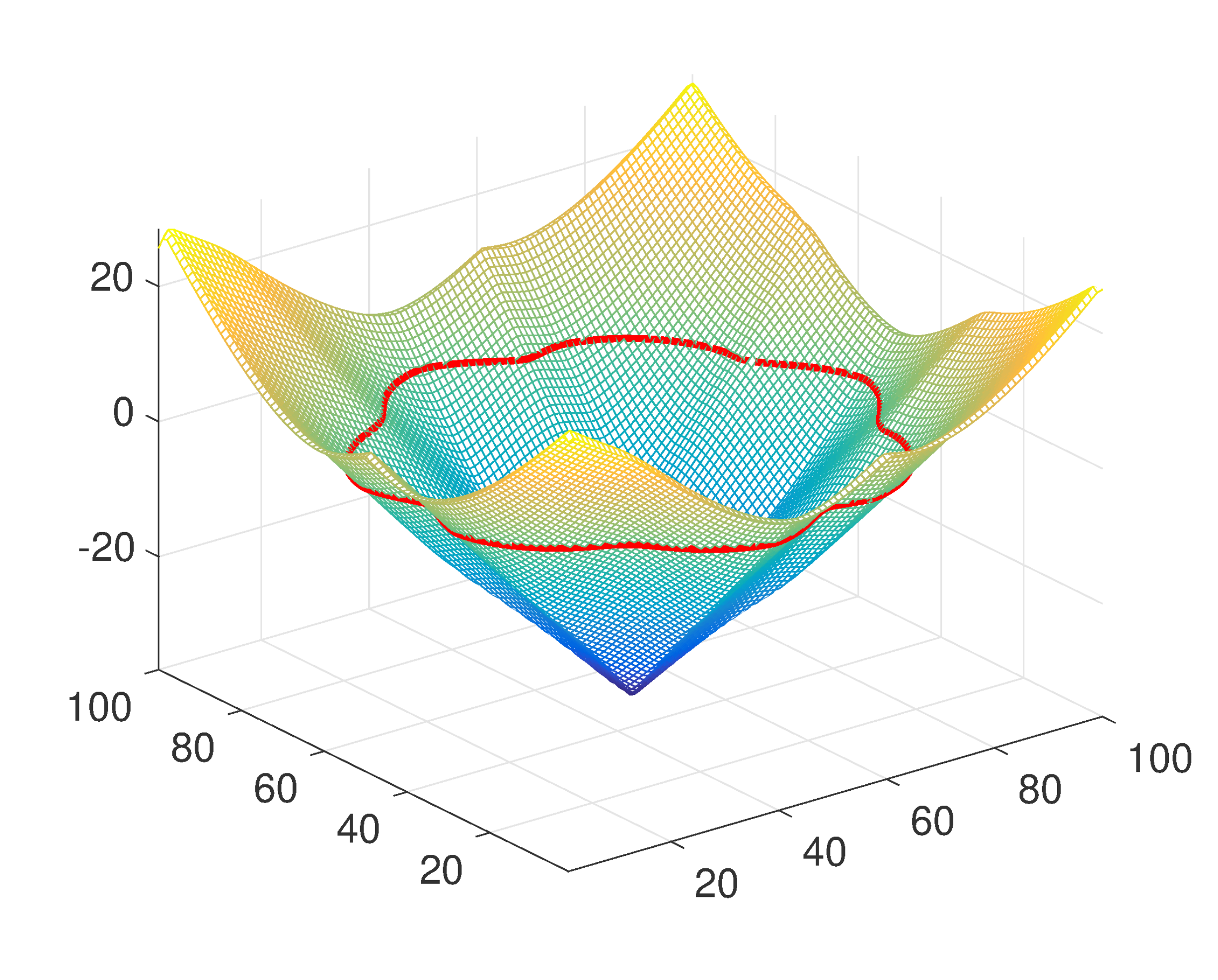}
  \put(-155,110){$(e)$}
  \includegraphics[scale=0.23]{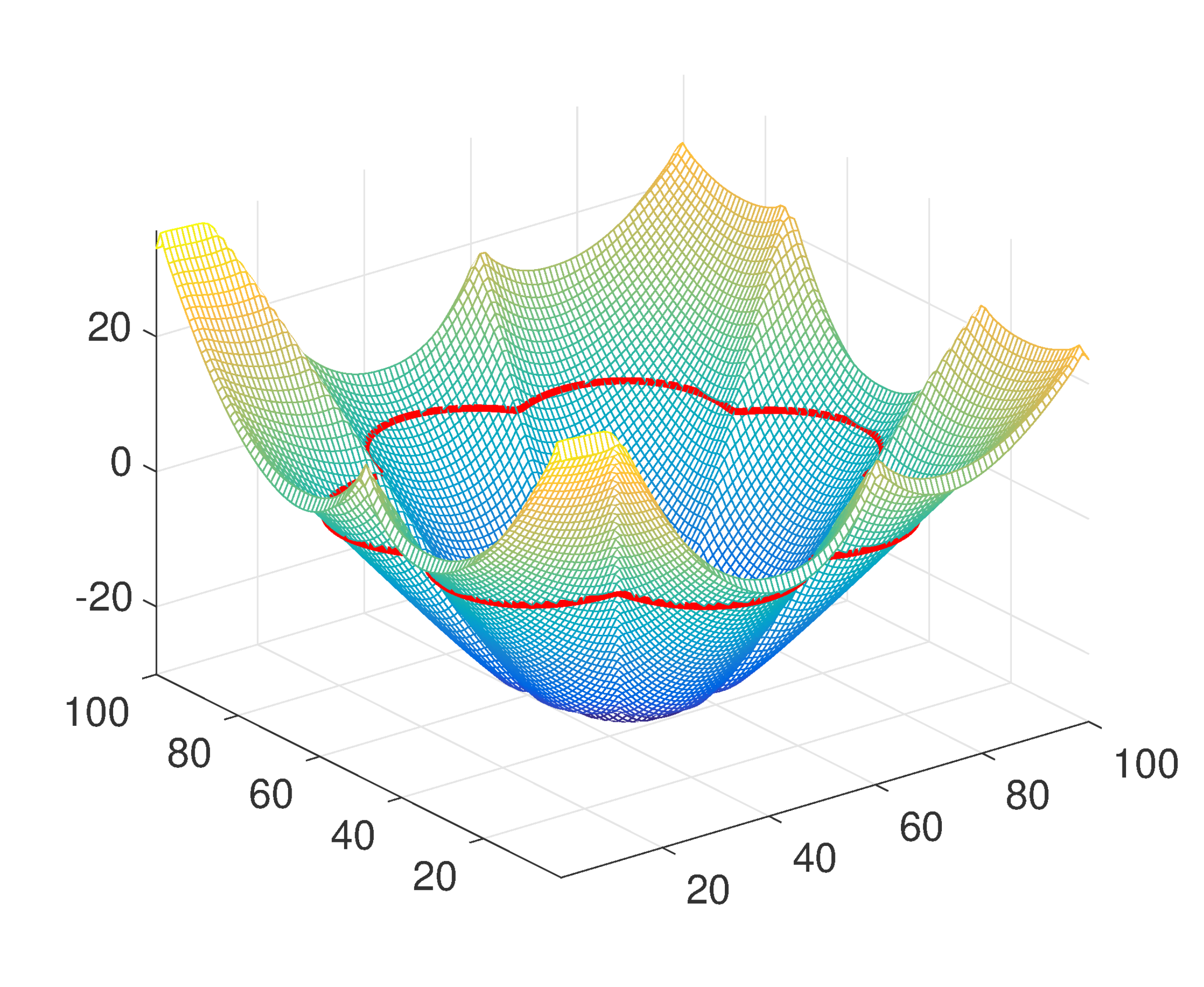}
  \put(-155,110){$(f)$}
 \end{center}
 \caption{A comparison of the outward motion in the normal direction of a star-shaped contour for the level set methodology without reinitialization, traditional reinitialization and DRLSE from left to right respectively. Top row is the evolution of the zero level set curves with time, solid red line is the initial condition and the solid blue lines represent the subsequent time-steps. Bottom shows the evolution of the level set surface with the solid red line highlighting the zero level set crossing.}
 \label{fig:onm_star}
\end{figure}
Consider the same implicit function described in \S\ref{subsubsec:sps_curv}. Figure \ref{fig:onm_star} shows the behavior of a more complicated shape for the case of the outward motion in the normal direction. Qualitatively, similar trends are observed when compared to the circular shape. This ensures that even more complicated shapes are handled effectively given our numerical scheme. Figure \ref{fig:onm_star}$(a)$ shows the ringing generated in the far field as the zero level set curve advects away from the initial star. Figure \ref{fig:onm_star}$(b)$ is well-behaved with reinitialization but still loses its symmetrical shapes around the star lobes. The curvature diffuses at a faster rate leading to an early flattening of the curve. Figure \ref{fig:onm_star}$(c)$ shows a symmetrical outgrowth of the seven pointed star. As mentioned in previous sections, figures \ref{fig:onm_star}$(d)$, $(e)$ and $(f)$ are the final level set surface showing the large numerical errors for the no reinitialization case, and a well-behaved final solution for the traditional level set and DRLSE. Similar to the conclusions made above, the flattening of the level set at $x_c$ is inconsequential to the final solution since the zero level curve is moving away from it.  

\subsection{Energy Minimization}
We take the LS formulation of the Gibbs free energy model presented in \S\ref{subsec:gibbs_levset} and solve the governing equations using the variational formulation presented in \S\ref{sec:num_imp}. The governing evolution equations are a combination of curvature-driven flow with motion in the normal direction. Longitudinal grooves are used first because of the well-defined corners. This causes the interface to pin at the edges and allows for direct comparison to the Young-Laplace equation. Then a cosine wavy substrate is simulated to compare to the analytic solution obtained by \cite{carbone2005}.

\subsubsection{$2D$ Longitudinal Grooves}
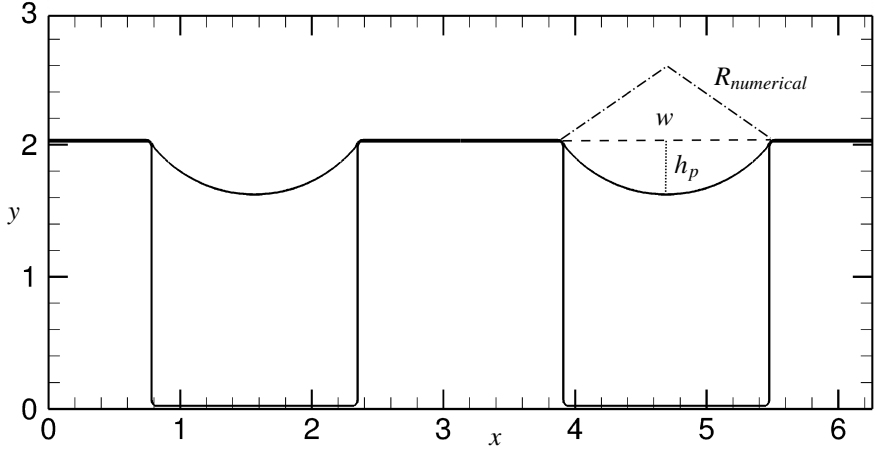
\begin{figure}
 \begin{center}
  \includegraphics[scale=0.4]{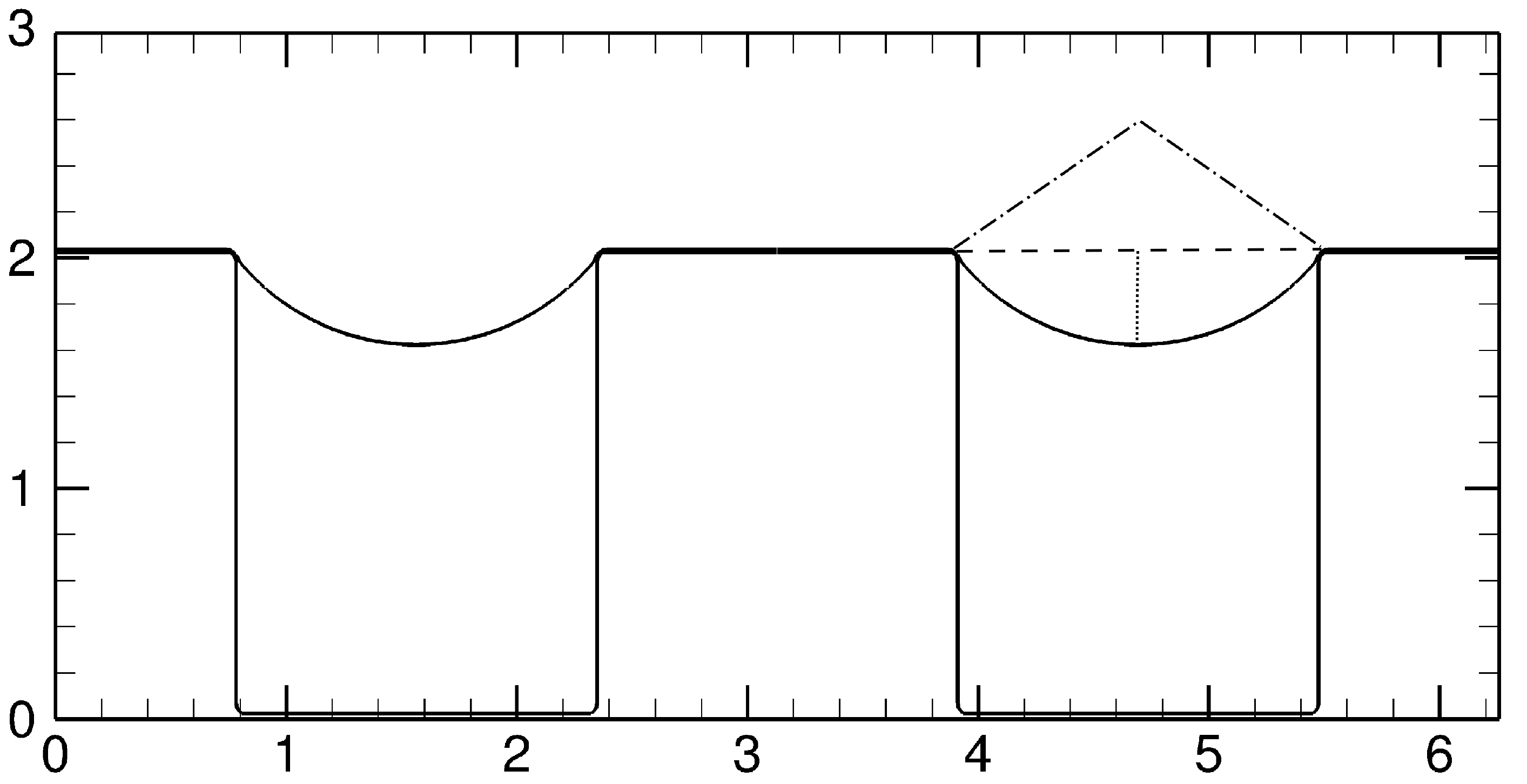}
  \put(-330,85){$y$}
  \put(-150,0){$x$}
  \put(-80,102){$h_p$}
  \put(-87,120){$w$}
   \put(-65,135){$R_{numerical}$}
 \end{center}
  \caption{Equilibrium interface location for a $2$D grooved substrate. The thick solid line represents the grooved geometry and the thin solid line the zero level set curve represents the interface. The dashed-dot black line represents the radius of curvature of the meniscus, the dashed line denotes the width of the groove and the dotted line the meniscus height (measured from the contact points).}
 \label{fig:gfe_grooves}
\end{figure}
Consider a rough substrate representing a longitudinal groove, the level set that represents the solid is given by: 
\begin{equation} 
\psi(\mathbf{x}) = (y-y_c) - \sign \bigg[h,\cos\big(k_x (x-x_c)\big)\bigg],
\end{equation}
such that $h=1$, with a wavelength of $\pi$ such that $k_x = 2$ to get multiple grooves and $sgn$ the sign function. The grooves are shifted by $\mathbf{x_c}=(0,1,0)$ so that the bottom of the surface corresponds to the origin. The domain extents are $[L_x,L_y,L_z] = [2 \pi, 3, 2 \pi]$ and grid size $n_x \times n_y \times n_z = 256 \times 123 \times 1$. The time-step is taken to be $dt = 10^{-4}$ and $\alpha=0.4$. 
For an external pressure of $\Delta p = 1$, and a non-dimensional $\tau^- = 1$, the Young-Laplace equation gives a radius of curvature $R_{analytic}=1$. From the current results, the radius of curvature is computed numerically from the contact points and penetration depth $h_p$ into the groove using $R_{numerical}={w^2}/(8h_p)+h_p/2=0.993$ where $w$ is the groove width. The error is $0.7\%$ showing good agreement with the theoretically predicted results. 

\subsubsection{$2D$ Wavy Substrate}
Consider a wavy substrate using a profile given by:
\begin{equation}
 \psi(\mathbf{x}) = (y-y_c) - h \cos \big(k_x (x-x_c)\big). 
\end{equation}
The peak to valley height $h$ of the wavy substrate is taken to be twice the critical height $h_{cr}$ where $h_{cr}=-\tan \theta_Y$ \citep{carbone2005} and $\theta_Y = 140^{\circ}$. The wavelength of the substrate is taken to be $2 \pi$ such that $k_x = 1$. The substrate is shifted by $\mathbf{x_c}=(0,h_{cr})$. The domain extents are $[L_x,L_y,L_z] = [2 \pi, 4, 2 \pi]$ with a spatial resolution of $n_x \times n_y \times n_z = 256 \times 164 \times 1$ and a time-step of $dt = 10^{-5}$. \cite{carbone2005} derived an analytical model to predict the interface equilibrium location given an external pressure field. They analyzed the asymptotic behavior of the system and proposed a simple criterion for the minimum $h_{cr}$ that prevents a Wenzel state. Their solution can predict the pressure value required to overcome the energy barrier to transition from a Cassie state to a Wenzel state. Using our model, we perform a parametric study of different values of pressure and compare the resulting interface shape and location to the analytic solution. The pressure value that leads to the Wenzel state is obtained using the analytic solution for the above conditions which comes out to be $0.25$. The values of external pressure prescribed are: $\Delta p = 0.05, 0.1, 0.15, 0.2, 0.26$ where the last value was picked to test if we can predict interface failure. The regularization coefficient for this particular problem was varied to gauge its effect on the interface location. For the first case we use $\alpha=0.4$, for the next two cases we use $\alpha=0.8$ and for the last two we use $\alpha=0.4$. We also tested $\alpha=0.2$ for the third case (not presented here); overall we did not see an appreciable effect, the error in interface location varied within $1\%-3\%$. Note that all selected values of $\alpha$ satisfy the CFL condition for a given time-step. For all the cases, the simulation is terminated when equilibrium is achieved. This is determined by keeping track of the Gibbs free energy over time until it tends to a steady state. Once the change in $|(G_{tot}^{n+1}-G_{tot}^{n})/G_{tot}^n| < \mathcal{O}(10^{-3})$ the simulation is stopped.  
\begin{figure}
 \begin{center}
  \includegraphics[scale=0.6]{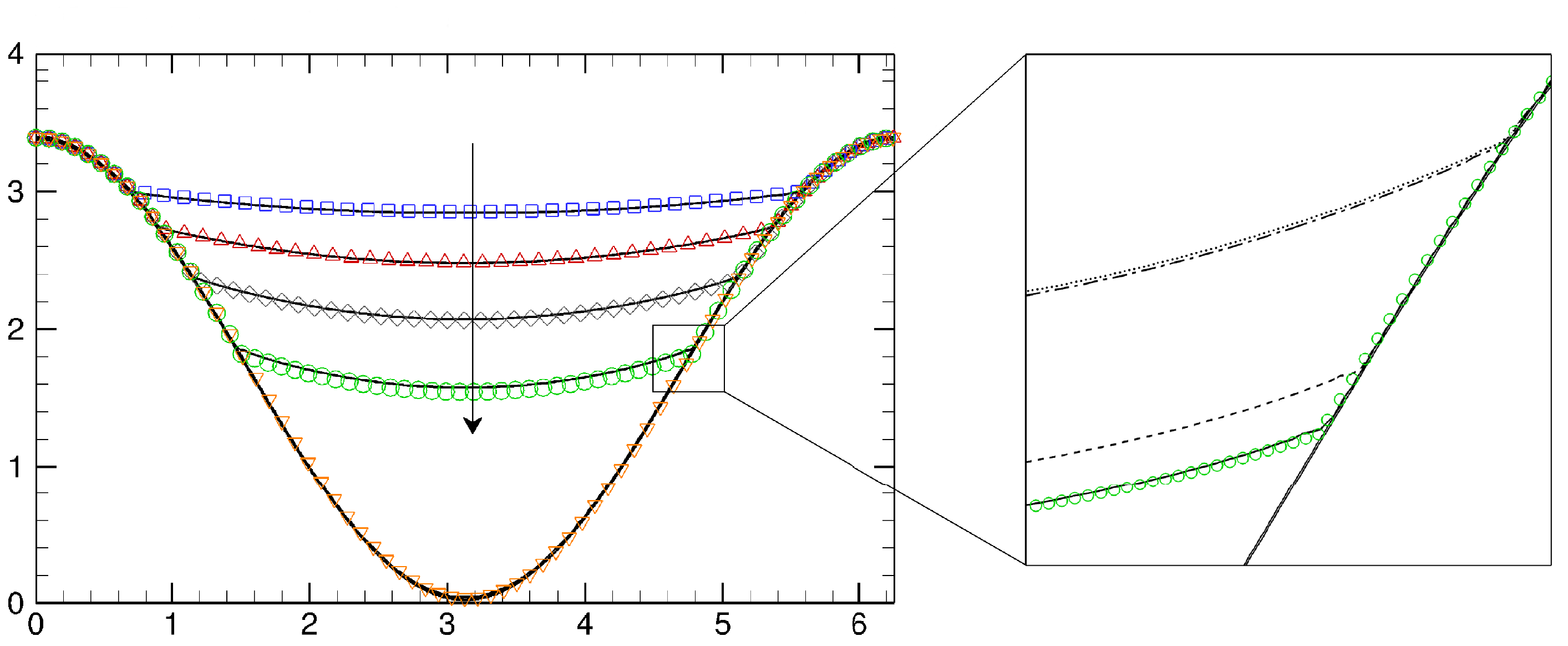}
  \put(-500,100){$y$}
  \put(-340,0){$x$}
  \put(-348,165){$\Delta p$}
 \end{center}
  \caption{Equilibrium interface location for a cosine substrate. The solid black lines represent the numerical simulations and the symbols are Carbone's results. The arrow indicates increasing external pressure. Symbols are: $\Delta p = 0.05$ ($\Box$), $0.1$ ($\vartriangle$), $0.15$ ($\Diamond$), $0.2$ ($\bigcirc$), $0.26$ ($\triangledown$). The inset shows a zoomed in plot for $\Delta p=0.2$ with different surface tension models: harmonic mean using $\tau^*$ (solid line), arithmetic mean using $\tau^*$ (dashed line), harmonic mean using $\tau^+$ (dashed-dotted line), and arithmetic mean using $\tau^+$ (dotted line).}
 \label{fig:carbone_cosine}
\end{figure}
Figure \ref{fig:carbone_cosine} shows a comparison between our numerical model (solid lines) and the analytic solution (symbols). Good agreement is observed. Note that for the value of $\Delta p = 0.26$, the interface fails and lies exactly on top of the cosine substrate. The inset in figure \ref{fig:carbone_cosine} shows the effect of different surface tension models. We fix $\Delta p$ at $0.2$ and change the method with which surface tension is computed. It is evident that using the harmonic mean distribution of Eq. (\ref{eq:tau_geom}) yields the most accurate results. Using an arithmetic distribution given in Eq. (\ref{eq:contin_tau}) with $\tau^*$ instead of $\tau^+$ underestimates the location by $7\%$. Note that using $\tau^+$ with either arithmetic or harmonic distribution significantly underestimates the location with error around $33\%$.

Table \ref{tab:table_1} shows quantitative results for the mean LG interface height, the triple line contact points and radius of curvature. These quantities are manually extracted from a visualization software.
\begin{table}
   \begin{center}
    \begin{tabular}{c|c|c|c|c|c|c|c|c|c}
      $\Delta p$ & $ \bar{y}_{analytic}$ & $\bar{y}_{numerical}$ & \textit{E}(\%) & $a_{analytic}$ & $a_{numerical}$ & \textit{E}(\%) & $R_{analytic}$ & $R_{numerical}$ & \textit{E}(\%)\\
      \hline
      0.05 & 2.901 & 2.906 & 0.17 & 0.69 & 0.71 & 2.3 & 19.93 & 19.316 & 3.1 \\
      0.1 & 2.57 & 2.59 & 0.77 & 0.91 & 0.92 & 1.3 & 9.93 & 9.78 & 1.55 \\
      0.15 & 2.16 & 2.17 & 0.55 & 1.173 & 1.17 & 0.085 & 6.66 & 6.56 & 1.54 \\
      0.2 & 1.63 & 1.66 & 1.4 & 1.50 & 1.49 & 0.86 & 4.98 & 4.944 & 0.81 \\
    \end{tabular}
  \end{center}
  \caption{Comparison between the analytic and numerical values obtained for: the mean LG interface height $\bar{y}$, interface contact point $a$, and radius of curvature $R$ for a given external pressure $\Delta p$.}
  \label{tab:table_1}
\end{table}

\subsection{$2D$ Droplet at a Wall}
\begin{figure}
  \begin{center}
    \includegraphics[scale=0.34]{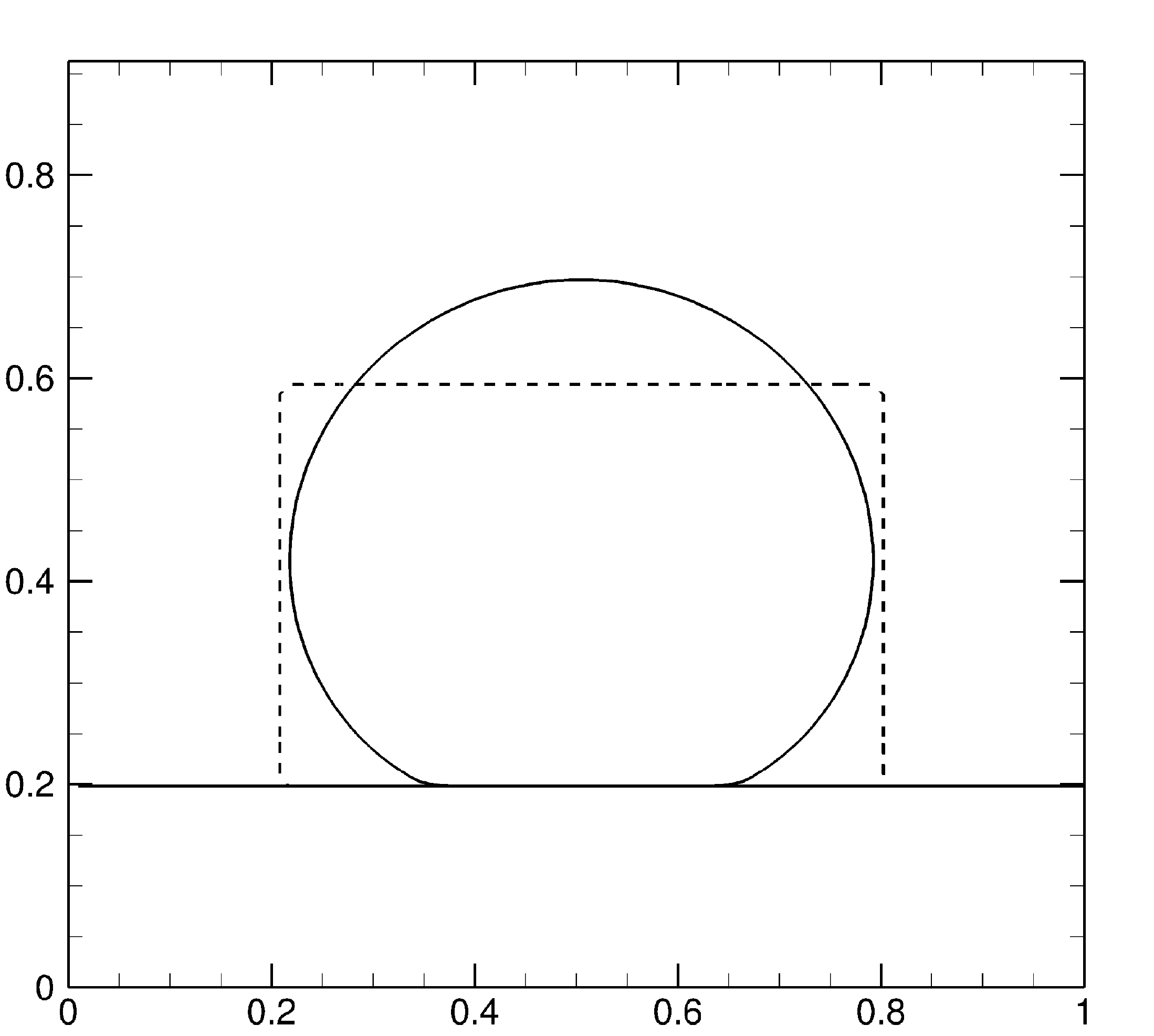}
    \put(-220,180){$(a)$}
    \put(-220,95){$y$}
    \put(-110,-7){$x$}
    \includegraphics[scale=0.33]{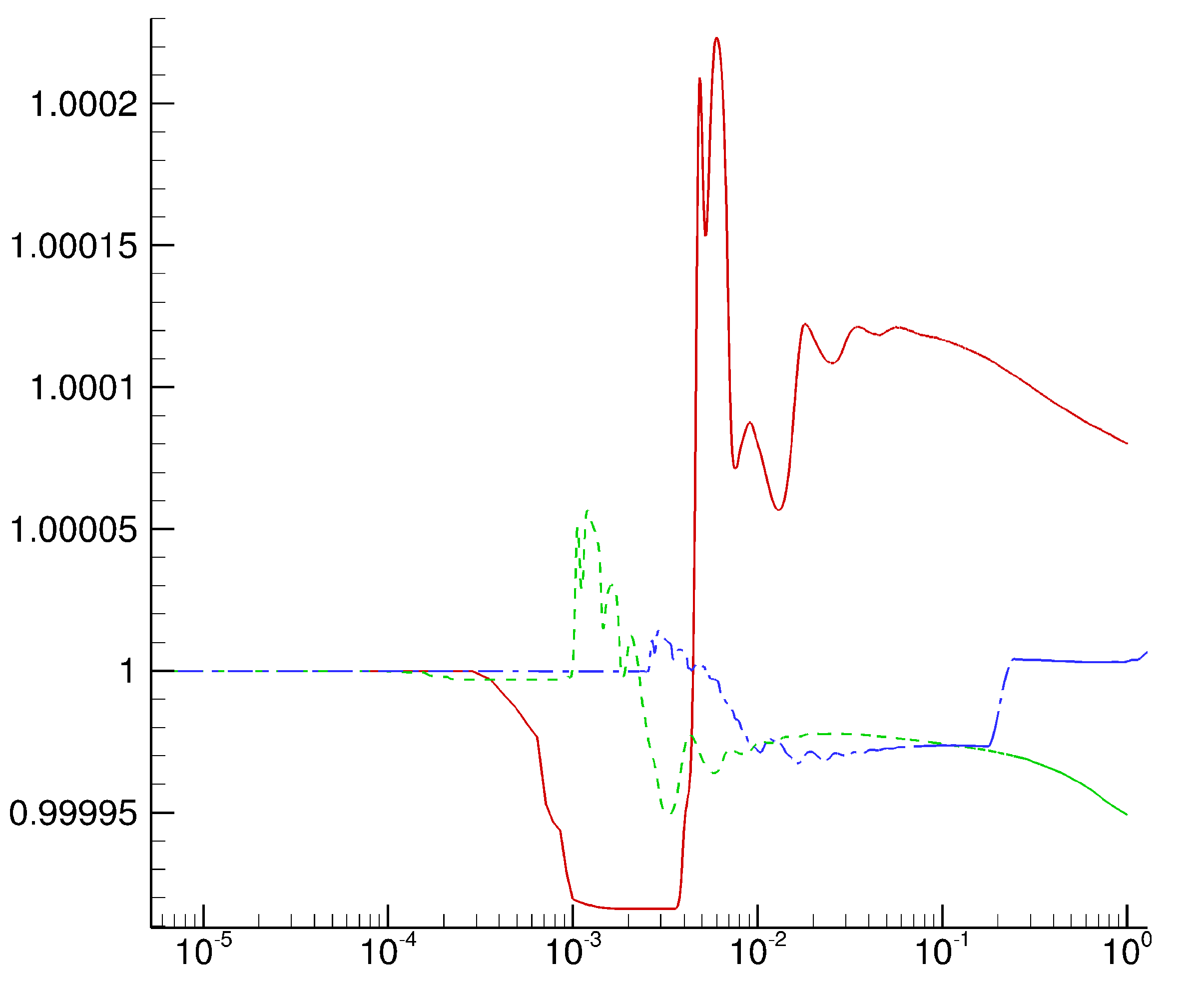} 
    \put(-220,180){$(b)$}
    \put(-230,98){$V/V_o$}
    \put(-100,-7){$t/t_f$}
    \\
  \end{center}
  \caption{Evolution of a $2$D bubble over an unwetted wall from an initial binary rectangle to its equilibrium position. $(a)$ The dashed black line is the initial condition of a binary rectangle, and the solid black line is the final equilibrium location. $(b)$ volume conservation over time for three different grid resolutions: red solid line for $50\times50$, green dashed line for $100\times100$ and blue dashed-dot line for $200\times200$. The time axis is shown on a log scale.}
  \label{fig:vol_cons}
\end{figure}
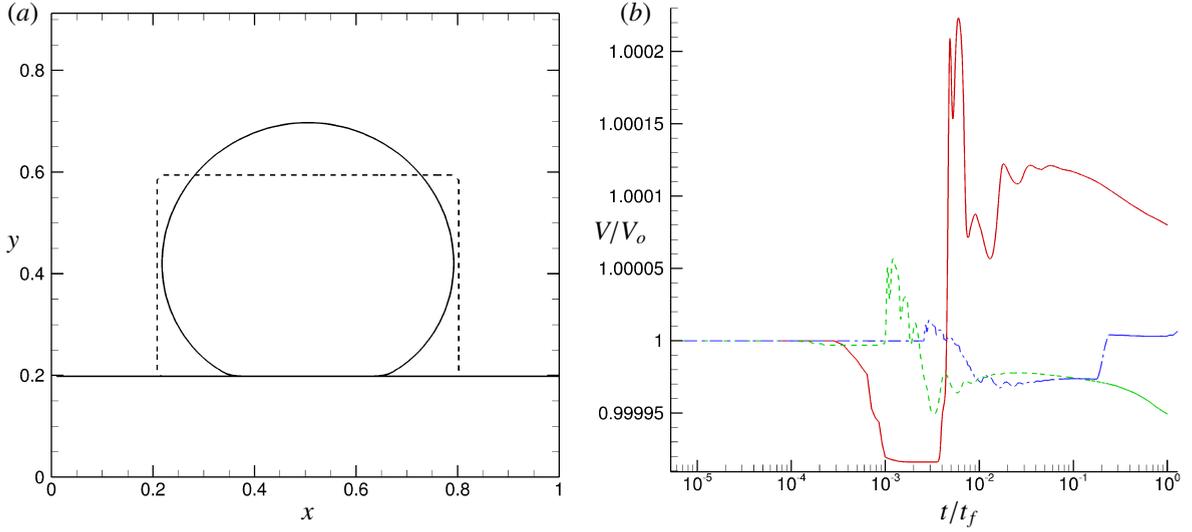

Consider a flat solid wall initialized as a binary function instead of an SDF given by:
\begin{equation}
 \psi(\mathbf{x}) = \begin{cases}
		       c_o, & \text{for $y<h$} \\ 
		      -c_o,  & \text{for $y \geq h$} \, , 	      
		      \end{cases}
\end{equation}
where $h=0.2$ and $c_o=3\epsilon$. A rectangular droplet of length $0.4$ and width $0.6$ centered at $(x_c,y_c)=(0.5,0.4)$ is initialized on the solid wall. The domain extents are $[L_x,L_y,L_z]=[1, 1, 1]$. The simulation is run using three different grid resolutions $n_x \times n_y \times n_z=$ $50\times50\times1$, $100\times100\times1$ and $200\times200\times1$. The time-step taken for each grid is $10^{-5}$, $10^{-6}$ and $10^{-7}$ respectively with $\alpha=0.4$. The bulk energy here is the gravitational potential energy, i.e. $h(\mathbf{x}) \cdot \mathbf{g}$ instead of $\Delta p$, where $\mathbf{g}$ is the magnitude and direction of the gravitational force which is taken to be unity in this example. The surface tension values are $\tau^-=0.6$ and $\tau^+=0.5$ which corresponds to $\theta_Y=140^{\circ}$. Figure \ref{fig:vol_cons}$(a)$ shows the evolution from the initial rectangular droplet (dashed line) into a droplet at equilibrium (solid line) that satisfies the contact angle of $\theta_Y=139.72^{\circ}$, which is within $0.2\%$ of the prescribed contact angle. Figure \ref{fig:vol_cons}$(b)$ shows the volume of the liquid normalized by the initial volume as it evolves in time (on a log scale) up until the equilibrium solution. Similar to the previous simulations, we track the Gibbs free energy until it reaches steady state (not shown here). As the interface evolves from an initial binary function, the volume experiences a sharp oscillation as the penalty term smears out the interface to satisfy an SDF. The jump in volume is significantly reduced as the grid resolution is increased. The volume then plateaus as the interface reaches equilibrium. The change in volume at equilibrium from the coarsest grid to the finest are: $8.046\times10^{-3}\%$, $5.037\times10^{-3}\%$ and $3.5\times10^{-4}\%$ respectively. For the finest grid, we ran the simulation $10$ times longer after it reached equilibrium to quantify the effectiveness of volume conservation, and the largest change in volume observed was around $1.7\%$.
\begin{remark}
  To test the effect of the Lagrange multiplier constraints, the no-penetration constraint is turned off and as expected the bubble leaks into the solid wall deforming its equilibrium shape. Similar behavior is observed for the longitudinal grooves and wavy substrate where the interface does not feel the effect of the wall. When the volume conservation constraint is turned off, the binary rectangular bubble evolves into a droplet shape since it behaves like a curvature-driven flow, however it keeps shrinking until the drop disappears.    
\end{remark}

\section{Numerical Experiments}\label{sec:results}
The numerical results for predicting equilibrium interfacial shapes over a variety of rough surfaces will be discussed. In the earlier section, we validated our results against the analytical solution of a cosine substrate obtained by \cite{carbone2005} for different loading conditions. In the following sections, we demonstrate the robustness of the algorithm for different geometries such as a $3$D wavy substrate, longitudinal grooves, posts, and random roughness. 

\subsection{$3D$ Longitudinal Grooves}
\begin{figure}
 \begin{center}
  \includegraphics[scale=0.35]{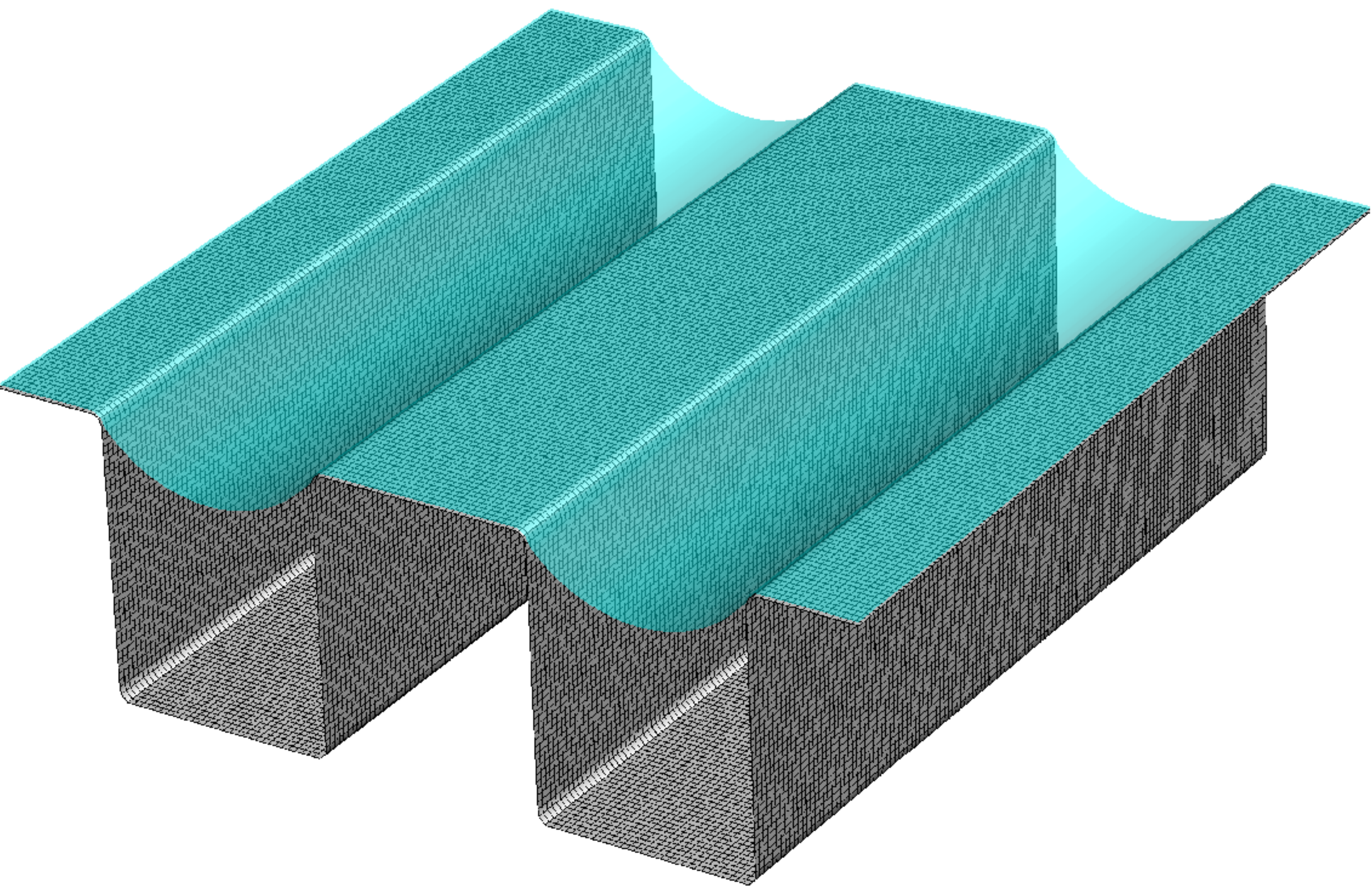}
 \end{center}
 \caption{Equilibrium interface location for a $3$D grooved substrate where the gray isosurface represents the rigid substrate and the cyan transparent isosurface of the zero level set represents the interface.}
 \label{fig:gfe_grooves_3d}
\end{figure}
Consider a rough substrate representing a $3$D longitudinal groove, the level set that represents the solid is given by: 
\begin{equation} 
\psi(\mathbf{x}) = (y-y_c) - \sign \bigg[h,\cos\big(k_x (x-x_c)\big)\bigg],
\end{equation}
such that $h=1$, where the wavelength is $\pi$ such that $k_x = 2$. The grooves are shifted by $\mathbf{x_c}=(0,1,0)$ so that the bottom of the surface corresponds to the origin. The domain extents are $[L_x,L_y,L_z] = [2 \pi, 3, 2 \pi]$ and grid size $n_x \times n_y \times n_z = 128 \times 62 \times 128$ for the $3$D case. The time-step is taken to be $dt = 10^{-5}$ and $\alpha=0.4$. 
For an external pressure of $\Delta p = 1$, and a non-dimensional $\tau^- = 1$, Young-Laplace give a radius of curvature $R_{analytic}=1$. Figure \ref{fig:gfe_grooves_3d} the error in the radius of curvature computed numerically is of similar order to the $2$D validation case.  

\subsection{Grooved Posts}
\begin{figure}
 \begin{center}
  \includegraphics[scale=0.3]{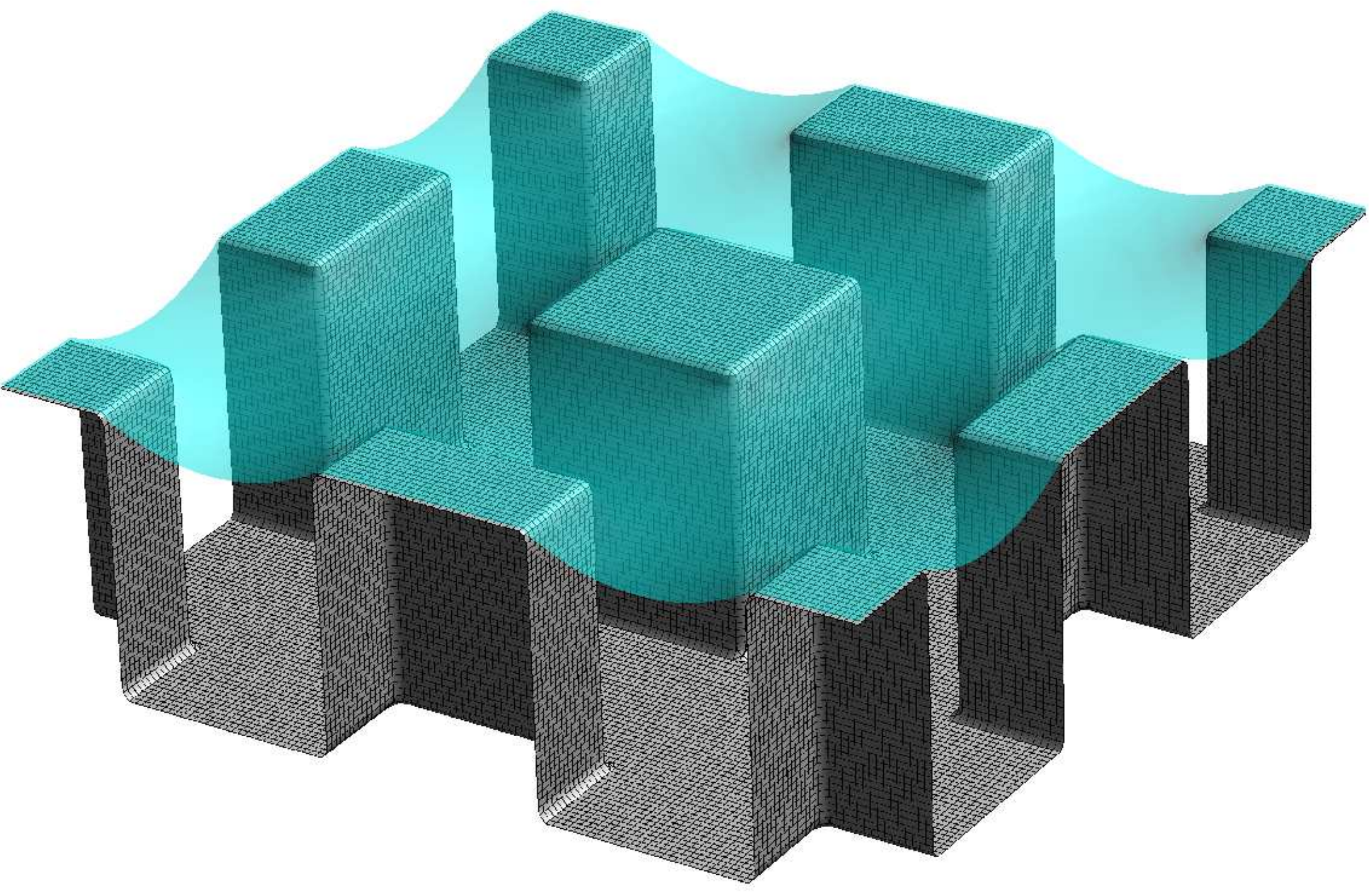}
  \put(-220,180){$(a)$}
  \includegraphics[scale=0.3]{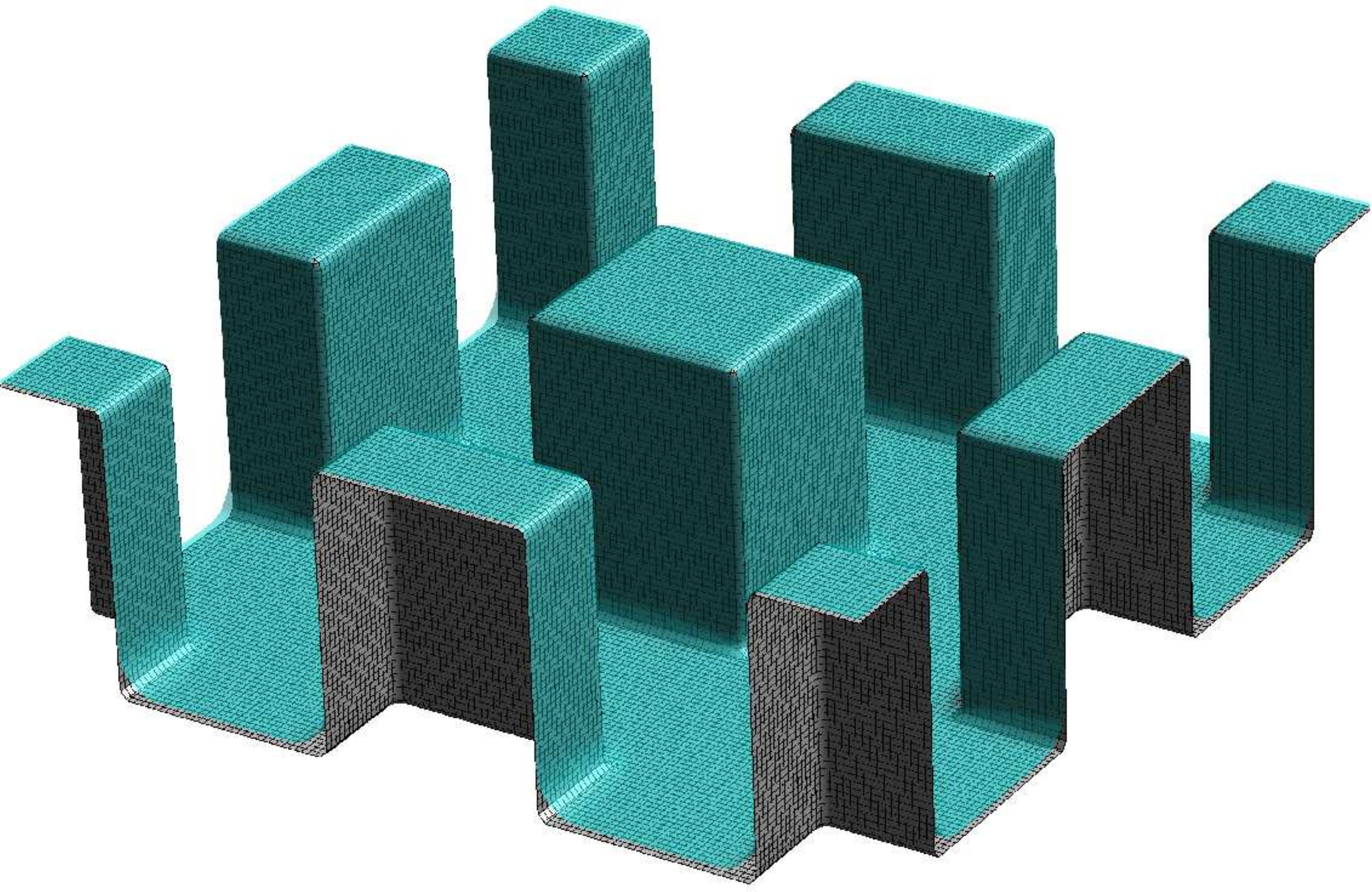}
  \put(-220,180){$(b)$}
 \end{center}
  \caption{Equilibrium interface location for post geometries, the gray isosurface represents the rigid substrate and the cyan transparent isosurface of the zero level set represents the interface. $(a)$ Equilibrium shape for $\Delta p = 0.5$ and $(b)$ interfacial failure at $\Delta p = 1$ where the liquid completely wets the cavities of the post.}
 \label{fig:gfe_posts}
\end{figure}
In order to represent a post geometry, we make use of the Boolean operations. Consider the level sets that represent the solid: 
\begin{equation} 
\psi_1(\mathbf{x}) = (y-y_c) - \sign \bigg[h,\cos\big(k_x (x-x_c)\big)\bigg],
\end{equation}
which represents longitudinal grooves in the $x-y$ plane extending in the $z-$direction, and 
\begin{equation} 
\psi_2(\mathbf{x}) = (y-y_c) - \sign \bigg[h,\cos\big(k_z (z-z_c)\big)\bigg],
\end{equation}
the longitudinal grooves in the $y-z$ plane extending in the $x-$direction. Take the intersection $\psi = \psi_1 \cap \psi_2$ such that
\begin{equation} 
\psi(\mathbf{x}) = \max(\psi_1,\psi_2).
\end{equation}
This gives the post geometry where $h=1$. The wavelength is $\pi$ such that we obtain $k_x = 2$ and $k_z = 2$. The domain extents are $[L_x,L_y,L_z] = [2 \pi, 3, 2 \pi]$ and grid size $n_x \times n_y \times n_z = 128 \times 62 \times 128$. The time-step is $dt = 10^{-5}$ and $\alpha=0.4$. Two different values for external pressure were simulated, $\Delta p = 0.5$ and $\Delta p = 1$. The aim was to demonstrate the capability of predicting failure. At $\Delta p =0.5$ the interface goes to equilibrium, when the value for pressure is doubled, the interface is no longer pinned and the interface fails and fills up the grooves. This process of failure is referred to as a depinned recession in \citep{xiang2017ultimate}.

\subsection{$3D$ Wavy Substrate}
\begin{figure}
 \begin{center}
  \includegraphics[scale=0.4]{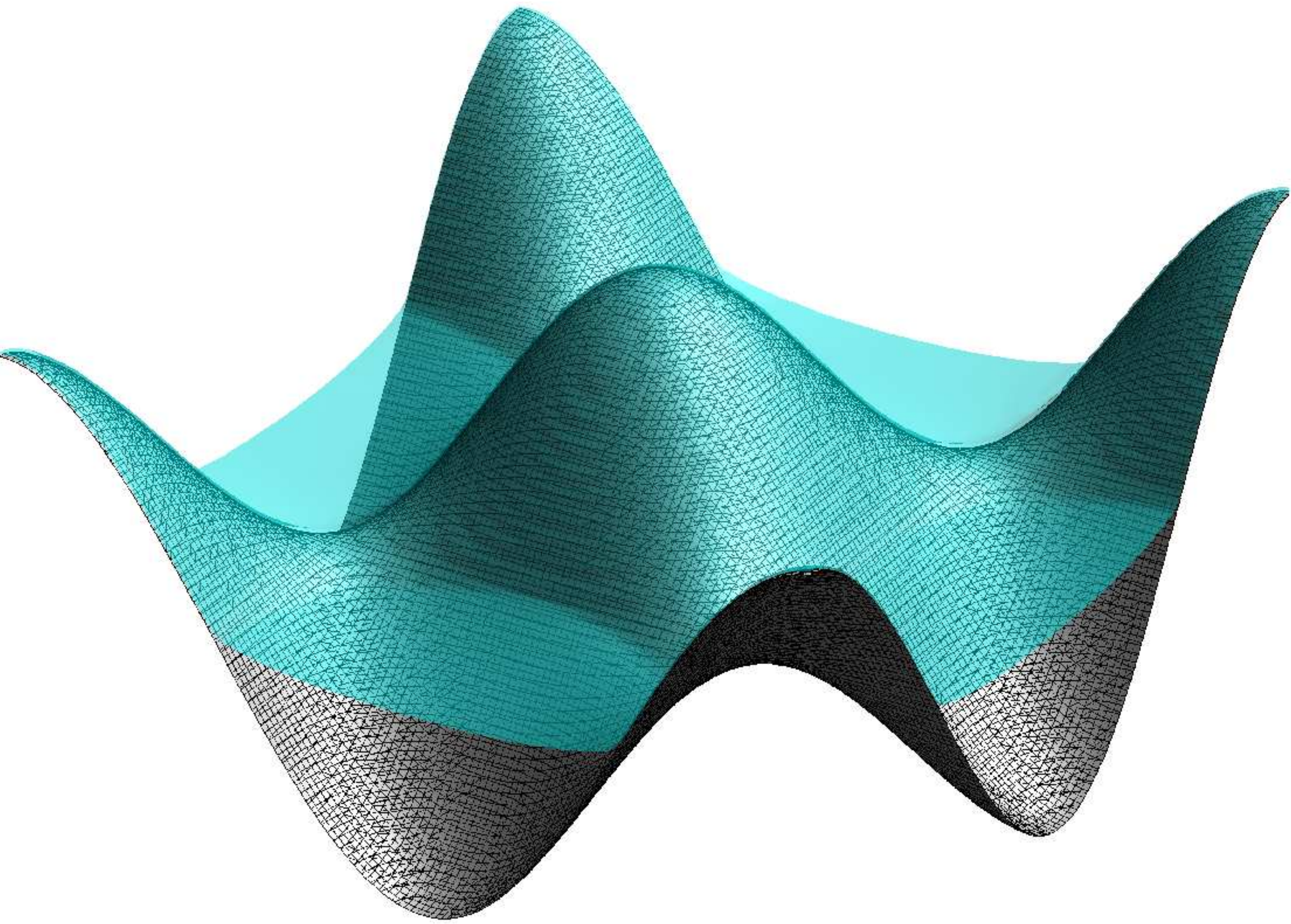}
 \end{center}
 \caption{Equilibrium interface location for a $3$D cosine substrate using a loading condition of $\Delta p = 0.2$. The gray isosurface represents the rigid substrate and the cyan transparent isosurface of the zero level set represents the interface.}
 \label{fig:gfe_3d_cosine}
\end{figure}
Consider a three-dimensional wavy substrate that is given by the following level set:
\begin{equation}
 \psi(\mathbf{x}) = (y-y_c) - h \cos \big(k_x (x-x_c)\big) \cos \big(k_z (z-z_c)\big),
\end{equation}
where the peak to valley height is $h=2 h_{cr}$ where $h_{cr} = - \tan \theta_Y$ and $\theta_Y = 140^{\circ}$. The wavelength in both direction is $2 \pi$ such that $k_x = 1$ and $k_z = 1$. The substrate is shifted by $\mathbf{x_c}=(0,h,0)$ so that the valley corresponds to the origin. The domain extents are $[L_x,L_y,L_z] = [2 \pi, 3, 2 \pi]$ and grid size $n_x \times n_y \times n_z = 128 \times 62 \times 128$. The time-step is $dt = 10^{-5}$ and $\alpha=0.4$. The interface is initialized at $y=2.5$ as a binary function, as the interface evolves and moves down towards the solid substrate, the Gibbs energy increases until it reaches steady state when the solution goes to an equilibrium position. The external pressure prescribed is $\Delta p = 0.2$, the equilibrium solution is consistent with the two-dimensional case presented in the validation section if a cut plane is taken at $z=0$. 

\subsection{Random Roughness}
\begin{figure}
 \begin{center}
  \includegraphics[scale=0.4]{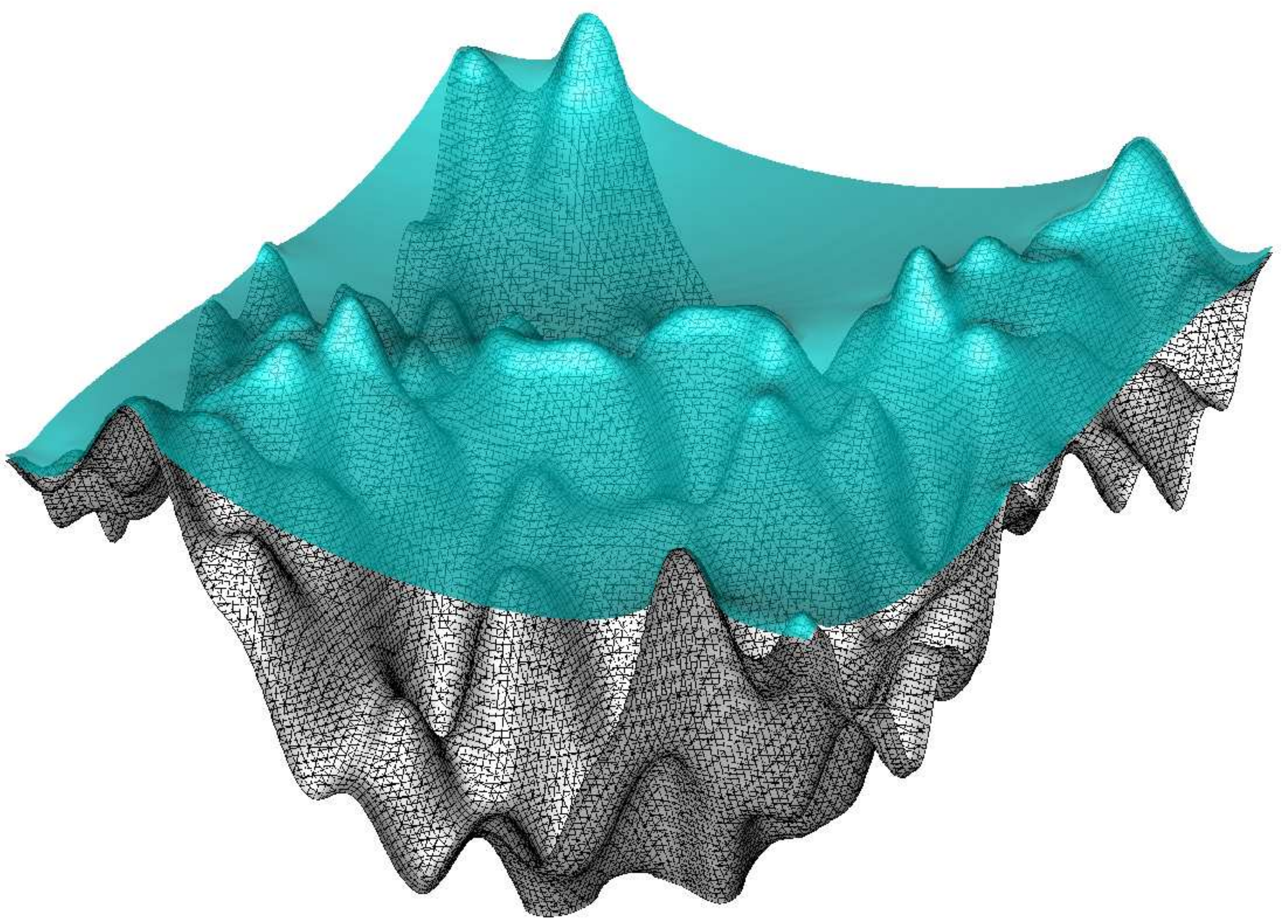}
 \end{center}
 \caption{Equilibrium interface location for a random rough substrate for a loading condition of $\Delta p = 0.2$. The gray isosurface represents the rigid substrate and the cyan transparent isosurface of the zero level set represents the interface.}
 \label{fig:gfe_random_roughness}
\end{figure}
A Fractal geometry is used to represent a random rough surface. The geometry is specified as a composition of many elementary waves in the form of $\cos ( \mathbf{k} \cdot \mathbf{x} + \varphi$). Define a range of amplitudes that tapers off based on a certain distribution:
\begin{equation}
 h_{mn} = \frac{1}{(m^2+n^2)^{\, \beta/2}},
\end{equation}
where $m$ and $n$ are the wavenumbers and $\beta$ the spectral exponent. A Gaussian distribution $g_{mn}$ is used to get a smooth random variation in its amplitudes such that  
\begin{equation}
 A_{mn} = g_{mn} h_{mn}.
\end{equation}
The phase angle is also sampled from a Gaussian distribution and is scaled such that it varies between $-\pi/2$ and $\pi/2$:
\begin{equation}
 \varphi_{mn} = \frac{\pi}{2} g_{mn}.
\end{equation}
The rough surface height distribution is then represented by a double sum over the wavenumbers in both spatial directions:
\begin{equation}
 h_f(\mathbf{x}) = \sum_{m,n} A_{mn} \cos( k_{mn} \cdot \mathbf{x} + \varphi_{mn} ),
\end{equation}
this is scaled such that the maximum height from peak to valley is $2 h_{cr}$. Then the level set that represents the solid is given by:
\begin{equation}
 \psi(\mathbf{x}) = (y-y_c) - h_f(\mathbf{x})
\end{equation}
For this simulation, $N=10$ for the spatial frequency resolution, and $\beta=1.5$ which represents a fairly rugged surface. The domain extents are $[L_x,L_y,L_z] = [2 \pi, 3, 2 \pi]$ and grid size $n_x \times n_y \times n_z = 128 \times 62 \times 128$. The time-step is $dt = 10^{-5}$ and $\alpha=0.4$. The external pressure is $\Delta p = 0.2$. Note that the algorithm reaches an equilibrium solution and is able to handle rough surfaces that have rapid variations in their structure.

\subsection{Drop Shapes}

\begin{figure}
  \begin{center}
  \includegraphics[scale=0.4]{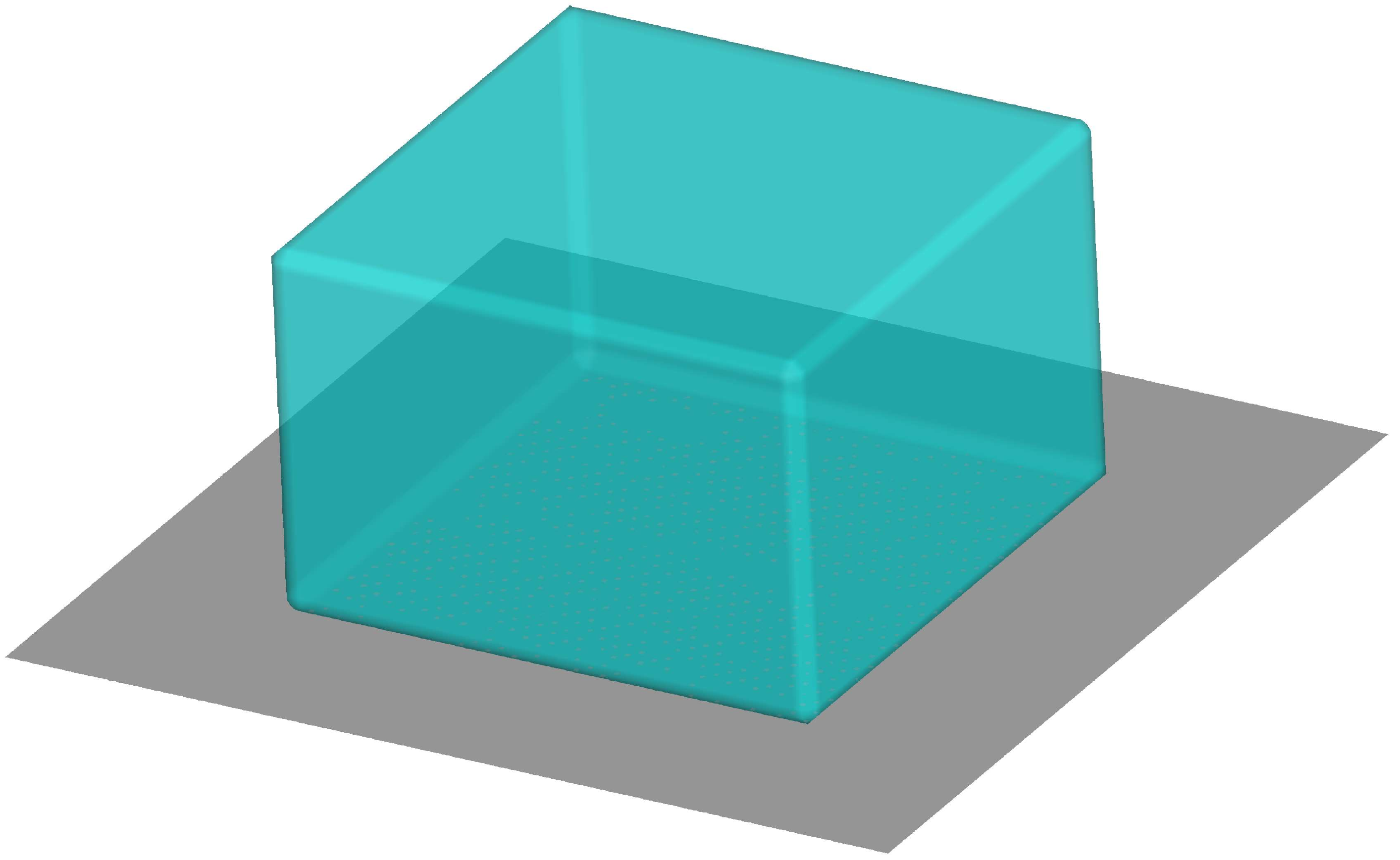} 
  \includegraphics[scale=0.4]{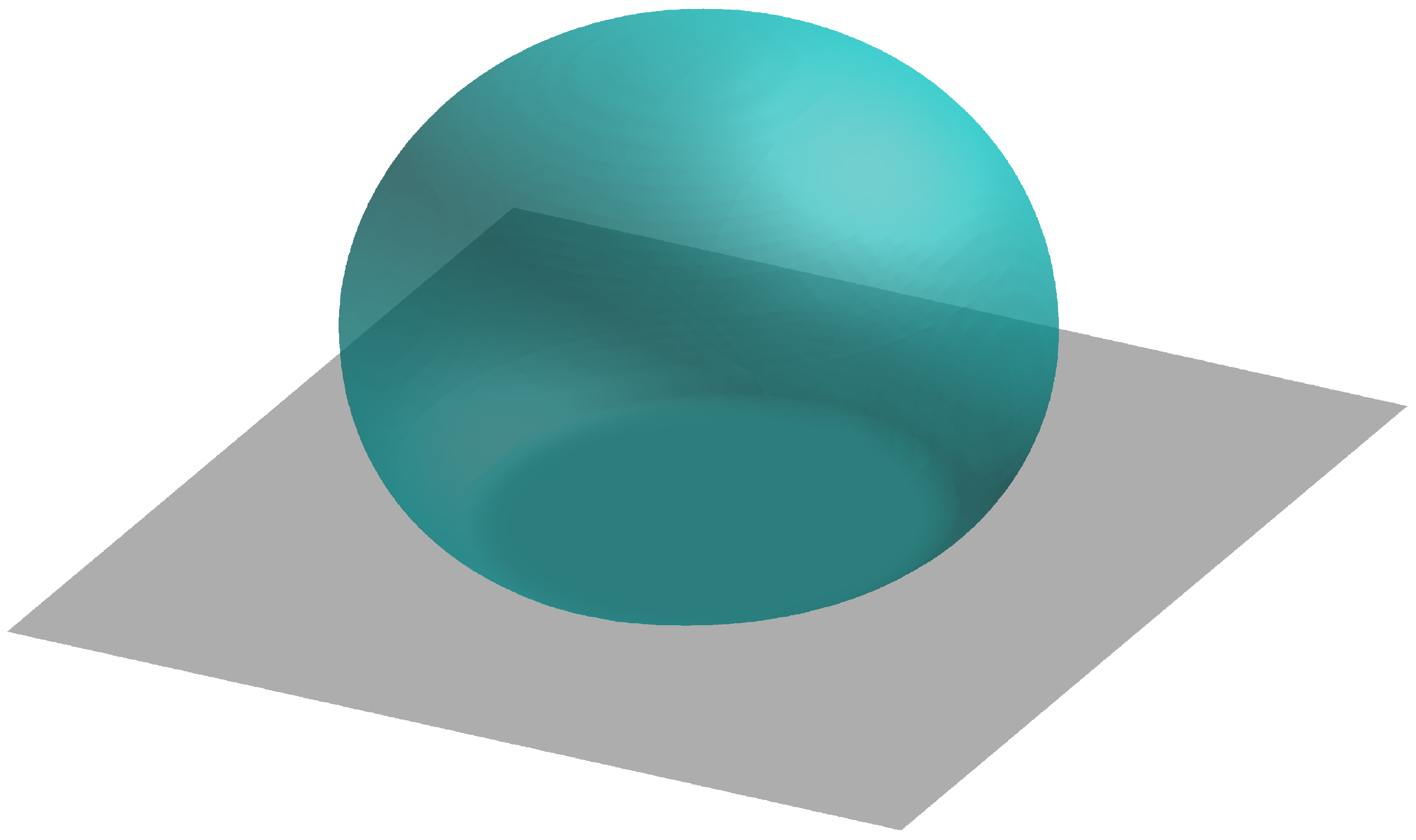} \\
 \end{center}
 \caption{Drop shape at equilibrium for a flat surface. Top shows the binary cuboid initial condition. Bottom shows the final equilibrium solution for the drop. Surface tension values used are $\tau^-=0.6$ and $\tau^+=0.5$ in non-wetted conditions.}
 \label{fig:bubble_flat}
 \end{figure}

\begin{figure}
   \begin{center}
   \includegraphics[scale=0.4]{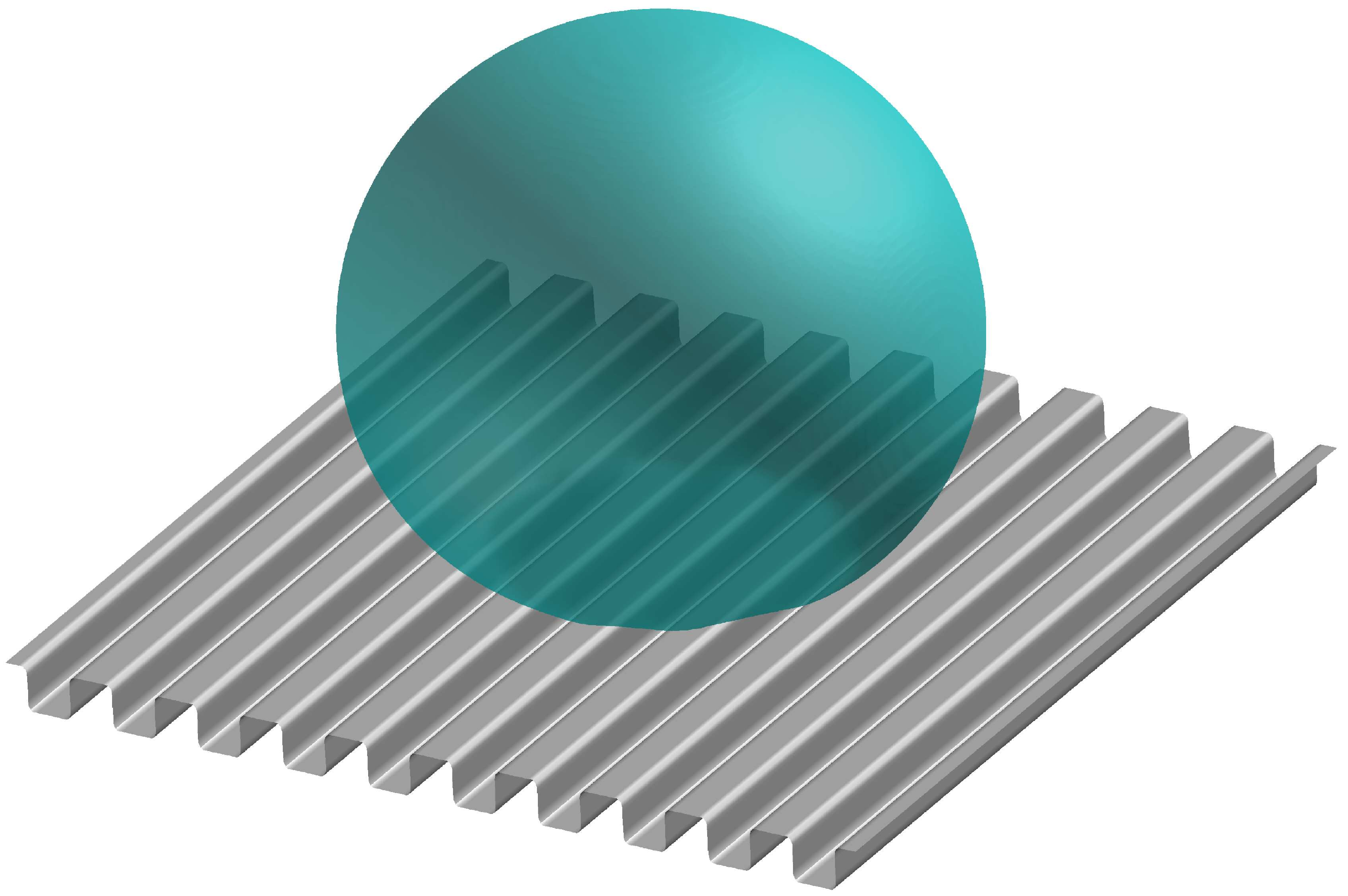} 
 \end{center}
 \caption{Drop shape at equilibrium for longitudinal grooves. Here $h=0.025$, $k_x = 20\pi$ and $x_c=(0,0.15,0)$. Figure shows the final equilibrium position for $\tau^-=0.6$ and $\tau^+=0.5$ in non-wetted conditions.}
 \label{fig:bubble_grooves}
 \end{figure}

\begin{figure}
   \begin{center}
   \includegraphics[scale=0.4]{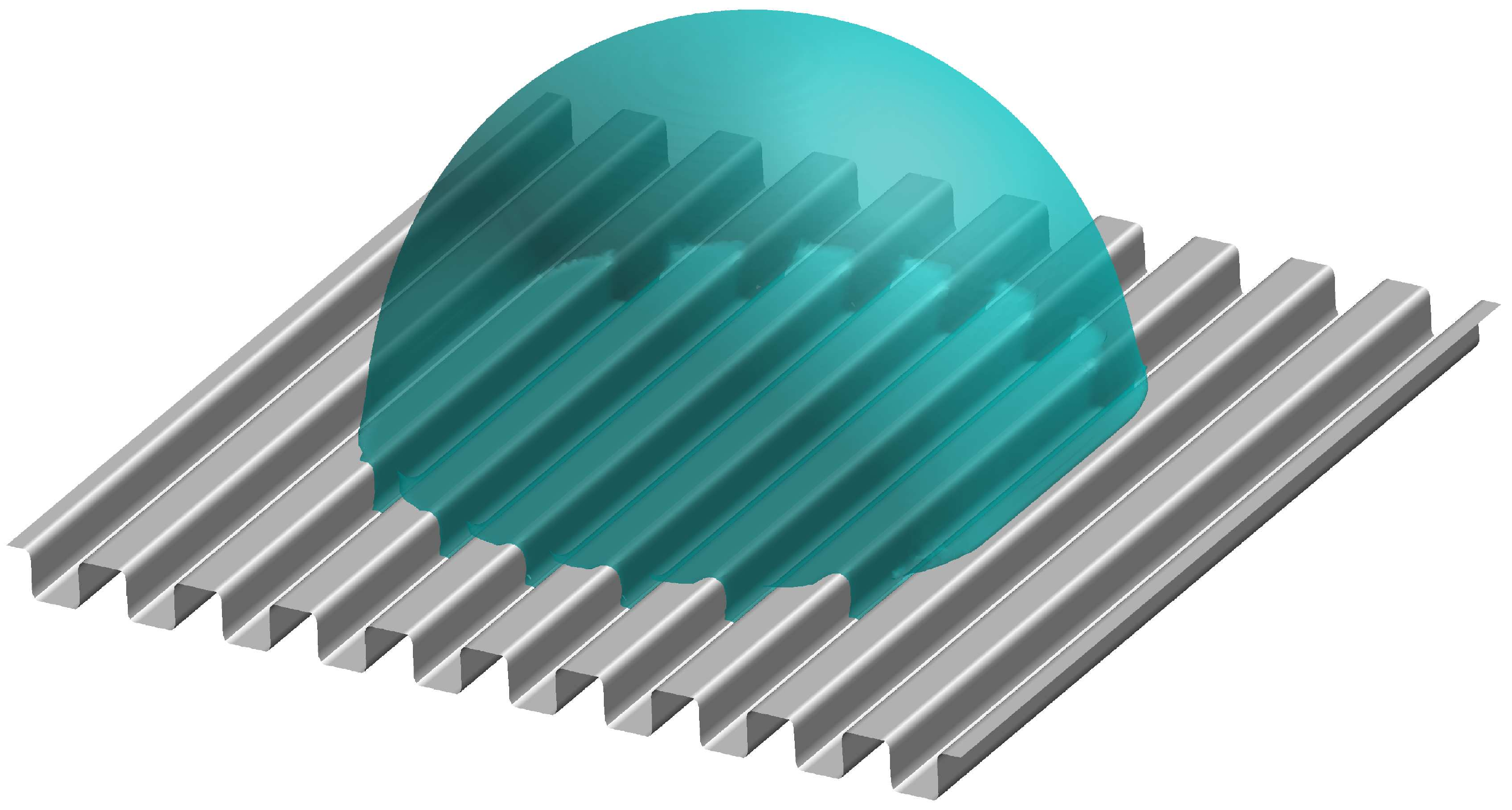} \\
 \end{center}
 \caption{Drop shape at equilibrium for longitudinal grooves. Here $h=0.025$, $k_x = 20\pi$ and $x_c=(0,0.15,0)$. Figure shows the final equilibrium position for $\tau^-=0.6$ and $\tau^+=10^{-3}$ to represent a Wenzel state where the liquid fills the grooves from the initial condition. Note the anisotropy in the drop shape due to wetting.}
 \label{fig:bubble_grooves_wenzel}
 \end{figure}

\begin{figure}
  \begin{center}
  \includegraphics[scale=0.4]{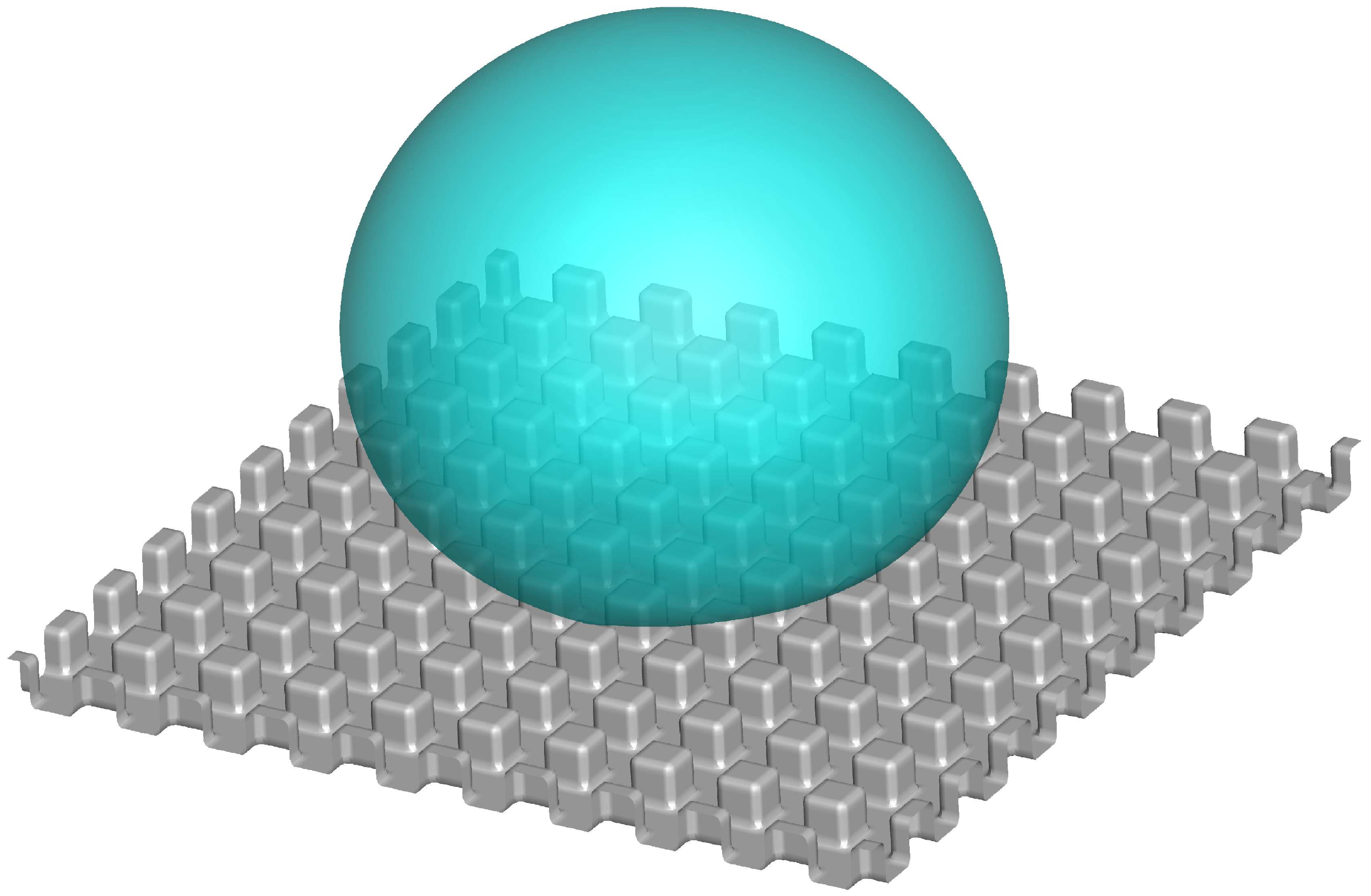}
   \end{center}
  \caption{Drop shape at equilibrium over posts. The geometrical parameters are the same as the longitudinal grooves but is applied in both planes with $\tau^-=1$ and $\tau^+=0.833$ in non-wetted conditions.}
 \label{fig:bubble_post}
 \end{figure}

\begin{figure}
 \begin{center}
  \includegraphics[scale=0.4]{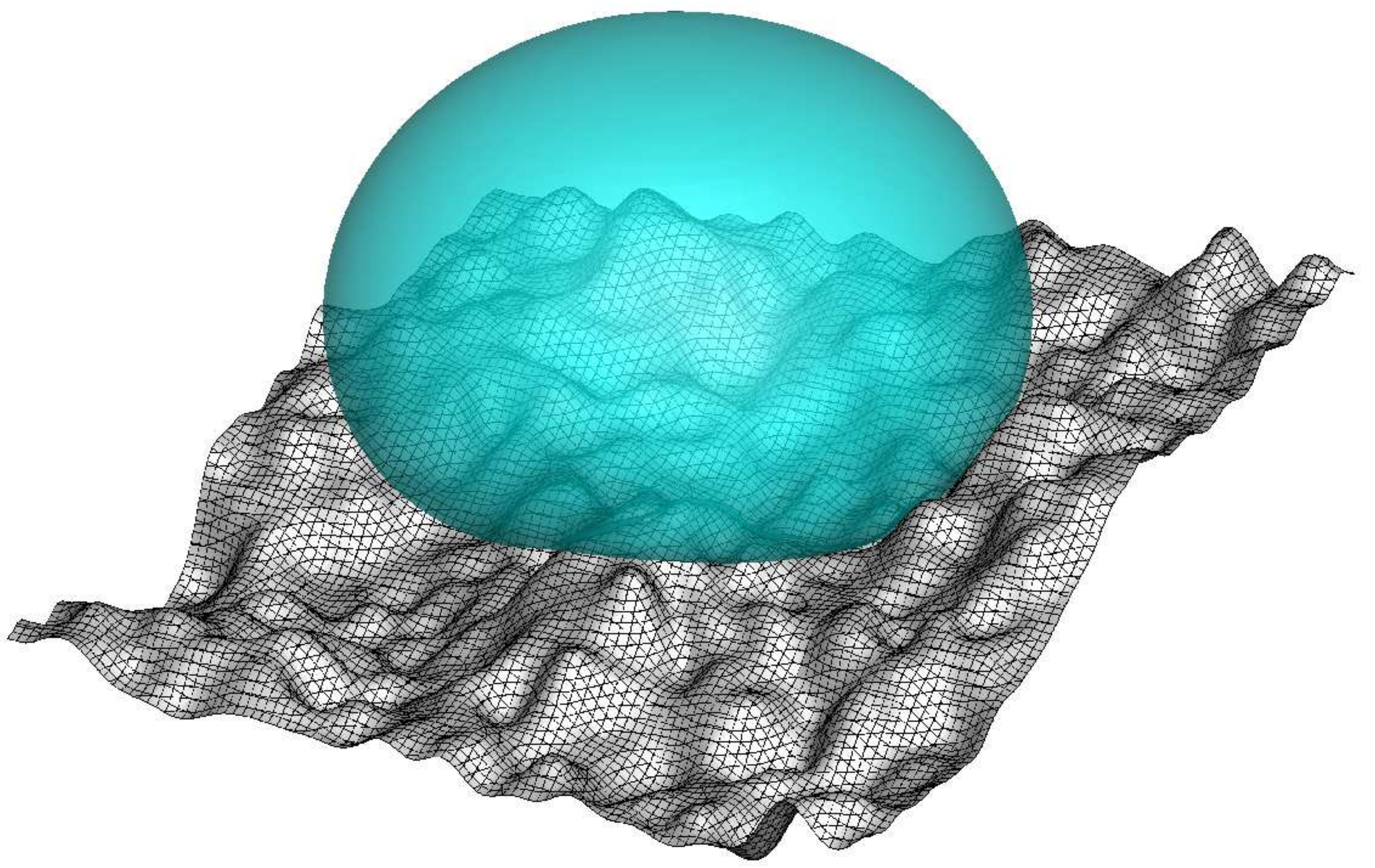}
 \end{center}
 \caption{Drop shape at equilibrium over random rough surface. Spatial frequency resolution $N=96$, $\beta=1.5$ and the mean height $h_{cr}=0.15$ with $\tau^-=0.6$ and $\tau^+=0.5$ in non-wetted conditions. Note the loss in spherical symmetry in the bubble shape.}
 \label{fig:bubble_roughwall}
\end{figure}

The drop shape was also investigated in the context of the macroscopic formulation of the equilibrium solution. The bubble is initialized as a binary cuboid over a flat surface, longitudinal grooves, posts, and random rough surface using the methods described in the previous section. The surface parameters are described in the captions of the corresponding figures. All simulations were run using $dt = 10^{-7}$, with domain extents of $[L_x,L_y,L_z] = [1, 1, 1]$ and grid size $n_x \times n_y \times n_z = 200 \times 200 \times 200$. Note that here the external pressure was substituted for $h(\mathbf{x}) \cdot \mathbf{g}$ where $\mathbf{g}$ is the non-dimensional gravitational constant taken to be unity in the wall-normal direction. When initializing a binary cuboid on rough surfaces, some regions might overlap between the interface and solid, therefore it is best to define $\phi(\mathbf{x},0)=\phi(\mathbf{x},0) \cup (-\psi(\mathbf{x}) -\epsilon)$. The key take-away point here is the ability of the algorithm to capture the different locus of contact points for different geometries as observed in figures (\ref{fig:bubble_flat}-\ref{fig:bubble_roughwall}), along with the anisotropy in the shape of the bubble as observed in figure \ref{fig:bubble_grooves_wenzel} for the Wenzel state.

\section{Summary} \label{sec:summary}
A variational level set methodology is developed in the framework of minimizing Gibbs free energy. The level set method avoids reinitialization by using a penalty term that regularizes the distance function. The method is able to reproduce canonical level set evolution cases such as motion in the normal direction and curvature-driven flow. The Gibbs free energy model is general in that it can incorporate an external pressure field with a dissolved gas concentration parameter, and a solid surface that can represent a wide range of geometries. The method is simple to implement, robust and parallel. The surface tension model uses a harmonic mean distribution modified to give the desired value of surface tension at the interface location. The modified harmonic mean distribution gave more accurate results when compared to other distributions. The algorithm is validated with canonical level set evolution test cases, the Young-Laplace equation and an analytic solution over a $2$D wavy substrate. Good agreement is observed. Several geometries that include $3$D longitudinal grooves, posts, a wavy substrate, and random roughness using fractal geometry were investigated and the algorithm is shown to handle complicated geometry. Simple solid surfaces like flat walls and grooves can be represented as a binary level set functions whereas more complicated geometry require smooth level sets. The effect of Lagrange multipliers is also explored; integrating the constraints over the entire domain is more robust as opposed to a local integration. Failure to enforce the no-penetration constraint causes the interface to leak into the solid wall. If the volume conservation constraint is violated, the drops behave like a curvature-driven flow; the interface shrinks/expands until it is completely out of the domain. In the context of microscopic problems, gravity effects can be ignored and an external pressure field can be applied to obtain an equilibrium meniscus shape over a wide range of rough surfaces. For macroscopic problems, the effects of gravity can be included. Liquid drops are able to reach an equilibrium shape while conserving volume. Different surfaces can lead to different equilibrium positions. We show that the algorithm is capable of capturing contact lines accurately and also the anisotropy in liquid drop shapes. The proposed algorithm provides a useful tool for studying properties of superhydrophobic surfaces and the inception of nucleation sites in the context of cavitation.

\section*{Acknowledgment}
This work is supported by the United States Office of Naval Research (ONR) under ONR Grants N00014-12-1-0874 and N00014-17-1-2676 with Dr. Ki-Han Kim as technical monitor. The computations were made possible through  the Minnesota Supercomputing Institute (MSI) at the University of Minnesota.

\bibliographystyle{model1-num-names}
\bibliography{submit}

\end{document}